%% file: sb9hermes.tex
\documentclass{aa}

\usepackage{natbib}
\usepackage{graphicx}
\usepackage{txfonts}
\usepackage[colorlinks, citecolor=blue, breaklinks=true]{hyperref}
\usepackage{longtable}
\usepackage{siunitx}
\usepackage[normalem]{ulem}

\bibpunct{(}{)}{;}{a}{}{,} 
\def\Msun{M$_\odot$}
\def\kms{km~s$^{-1}$}
\newcommand{\Gaia}{\emph{Gaia}\xspace}

\begin{document}
\title{An update of SB9 orbits using HERMES/Mercator radial velocities\thanks{Based on observations carried out with the Flemish Mercator Telescope at the Spanish Observatorio del Roque de los Muchachos (La Palma, Spain).}
}
\author{
T.~Merle\inst{1,2}
\and D.~Pourbaix\inst{1} \fnmsep\thanks{Deceased, Nov. 14, 2021}\,
\and A.~Jorissen\inst{1}
\and C.~Siopis\inst{1}
\and S.~Van Eck\inst{1}
\and H. Van Winckel\inst{3}
}
\institute{Institut d'Astronomie et d'Astrophysique, Universit\'e Libre de Bruxelles (ULB), Belgium\\
\email{tmerle@ulb.ac.be}
\and Royal Observatory of Belgium, Avenue Circulaire 3, 1180 Brussels, Belgium
\and Instituut voor Sterrenkunde, Katholieke Universiteit Leuven, Belgium
}
\date{Received month, day, year; accepted month, day, year}

\titlerunning{SB9 orbits with HERMES RV}
\authorrunning{T. Merle et al.}

\abstract{}
{The \Gaia mission is delivering a large number of astrometric orbits for binary stars. By combining these with spectroscopic orbits for systems with two observable spectra (SB2), it is possible to derive the masses of both components. However, to get masses with a good accuracy requires accurate spectroscopic orbits, which is the major aim of the present paper.  A  subsidiary aim 
is to discover SB2 systems hiding among known SB1, and even though this search may often prove unsuccessful, the acquired radial velocities may be used anyway to improve the existing spectroscopic orbits.}
{New radial velocities for 58 binary systems from the {\it Ninth Catalogue of Spectroscopic Binary Orbits} (SB9), obtained using the high-resolution  HERMES spectrograph installed on the 1.2 m Mercator telescope, were used to possibly identify hitherto undetected SB2 systems. For SB1 systems with inaccurate orbits, these new radial-velocity measurements were used to improve the orbital accuracy.}
{This paper provides 51 orbits (41 SB1 and 10 SB2) that have been improved with respect to the solution listed in the SB9 catalogue, out of the 58 SB9 orbits studied, which belong to 56 stellar systems. Among them, there are five triple and  four quadruple systems.  Despite the high resolution of HERMES, HIP~115142~A is the only system which we detected as a new SB2 system. The B component of the visual binary HIP~92726 has now been found to be a spectroscopic system as well, which makes HIP~92726 a newly discovered quadruple system (SB1+SB1). The high HERMES resolution allowed us moreover to better isolate the signature of the secondary component of HIP~12390, HIP~73182 and HIP~111170. More accurate masses have thus been derived for them. Among the 30 SB also present in \Gaia Data Release 3 (DR3) and with periods shorter than the \Gaia DR3 time span ($\sim 1000$~d), only 5 were flagged as binaries by DR3. Various DR3 selection criteria are responsible for this situation.}
{}

\keywords{(stars:) binaries: spectroscopic -- (stars:) binaries (including multiple): close -- techniques: spectroscopic -- techniques: radial velocities -- Stars: fundamental parameters}

\maketitle

\begin{table*}
\centering
\scriptsize
\caption{Programme stars and the behaviour of their old and new radial velocities with respect to the reference orbital solution, as listed in column `Reference orbit'.  The column labelled `DR3' lists the type of binary solution provided by \Gaia DR3 \citep[][`astrom' stands for `astrometric orbital solution', `Acc7' and `Acc9' for acceleration models with 7 and 9 parameters]{Arenou2022}. An hyphen in column DR3 means that there is no DR3 entry for the corresponding SB9 system, whereas 'noRV' means that no RV is available from \Gaia DR3, 'brightG' that $G_{\rm RVS} < 5.5$ (see details in Sect.~\ref{Sect:GaiaDR3}), 'nRV$<10$' that the number of RV was not large enough for DR3 to look for a spectroscopic orbit. The column labelled `Mask' lists the mask adopted for the cross correlation: Bal stands for Balmer, F0ROT for a rotationally-broadened F0 spectrum, G2 for solar-type, Arc for Arcturus, and M4 for a M4V spectral template. $\langle$O$-$C$\rangle$ (new) and $\sigma (O-C)$ (new) are the mean and standard deviation of the residuals of the Hermes RV with respect to the reference orbit.  $\sigma (O-C)$ (old) is the standard deviation of the original residuals against the reference orbit [when listed by the author(s)]. The column labelled SB1/2/c specifies whether or not the star has been detected as SB2 by either the reference orbit or by the HERMES measurements. The label 'SBc' in that column indicates that the HERMES spectra reveal the composite nature of the system, and that RV from the two components could be derived by focusing at the violet part of the spectrum (revealing the Balmer lines in the warm component) or at the yellow-red part of the spectrum (revealing lines from the cool component).
The flag `R' in column Remark means that a revised orbit combining old and new (HERMES) measurements has been computed. The revised orbit and the corresponding $\sigma (O-C)$ are then listed in Table~\protect\ref{tab:revorbits_old_new}. Appendix~\ref{sect:Star} provides detailed explanations as to why a revised orbit is not published for systems without flag `R'.  The column labelled $\Delta RV$ lists the offset $RV_{\mathrm {HERMES}} - RV_{\mathrm {SB9}}$ (see Sect.~\ref{sect:Observations}) between the SB9 and HERMES RV zero-points, along with its uncertainty.} 
\label{tab:ProgrammeStars}
\renewcommand{\tabcolsep}{3pt}
\begin{tabular}{rrccccccccccccc}
  \hline\hline
HIP & SB9 & Sp & DR3 & SB1/2/c & Mask & $N$ & $N$ & $\langle$O$-$C$\rangle$ & $\sigma (O-C)$ & $\sigma (O-C)$ & Reference  & Remark & Old  & $\Delta RV$\\
    &     &    &         &      & &(new)&(old)&  (new)  &    (new)       &    (old) & orbit & & system & (km/s)\\ 
    \hline
\input{tab/table1data}
\end{tabular}
\tablebib{
(1) \citet{2009A&A...507..541G}; (2) \citet{Harper-1926}; (3) \citet{2000A&AS..145..215P};
(4) \citet{Griffin-1980}; (5) \citet{Halbwachs-2016}; (6) \citet{2000A&A...355.1015D};
(7) \citet{1988AJ.....96.1040P}; (8) \citet{Abt-1965:a}; (9) \citet{2006MNRAS.371.1159G};
(10) \citet{1995AJ....109..326W}; (11) \citet{1998A&AS..131...43U}; (12) \citet{2007A&A...473..829M};
(13) \citet{2005Obs...125...81G}; (14) \citet{2003MNRAS.342.1271C}; (15) \citet{1999A&A...347..164W};
(16) \citet{2008AN....329...44C}; (17) \citet{1998A&AS..131...25U}; (18) \citet{Massarotti-2008};
(19) \citet{2007AN....328..527C}; (20) \citet{2004Obs...124...97G}; (21) \citet{2001AJ....122.3419C};
(22) \citet{1928AnCap..10....8S}; (23) \citet{2006AJ....131.1022T}; (24) \citet{1995ApJ...452..870T};
(25) \citet{2010Obs...130..125G}; (26) \citet{1969MNRAS.142..543H}; (27) \citet{2007ApJ...659..626Z};
(28) \citet{1999A&A...351..619F}; (29) \citet{2013AJ....145...41K}; (30) \citet{1992A&A...254L..13D};
(31) \citet{1977Obs....97..173R}; (32) \citet{1992MNRAS.256..575S}; (33) \citet{2002AstL...28..393S}; (34) \citet{1952ApJ...116..383F}; 
}
\end{table*}

\addtocounter{table}{-1}
\begin{table*}
\caption{Continued.}  
\tablebib{
(35) \citet{1931PAAS....6..278S}; (36) \citet{2016AJ....152...46W};
(37) \citet{McClure-1997:a}; (38) \citet{2009A&A...498..627F}; (39) \citet{2009Obs...129..147G};
(40) \citet{2009Obs...129..147G}.
}
\tablefoot{
\tablefoottext{a}{It is unclear whether that spectral combination corresponds to the pair AB or Aa,Ab (see text)}
\tablefoottext{b}{Composite nature of this spectrum not confirmed by HERMES (see Sect.~\ref{sect:Star})} \tablefoottext{c}{Not seen as SB2 by HERMES} \tablefoottext{d}{Component spectral type derived from 0.6~M$_{\odot}$ mass from \citet{2022AJ....163..118A}} \tablefoottext{e}{The violet Balmer lines from the hot component are too wide to derive a precise RV from the HERMES spectra} \tablefoottext{f}{The orbit listed in Table~\ref{tab:revorbits_old_new} is computed from HERMES RV only.} \tablefoottext{g}{Based on lines from the cool component.} \tablefoottext{h}{Although HIP 115142B is not an entry of the SIMBAD database, it is used here for convenience. Use HD 219877B in SIMBAD instead.}
\tablefoottext{i}{See HIP~91751 in Appendix~\ref{sect:Star}}
}
\end{table*}

\section{Introduction}\label{sect:Intro}
In December 2013, the European Space Agency launched \Gaia \citep{Lindegren-2008}, which  accurately and repeatedly observes about two billion stars, including a substantial fraction of binaries.  
About 508\,000 astrometric binaries and 277\,000 spectroscopic binaries (with one -- SB1 -- or two -- SB2 -- observable spectra) were already provided in the \Gaia third data release \citep[hereafter \Gaia DR3;][]{Arenou2022} including binaries with partial solution (radial-velocity trends and astrometric acceleration solutions), and many more are expected in \Gaia DR4.

In order to provide valuable constraints for any cutting-edge study involving stellar evolution, stellar masses should be known to better than a few percent \citep[\emph{e.g.},][]{Andersen-1991:a}.  Whereas such an accuracy is now routinely reachable with eclipsing binaries \citep[\emph{e.g.},][]{Torres-2010, 2021A&ARv..29....4S}, non-eclipsing systems seldom yield such a high accuracy.  Indeed, their double-lined spectroscopic (SB2) orbit has to be complemented with an interferometry- or astrometry-derived orbital inclination. The astrometric orbits provided by the \Gaia data releases now offer access to the orbital inclinations, as was the case for the systems with an astrometric orbit derived from Hipparcos data \citep{Hipparcos,Lindegren-1997:a,Soderhjelm-1999:b}.   
Anticipating the availability of these inclinations, a team led by Halbwachs \citep{Halbwachs-2009,Halbwachs-2014:a,Halbwachs2017} has been seeking for several years a possible spectral signature of the secondary component of known single-lined spectroscopic binaries (SB1) from the {\it Ninth Catalogue of Spectroscopic Binary Orbits} \citep[hereafter SB9\footnote{http://sb9.astro.ulb.ac.be}, version 2021-03-02]{2004A&A...424..727P}.  As a result, some of these SB1 have already been changed into SB2. Furthermore, irrespective of the number of detected components, the orbits resulting from this observational effort are often better than the original ones \citep{Halbwachs-2011,Halbwachs-2012,Halbwachs-2013,Halbwachs-2014:a,Halbwachs-2016,Halbwachs2017}. 
Indeed, among systems currently listed in the SB9, few have orbits that are accurate enough to yield stellar masses to better than a few percent. \cite{Fekel-2004} have already shown the benefit of such a long-term endeavour, by re-observing some known systems to improve their orbit.
In addition, numerous works \cite[\emph{e.g.},][]{2009AJ....137.3442S,2010EAS....45..425R, 2013ARA&A..51..269D,2017ApJS..230...15M,2020A&A...635A.155M,2022AJ....163..118A,2022AJ....163..220V} are based on the SB9 catalogue serving as a reference,  thus underlining the usefulness of improved spectroscopic 
 orbits, in particular because many SB9 orbits come from the first generation of spectrographs that were not so precise.
Moreover, SB9 orbits may serve as benchmarks to which \Gaia spectroscopic orbits may be compared \citep[][]{Arenou2022}. 

Here we report on an effort to  discover SB2 systems hiding among known SB1. Even though this search may often turn out to be unsuccessful, the acquired radial velocities may be used anyway to improve SB9 orbits. The HERMES spectrograph \citep{Raskin-2011} attached to the Mercator 1.2-m telescope located at the {\it  Observatorio del Roque de los Muchachos}, on the island of La Palma (Canary Islands), was thus used to obtain radial velocities (RV) of 58 SB9 systems, with the aim of
\begin{itemize}
    \item adding  more accurate HERMES RV to the set of older RV used to compute the SB9 orbit, in order to improve its accuracy;
    \item finding new SB2 systems;
    \item computing combined visual (or astrometric) + spectroscopic orbits whenever possible.
\end{itemize}

The paper is organized as follows. The observations are presented in Sect.~\ref{sect:Observations}. The most interesting targets, for which a new or more accurate SB2 spectroscopic orbit could be combined with a visual orbit, are discussed in Sect.~\ref{Sect:New_SB2_systems}, and the masses so obtained  are presented in Sect.~\ref{Sect:masses}.  The revised SB1 orbits are presented in Sect.~\ref{Sect:SB1}, and the combined spectroscopic/astrometric (Hipparcos) orbits in Sect.~\ref{sect:Hipparcos}, with a discussion on individual systems in Appendix~\ref{sect:Star}. In the absence of astrometric epoch data from \Gaia, the use of Hipparcos astrometric data is still useful, in combination with our improved spectroscopic orbits \citep[see also][]{Leclerc-2023}. Some of the reasons why not all binaries from this paper are included in the Gaia DR3 binary tables are discussed in Sect.~\ref{Sect:GaiaDR3}.

\section{Observations and orbit computations}\label{sect:Observations}

The programme stars were selected on  the same criteria as for the original sample observed at the {\it Observatoire de Haute Provence} \citep[OHP;][]{Halbwachs-2009,Halbwachs-2014:a}, adjusting for the limiting declination of La Palma ($\delta \ge -30^{\circ}$), resulting in a sample of 58 objects (slightly overlapping with the OHP sample; Table~\ref{tab:ProgrammeStars}).

The radial velocities (RV) were measured in 2014 -- 2016 with the HERMES spectrograph \citep{Raskin-2011} attached to the Mercator 1.2-m telescope. HERMES RV are tied to the IAU system as defined by \citet{1999ASPC..185..367U}. The precision on a long-term data set is about $\epsilon = 55$~m\,s$^{-1}$, as derived from a 5-year-long series of observations of radial-velocity standards \citep[see, \emph{e.g.},][]{2016A&A...586A.158J}, except for A stars whose RV is derived from the Balmer H$\alpha$ line. For F, G, K, and M stars, appropriate masks are available, 
as indicated in column `Mask' in Table~\ref{tab:ProgrammeStars}. 

The column `$\sigma$(O$-$C) (old)' lists the standard deviation of the `Observed minus Calculated' (O$-$C) residuals (retrieved from SB9) against the reference orbit listed in column `Reference orbit'. That reference orbit is the most recent orbit from SB9. In that respect, we must note that more recent orbits than those listed in Table~\ref{tab:ProgrammeStars} are available for HIP~95176 \citep{2006A&A...459..849K} and HIP~111170, HIP~115126 \citep{2013AJ....145...41K}. They were however not used in the table because HIP~95176 is a complex system  (as discussed in Appendix~\ref{sect:Star}) for which the systemic velocity  is not easy to define (\citealt{2006A&A...459..849K} provide many), and \citet{2013AJ....145...41K} used an iodine cell which provides relative velocities not easily tied to an absolute reference system.

In Table~\ref{tab:ProgrammeStars}, $\langle$O$-$C$\rangle$ (new) and `$\sigma (O-C)$ (new)' are the mean and standard deviation of the O$-$C residuals of the Hermes RV with respect to the reference orbit.  
These two quantities were used to estimate the necessity of revising the existing orbit. If $\sigma (O-C)$ (new) $> 2 \sigma (O-C)$ (old) [and assuming of course $\sigma (O-C)$ (new) $> \epsilon$], it is an indication that the old orbit does not correctly fit the new HERMES RV and that a new revised orbit is mandatory. This may be caused by errors building up over time from a too uncertain orbital period (an orbit combining the old and new HERMES RV could then usefully improve the orbital period -- as will be shown below in Table~\ref{tab:revorbits_old_new}), or by an offset between the zero-points of the spectrographs. For instance, the orbits from \cite{1998A&AS..131...25U} were still in the old radial velocity system tied to Coravel. 

To handle these situations, the code of \citet{Pourbaix-1998:a} used to derive the orbits presented in this paper has been adapted to allow the possibility to compute, along with the orbital elements, a possible RV offset between different instrumental systems.\footnote{In practice, it was never necessary to separate the RV data set in more than two distinct instrumental groups (namely the SB9 RV and the HERMES RV), because the SB9 set was found to be already internally consistent. This may be assessed {\it a posteriori} by the fact that the global average {\it O-C} residuals is consistent with 0 (see the insert in the lower panels of Fig.~\ref{fig:revorbits1}).} When dealing with SB2 systems, the code is applied in two steps. First, the offset between instrumental systems is derived by processing (as an SB1) the component having the largest set of RV. The RV are then brought to a common zero-point (\emph{i.e.,} the HERMES/IAU RV system) by applying the offset derived in the first step, and the two SB2 components treated simultaneously to derive a possible offset between their $\gamma$-velocities \citep[in the case, \emph{e.g.}, of a substantial difference between their surface gravitational relativistic acceleration; see][]{1999A&A...344..172P}. The uncertainties on the RV for SB2 components could often be underestimated, especially in the case where the two velocity peaks are not well separated and the uncertainties difficult to estimate. In such cases, the code had difficulties to converge, and it was then necessary to empirically increase the uncertainties  (this was the case for HIP~22000 for instance). 

The zero point offsets between instrumental systems minimizing the orbital residuals are listed in Table~\ref{tab:ProgrammeStars}. Even though one would expect that a single value should characterize the offset between a given pair of instrumental systems\footnote{For instance, HERMES $-$ CfA = +0.139~\kms\
\citep[as given by \emph{e.g.},][]{Massarotti-2008}, and HERMES $-$ Cambridge CORAVEL = $-0.8$~\kms, as derived from \citet{2006MNRAS.371.1140G}.}, Table~\ref{tab:ProgrammeStars} reveals that these offsets vary from star to star. There are several possible causes for these star-to-star variations.
First, when HERMES adds only a couple of data points to former (\emph{e.g.}, CORAVEL) observations  (like for HIP~34935, 38217, 43041, 53717...), the shift between the HERMES and CORAVEL systems is not accurately constrained, as revealed by its large uncertainty. Second, the zero-point correction is likely to depend upon both the stellar colour and velocity, as shown by \citet{1999ASPC..185..367U}. Despite the fact that this offset parameter is often not robustly determined (its uncertainty being as large as its value), it has nevertheless been derived in all cases for the sake of homogeneity of the numerical processing of all target systems. As a drawback, the uncertainty on the zero-point offset gets reflected on the uncertainty of the centre-of-mass velocity.

The full set of HERMES RV are available at the Centre de Donn\'ees Stellaires (CDS, Strasbourg), with Table~\ref{Tab:RVs} presenting the first 25 lines as an example of its content. The old RV used to compute the former SB9 reference orbit are directly available from the SB9 database,  where the new HERMES RV will be listed as well, along with the newly derived orbit. 

\begin{table}[]
\footnotesize
    \centering
        \caption{ The first 25 lines of the list of HERMES RV. The column labelled `comp.' provides the system component (a or b) in the case of SB2 systems. The full table is available at CDS, Strasbourg and in the SB9 database (http://sb9.astro.ulb.ac.be).  }
    \begin{tabular}{crrlc}
    \hline
HIP & HJD$-2\;400\;000$ & RV    & \multicolumn{1}{l}{$\pm\epsilon$}  & comp. \\
     &                  &\multicolumn{2}{c}{\phantom{RV}(km/s)}\\
    \hline
HIP000183 & 56878.6448  &   13.0  &  1.1   & a\\
HIP000183 & 56907.5977  &   11.8  &  1.0   & a\\
HIP000183 & 56934.5421  &    9.2  &  1.1   & a\\
HIP000183 & 56954.5527  &    8.5  &  1.0   & a\\
HIP000183 & 56995.3527  &    7.4  &  1.2   & a\\
HIP000183 & 57221.7092  &   $-3.8$  &  1.0   & a\\
HIP000183 & 57336.3940  &   $-4.4$  &  1.0   & a\\
HIP000183 & 57595.7256  &   $-4.0$  &  1.7   & a\\ 
HIP000443 & 55030.7265  &  $-19.174$&  0.050 & a\\
HIP000443 & 55034.6805  &  $-19.418$&  0.050 & a\\
HIP000443 & 55037.6589  &  $-18.973$&  0.050 & a\\
HIP000443 & 55056.5959  &    4.233&  0.050 & a\\
HIP000443 & 55079.6424  &   $-2.730$&  0.050 & a\\
HIP000443 & 55100.5538  &  $-18.513$&  0.050 & a\\
HIP000443 & 55204.3550  &    7.752&  0.050 & a\\
HIP000443 & 55216.3635  &    8.964&  0.050 & a\\
HIP000443 & 55398.7366  &  $-19.395$&  0.050 & a\\
HIP000443 & 55416.7197  &   $-5.075$&  0.050 & a\\
HIP000443 & 55420.6096  &    2.721&  0.050 & a\\
HIP000443 & 55467.6473  &  $-18.948$&  0.050 & a\\
HIP000443 & 55468.5502  &  $-19.110$&  0.050 & a\\
HIP000443 & 55470.6045  &  $-19.308$&  0.050 & a\\
HIP000443 & 55470.6082  &  $-19.304$&  0.050 & a\\
HIP000443 & 55500.4969  &   13.374&  0.050 & a\\
HIP000443 & 57053.3279  &   $-7.510$&  0.050 & a\\  
...\\
\hline
    \end{tabular}
    \label{Tab:RVs}
\end{table}

\section{Analysis}\label{sect:Analysis}

In this main section of the paper, we address four distinct issues: (i) In Sect.~\ref{Sect:New_SB2_systems}, we report on the detection of the SB2 nature of a system previously known as SB1 (HIP~115142 A), and on the substantial improvement of the orbits of three previously known SB2 systems (HIP~12390, HIP~73182 and HIP~111170). (ii) For the latter  three systems, there is moreover the possibility of computing a combined (spectroscopic SB2)+(astrometric or visual) orbit, giving access to the component masses (Sect.~\ref{Sect:masses}). This is also possible for the SB1 + visual system HIP~28816, adopting the Gaia DR3 parallax. (iii) In Sect.~\ref{Sect:SB1} we improve the reference SB9 orbit by combining its data with the new HERMES data. (iv) These new reference orbits are combined with Hipparcos astrometric data in Sect.~\ref{sect:Hipparcos} to get a revised spectro-astrometric orbit. 

\begin{figure*}
\centering
\includegraphics[clip, width=\linewidth]{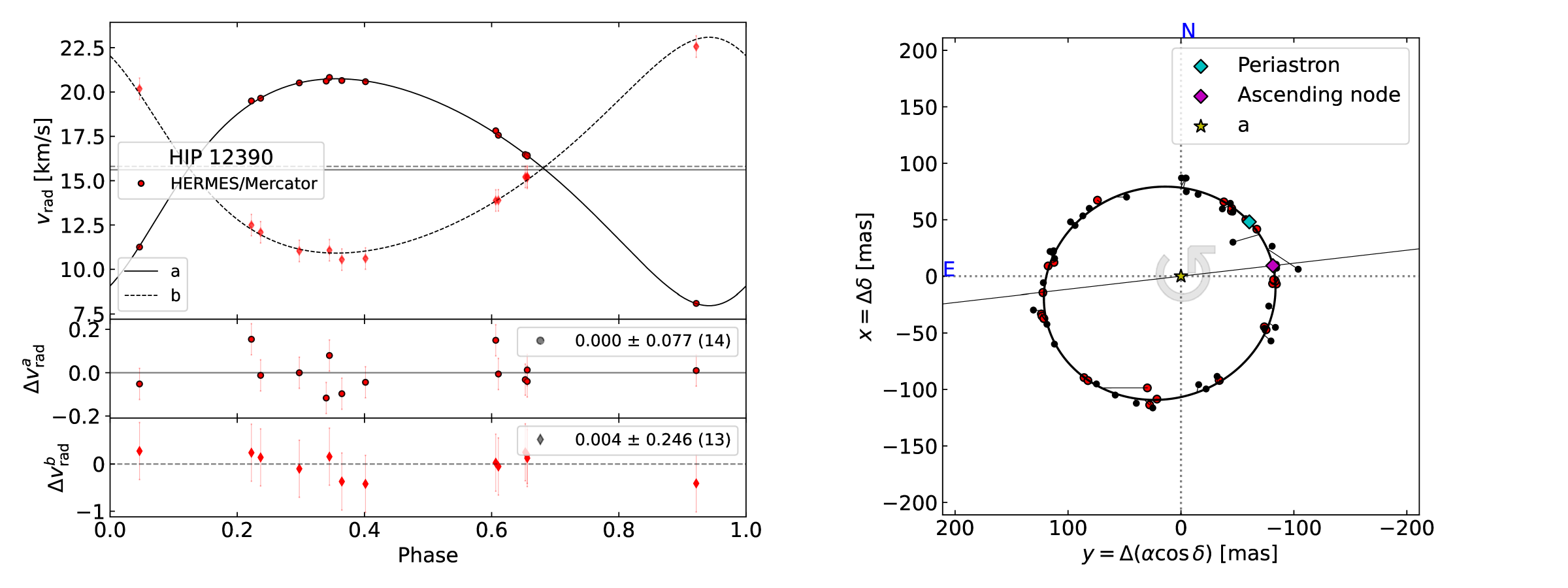}
\caption[]{\label{Fig:12390_HERMES}
The HIP~12390 spectroscopic orbit (left panel) with HERMES data points alone, which allow a better separation between the two components than the earlier, lower-resolution  data from CORAVEL \citep{1988A&A...195..129D}. The right panel shows the simultaneous visual solution. The red dots represent measurements used in the derived solution which is mostly  based on data acquired after 1975. The red diamonds in the left panel correspond to the component listed as 2 in Table~\ref{tab:revorbits_old_new}, which has a wider (7.4~\kms, after correction for the instrumental width of 3~\kms) but less contrasted contribution to the cross-correlation profile.   
Component A has a cross-correlation profile consistent with the instrumental width.}
\end{figure*}
\begin{figure*}
\centering
\includegraphics[clip, width=\linewidth]{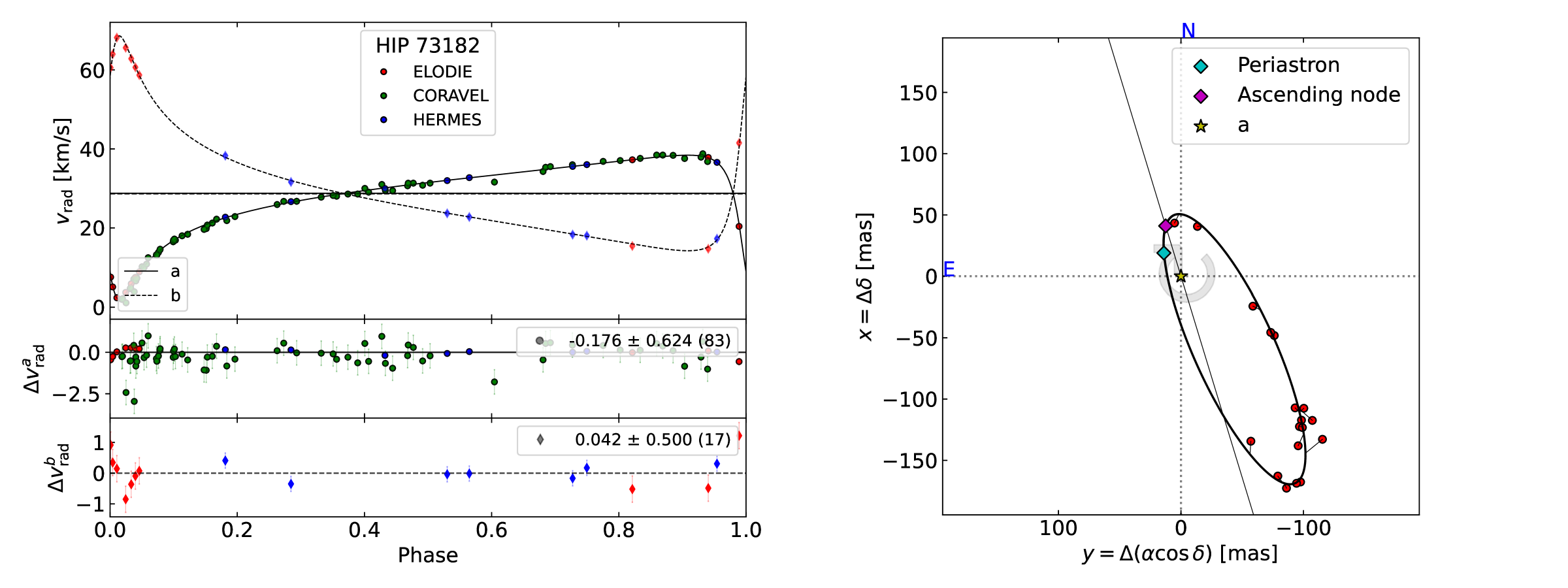}
\caption[]{\label{fig:HIP073182Comb}The HIP~73182 spectroscopic orbit (left panel), with the simultaneous visual solution (right panel).}
\end{figure*}

\subsection{SB2 systems}
\label{Sect:New_SB2_systems}

Despite the high resolution of HERMES, HIP~115142 A is the only star which we detected as a new SB2 system.  The high resolution of the HERMES spectrograph  allowed us moreover to better isolate the signature of the secondary component of HIP~12390, HIP~73182 and HIP~111170.  
\medskip\\
\noindent {\bf  HIP 12390} (HD 16620, $\epsilon$ Cet). Initially detected as a single-lined spectroscopic binary by \citet{Abt-1976} with a somewhat  uncertain period \citep{Morbey-1987}, this system was later described as a "line-width spectroscopic binary" by \citet{1988A&A...195..129D}.  These authors could see the spectra of both components  and  derived  an  SB2  orbit, but the two cross-correlation profiles (even at maximum  separation)  largely  overlap.  
No radial velocities from spectrographs other than CORAVEL have been published over the past thirty years \citep[][revision 2014.5]{2000A&AS..144....1M}. Our recent HERMES measurements, thanks to their better spectral resolution, allow us to derive the velocities of the two components with a much better accuracy. The corresponding orbit is listed in Table~\ref{Tab:VBSBorbits} and shown in Fig.~\ref{Fig:12390_HERMES}.
The combined CORAVEL + HERMES orbit has not been used; because of the low resolution of the CORAVEL data, the component spectra were not properly disentangled, leading to a $K_2$ amplitude larger than that of HERMES. Using these CORAVEL data would thus bias the solution. We note, though, that the HERMES orbit displays a slight anomaly as well, in that there is a shift of $0.18\pm0.08$~\kms\ between the  reference velocities of components 1 and 2, when these are left free in the minimisation \citep[see][for details]{2016A&A...586A..90P}. This may be due to the difficulty in disentangling the component velocities. The mass ratio of this system is $1.05\pm0.05$ resulting in similar apparent brightness for the two components, and making the velocity assignment difficult.
A combined astrometric / spectroscopic solution has been computed, using the large astrometric data set spanning the years 1977 -- 2015 (speckle and interferometric data) provided by the Fourth Catalogue of Interferometric Measurements of Binary Stars\footnote{http://ad.usno.navy.mil/wds, version October 2016} \citep{Hartkopf-2001:b}, thus updating older combined solutions \citep{Hartkopf-1989,2000A&AS..145..215P,2013MNRAS.428..321D}. 
We stress that most of the measurements provided by the WDS are not in agreement with the solution computed by \cite{2000A&AS..145..215P}, whose $\Omega$ differs by $\sim 180^\circ$ from the value provided by \cite{2013MNRAS.428..321D}. This situation most likely results from the ambiguity in the identification of the components which have similar magnitudes. To make the WDS measurements fit on the \cite{2000A&AS..145..215P} solution, we needed to flip the position angles of about 1/3 of them.
For this reason, the Pourbaix solution (which is supported by \citealt{Hartkopf-1989}) has been favoured over the \cite{2013MNRAS.428..321D} one, but there is no guarantee that this is the correct choice. Nevertheless the combined astrometric/spectroscopic solution from the present analysis lifts the ambiguity on the ascending node thanks to the availability of RV. The present combined solution provides a good simultaneous fit to both the astrometric and spectroscopic data, yielding a dynamical parallax $\varpi = 37\pm3$~mas, identical to the Hipparcos parallax of  $37.0\pm1.8$~mas. This case thus favours the original Hipparcos parallax \citep{Hipparcos} over its revision \citep[][$46.6\pm2.5$ mas]{Hip2},  which is however close to the \Gaia DR2 value ($42.49\pm0.76$~mas). No \Gaia DR3 parallax for this system has been published, due to the very large value of the excess-noise standard deviation, indicative of a disturbing signal (due to an extra component in the system?) in the \Gaia astrometric data. The system is further discussed in Sect.~\ref{Sect:masses} devoted to the derivation of the masses, where we compare with the results formerly obtained in the literature.
\medskip\\
\noindent {\bf  HIP 73182} (HD 131976). The HIP identifier refers to component B of Gliese 570 which is the SB2 system studied here. The A component ($V = 5.5$) has constant velocity, and is located 22" away, at a position angle of $122^{\circ}$.  With component G, a brown T-dwarf close to B, they form a quadruple system with the architecture 1+3, according to MSC \citep{2018ApJS..235....6T}. The new HERMES velocities bring a moderate improvement with respect to the reference orbit \citep{1999A&A...351..619F}.  However, new interferometric data have also been published lately \citep{2016AJ....151..153T}.  The simultaneous adjustment of the visual and spectroscopic observations yields an orbital parallax ($169.1 \pm 0.9$~mas) in excellent agreement with the revised Hipparcos solution \citep{Hip2}. 
The system complexity is reflected in \Gaia DR3 by the fact that \Gaia DR3 data did not make it possible to do better than a `two-parameter' solution (no parallax neither proper motion, and the standard deviation of the excess noise is 2.98~10$^5$~mas!).
Stellar masses   are derived in Sect.~\ref{Sect:masses} (also Table~\ref{Tab:VBSBorbits}).
\medskip\\
\begin{figure}
\includegraphics[width=\linewidth]{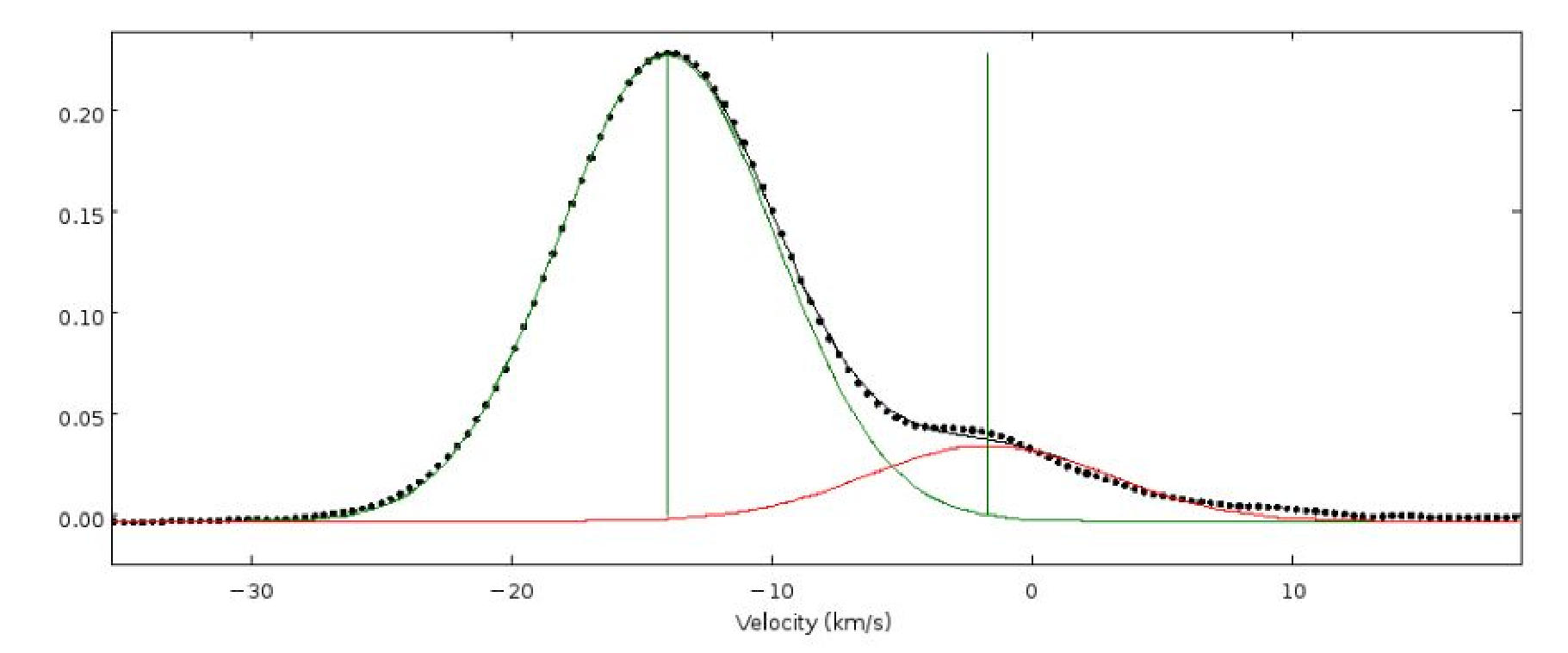}
\caption[]{\label{Fig:CCF_111170}
The CCF of HIP~111170 (at JD~$2\,457\,201$), fully revealing its SB2 nature. The widths of the Gaussians fitted to the  CCF are 3.8~\kms\ for the main peak (close to the instrumental value) and 8.7~\kms\ for the secondary star.   
}
\end{figure}
\begin{figure*}
\centering
\includegraphics[clip, width=\linewidth]{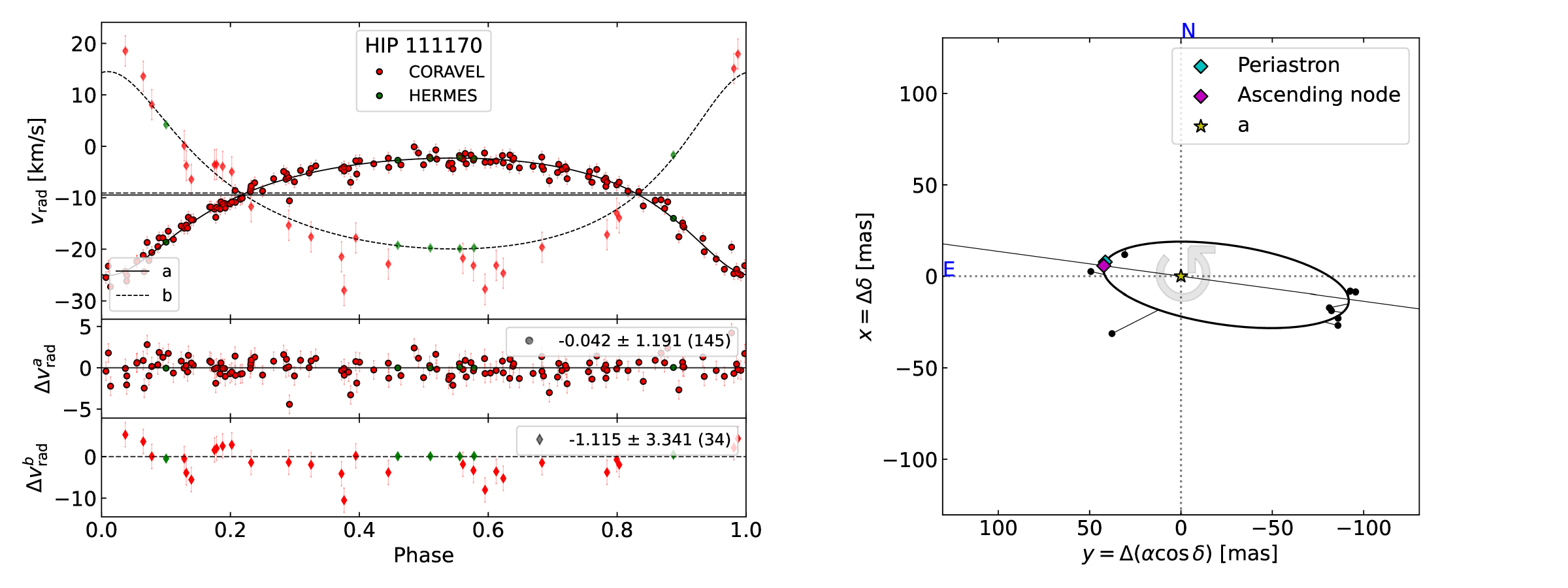}
\caption[]{\label{fig:HIP111170Comb}The HIP~111170 spectroscopic orbit (left panel), with the simultaneous visual solution (right panel).}
\end{figure*}

\begin{figure}
\centering
\includegraphics[width=\linewidth]{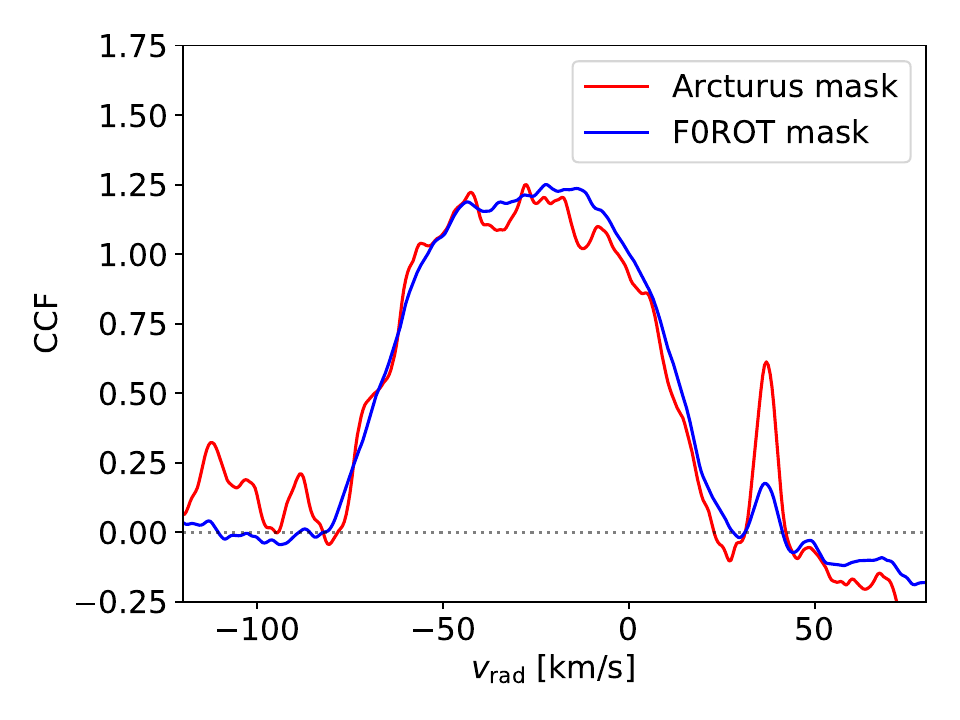}
\caption[]{\label{Fig:CCF_115142}
The CCF of HIP~115142A, fully revealing its SB2 nature, especially with the use of the Arcturus mask. The primary star is rapidly rotating ($v_{\rm rot} \sin i = 49.2$~\kms).
}
\end{figure}
\begin{figure*}
\centering
\includegraphics[clip, width=\linewidth]{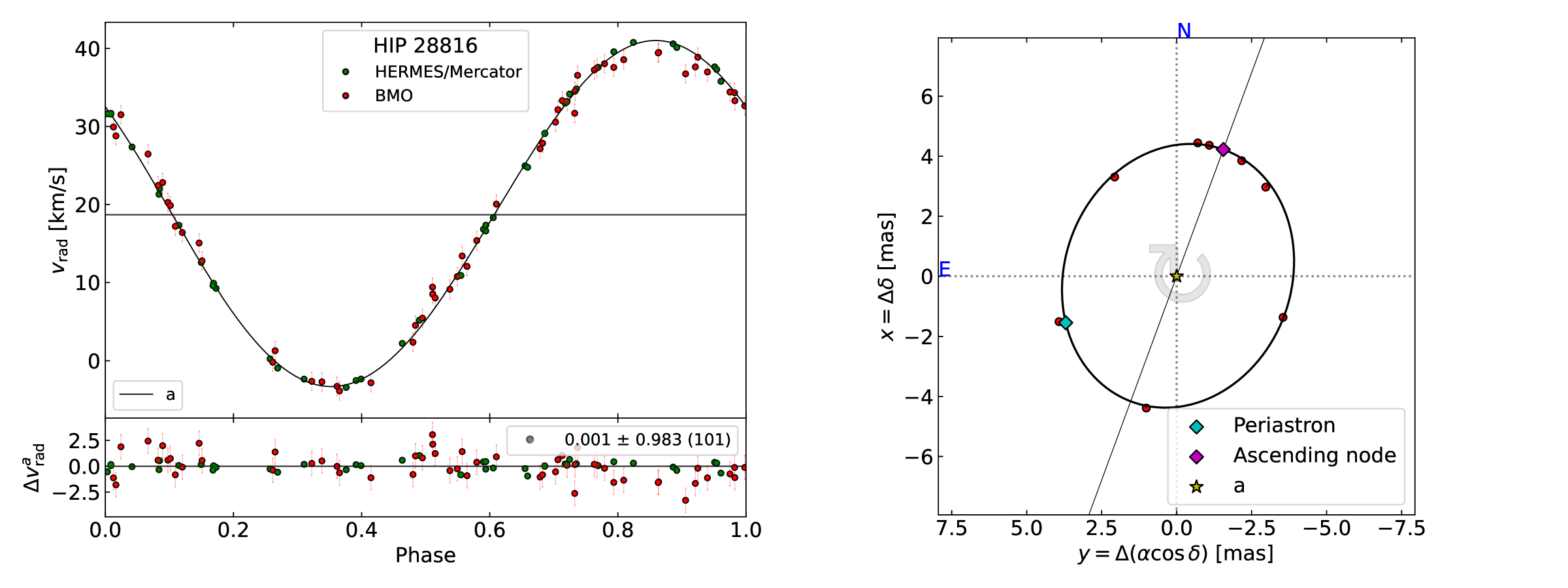}
\caption[]{\label{fig:HIP028816Comb}The HIP~28816 spectroscopic orbit (left panel), with the simultaneous visual solution (right panel).}
\end{figure*}
\noindent{\bf HIP 111170} (HD 213429, HR 8581). This system belongs to the $\beta$ Pic moving group. It was already processed as a combined SB2 + visual binary by \citet{2000A&AS..145..215P} but the velocity curve of component B used in that study \citep[and originally derived by][]{1988A&AS...75..167D} was noisy.  Indeed, the CCF of HIP~111170 was not fully resolved by the  spectrograph used by \citet{1988A&AS...75..167D}, as shown by their figures 1 and 2.  The object was therefore flagged as a Line Width Spectroscopic Binary (rather than a genuine SB2).  Our HERMES observations considerably improve the velocity curve of the secondary component, since the higher resolution of HERMES makes the secondary peak well distinct from the primary's (Fig.~\ref{Fig:CCF_111170}). The second component in the SB2 orbit is also component B of the visual pair WDS~J22313-0633, having a separation of 0.1~arcsec \citep{2001AJ....122.3466M} and a magnitude difference of  2.3~mag.  \citet{2013AJ....145...41K} provided an accurate orbit for component A of that system, as well as the amplitude $K_2$ of the velocity curve for the secondary component. Unfortunately, the individual velocities for component B were not provided in their publication. The HERMES measurements of the primary component confirm and do not improve the reference orbit  \citep{2013AJ....145...41K}. This system was also recently analyzed by \cite{2022AJ....163..220V} using Bayesian inference. 

A combined visual and spectroscopic SB2 orbit has been recomputed for this system (Fig.~\ref{fig:HIP111170Comb}), using visual measurements listed in the Fourth Catalogue of Interferometric Measurements of Binary Stars \citep{Hartkopf-2001:b} and leading to the orbital parameters as reported in Table~\ref{Tab:VBSBorbits}. The orbital parameters (especially the period and velocity semi-amplitudes) are slightly more precise than the values derived by \citet{2000A&AS..145..215P}. The orbital parallax ($41.3 \pm 1.6$~mas) is in good agreement with the \Gaia DR2 parallax ($39.2\pm 0.6$~mas; there is no \Gaia DR3 parallax). the component masses are discussed in Sect.~\ref{Sect:masses}.
\medskip\\ 
 \noindent {\bf  HIP 115142 A} (96 Aqr A, HD 219877 A) is the primary of a visual double star (ADS 16676, WDS 23194-0507), with 10 arcsec separation. 
Each component is a spectroscopic binary \citep{2007A&A...465..257T,2008Obs...128...89T}, with periods of 21.2~d (A) and 659.9~d (B). As indicated by \citet{2007A&A...465..257T}, the visual pair is probably physical since the two stars of the pair  share the same large proper motion ($0.2$~arcsec~yr$^{-1}$).

Component A is a rapidly rotating ($v_{\rm rot} \sin i = 49$~\kms) F star, which we detect as SB2 for the first time with HERMES \citep[Fig.~\ref{Fig:CCF_115142}, and compare with Fig.~1 of][where it is not visible]{2009Obs...129..147G}. The HERMES observation was luckily performed at an epoch close to quadrature, implying a maximum separation between the two velocity components. Figure~\ref{fig:revorbits1} presents a newly derived SB2 orbit for that star, yielding a (preliminary) mass ratio of $q = M_{A,1}/M_{A,2} = 2.0$.

\subsection{Masses}\label{Sect:masses}

Masses can be  exquisitely constrained by combining SB2 and visual orbits, as recently done by,  \emph{e.g.}, \citet{Piccotti2020} for 138 systems from ORB6 \citep{ORB6} and  SB9. We report here and in Table~\ref{Tab:VBSBorbits} a new derivation of masses for three SB2 systems with astrometric data from the \emph{Washington Double Star} catalogue \citep[WDS,][]{2001AJ....122.3466M}. 
 Masses for the components of the astrometric-SB1 system HIP~91751 can be obtained with a different, model-dependent approach that will be  discussed in Sect.~\ref{Sect:GaiaDR3} (see also Table~\ref{Tab:astromSB9}).

\begin{table*}
\centering
\caption[]{\label{Tab:VBSBorbits}Elements of the simultaneous adjustment of the visual and spectroscopic observations of HIP~12390, 28816, 73182 and 111170.  The value of $\omega$ in this Table differs by 180$^\circ$ from that in Table~\ref{tab:revorbits_old_new}, as usual when shifting from spectroscopic (component 1 with respect to the centre of mass) to visual (relative orbit of component 2 around component 1) orbit. For HIP~12390, component 2 here refers to the fast rotator.  The $N_1$ and $N_2$ values are each expressed as the sum  of spectroscopic + astrometric measurements. }

\begin{tabular}{lrrrr}
  \hline\hline
                         & HIP~12390    & HIP~28816  & HIP~73182           & HIP~111170\\
  \hline
$a$ (mas)                & $103.9\pm0.8$  & $4.47 \pm 0.05$& $148.1 \pm 0.5$     & $67.9\pm1.9$ \\
$i$ ($^\circ$)           & $23\pm2$     & $147.8\pm2.1$& $106.1 \pm 0.1$     & $69.8\pm2.5$\\
$\omega_2$ ($^\circ$)    & $34.1\pm0.8$   &$232\pm11$ & $308.3 \pm 0.2$     & $351.5\pm1.1$\\
$\Omega_2$ ($^\circ$)    & $276.6\pm0.8$\tablefootmark{$^a$}  & $339.8\pm0.8$& $16.9 \pm 0.2$      & $82.3\pm0.7$\\
$e$                      & $0.237\pm0.004$& $0.012\pm0.002$& $0.758 \pm 0.001$   & $0.373\pm0.007$\\ 
$P$ (d)                  & $968.3\pm0.2$  & $260.38\pm0.01$& $308.86 \pm 0.01$ & $632.485\pm0.056$\\ 
$T_0$ (HJD$-$2\,400\,000)& $54103\pm2$   & $50762 \pm 8$& $50270.86 \pm 0.04$ & $57905.6\pm1.3$\\
$V_0$                    & $+15.61\pm0.02$ & $+18.69\pm0.16$& $+28.8 \pm 0.8$   & $-9.48\pm0.10$\\ 
$K_1$ (\kms)             & $6.39\pm0.04$  & $22.15\pm0.06$& $18.23 \pm 0.06$    & $11.41\pm0.08$\\
$K_2$ (\kms)             & $6.1\pm0.3$    & -& $27.2 \pm 0.1$      & $17.24\pm0.18$ \\
$N_1$                    & $14 + 57$       & $101 + 8$ & $83 + 16$            & $145 + 9$\\
$N_2$                    & $13 + 57$       & - & $17 + 16$            & $34 + 9$\\
\medskip\\  
$\varpi$ (mas)           & $37\pm3$      & (DR3) $3.57\pm0.10$ & $169.1 \pm 0.9$        & $41.3\pm1.6$\\
$M_1$ (\Msun)            & $1.5\pm0.4$   & $0.79\pm0.03$& $0.562 \pm 0.006$   & $0.89\pm0.06$\\
$M_2$ (\Msun)            & $1.6\pm0.4$   & $3.06\pm0.47$ & $0.377 \pm 0.003$   & $0.59\pm0.04$\\ 
\hline\\
\end{tabular}
\tablefoot{
\tablefoottext{$^a$}{This argument of the ascending node differs by $\sim 180^\circ$ compared to the one from \citet{2013MNRAS.428..321D}. See text for more explanations.}
}
\end{table*}

\noindent{\bf HIP 12390.}
Our present solution derives masses of $1.5\pm0.4$ M$_\sun$ and $1.6\pm0.4$ M$_\sun$ for the components (Fig.~\ref{Fig:12390_HERMES}). HIP~12390 thus appears as a twin system of dwarf F5 stars, in agreement with the spectral types provided by \citet{Martin-1998:a}. The components are not exact twins, though, since component 2, with a CCF Gaussian width of 7.6~km\,s$^{-1}$ (as compared to the instrumental value of 3.0~km\,s$^{-1}$),  is rotating faster than component 1 (CCF Gaussian width of 3.4~km\,s$^{-1}$). The Fourth Catalogue of Interferometric Measurements of Binary Stars similarly mentions a $V$ magnitude difference of about 0.7.

The masses obtained here are substantially different from the values obtained by \citet{Martin-1998:a} ($M_1 =1.886\pm0.171$~\Msun, $M_2 =0.990\pm0.092$~\Msun), \citet{1988A&A...195..129D} ($M_1 =1.10\pm0.21$~\Msun, $M_2 =0.74\pm0.22$~\Msun), and \citet{2000A&AS..145..215P} ($M_1 =2.39\pm0.74$~\Msun, $M_2 =1.55\pm0.48$~\Msun), but the masses obtained in the present work are now fully  consistent with the spectral types, and not too different from those derived by \citet{Piccotti2020}, namely $M_1 = 1.6 \pm 0.1$~M$_\odot$, $M_2 = 1.2 \pm 0.1$~M$_\odot$. The complete orbital solution is listed in Table~\ref{Tab:VBSBorbits}. 
\medskip\\
\noindent {\bf  HIP 28816.} (HD 41511, 17 Lep) is the well-studied symbiotic system SS~Lep, for which rather accurate  masses ($M_1 = 1.30\pm0.33$~\Msun\ -- the M-type component, $M_2 = 2.71\pm0.27$~\Msun\ -- the A-type component) were already derived by \citet{2011A&A...536A..55B} from their interferometric orbit, the spectroscopic mass function from \citet{1995AJ....109..326W} and the Hipparcos parallax \citep{Hip2}.  Combining the HERMES velocities (probing the M-type component) with older velocities from \citet{1995AJ....109..326W} yields a revised orbit with a more precise period and a very small (but strictly positive) eccentricity which makes the argument of periastron $\omega$ not well constrained. 
We have recomputed a combined astrometric/spectroscopic orbit using the visual positions provided by \citet{2011A&A...536A..55B}. With respect to Table A.1 of that paper, the four AMBER measurements had to be rotated by 180$^\circ$.  The corresponding orbit is displayed in Fig.~\ref{fig:HIP028816Comb}, the combined orbital elements are listed in Table~\ref{Tab:VBSBorbits} and the spectroscopic elements in Table~\ref{tab:revorbits_old_new}. The masses listed in Table~\ref{Tab:VBSBorbits} are derived using the spectroscopic mass function $0.293\pm0.003$~\Msun, the orbital inclination, and the Gaia DR3 parallax combined with the astrometric semi-major axis yielding the sum of the masses. The new masses listed in Table~\ref{Tab:VBSBorbits} are in much better agreement with the spectral types than the previous estimates from \citet{2011A&A...536A..55B}.
\medskip\\
\noindent{\bf HIP 73182.} This star, with spectral type M1.5V, is component B of Gliese 570. Our combined astrometric and spectroscopic solution 
(Fig.~\ref{fig:HIP073182Comb}) yields masses $M_1 = 0.562 \pm 0.006$~M$_\odot$ and $M_2 = 0.377 \pm 0.003$~M$_\odot$ (Table~\ref{Tab:VBSBorbits}). 
\medskip\\
\noindent{\bf HIP 111170.}
The orbit presented in Table~\ref{Tab:VBSBorbits} yields masses of $0.89\pm 0.06$~\Msun\  for component 1 and   $0.59\pm0.04$~\Msun\   for component 2, to be compared with $1.4\pm 0.14$~\Msun\ and $0.78\pm0.05$~\Msun\ from \citet{2000A&AS..145..215P}, or $1.08^{1.21}_{0.88}$~\Msun\ and $0.60^{0.68}_{0.56}$~\Msun\ from \citet{2022AJ....163..220V}. 

\onecolumn
\scriptsize
\begin{longtable}{
l
@{}S[table-format=-3.2]@{}
@{}S[table-format=2.2]@{}
@{}S[table-format=-2.5]@{} 
@{}S[table-format=1.4]@{} 
@{}S[table-format=5.5]@{} 
@{}S[table-format=3.5]@{} 
@{}S[table-format=-6.3]@{} 
@{}S[table-format=3.3]@{} 
@{}S[table-format=-4.3]@{} 
@{}S[table-format=3.3]@{} 
@{}S[table-format=-3.3]@{} 
@{}S[table-format=2.3]@{} 
@{}S[table-format=-3.3]@{} 
@{}S[table-format=1.3]@{} 
@{}S[table-format=1.4]@{} 
@{}S[table-format=1.3]@{} 
}
\caption{\label{tab:revorbits_old_new}Revised spectroscopic orbits, computed by combining the HERMES measurements with older RV (except when otherwise indicated).  The first line of each entry lists the revised orbit, and the second one lists the old SB9 orbit, as referenced in Table~\ref{tab:ProgrammeStars}.   The systemic velocity $V_0$ for the revised orbit (first line of each entry) is given in the HERMES/IAU RV system. Hence the latter may be offset with respect to $V_0$ from the SB9 orbit by the zero-point offset listed in Table~\ref{tab:ProgrammeStars}. When available, the \Gaia DR3 SB orbit is listed on the third line.}\\
  \hline\hline
{HIP} & {$\omega_1$} & {$\sigma_{\omega_1}$} & {$e$} & {$\sigma_e$} & {$P$} & {$\sigma_P$} & {$T_0$} & {$\sigma_{T_0}$} & {$V_0$} & {$\sigma_{V_0}$} & {$K_1$} & {$\sigma_{K_1}$} & {$K_2$} & {$\sigma_{K_2}$} & {$\sigma_1$} & {$\sigma_2$}\\
 \cline{10-17}
& {($^\circ$)} & {($^\circ$)} & & & {(d)} & {(d)} & {(JD}& {(d)} & \multicolumn{8}{c}{(\kms)}\\ 
 &                 &                &  &  &    &       & {$-$2\,400\,000)} 
 \\ \hline
\endfirsthead
\caption{\emph{continued.}}\\
\hline\hline
{HIP} & {$\omega_1$} & {$\sigma_{\omega_1}$} & {$e$} & {$\sigma_e$} & {$P$} & {$\sigma_P$} & {$T_0$} & {$\sigma_{T_0}$} & {$V_0$} & {$\sigma_{V_0}$} & {$K_1$} & {$\sigma_{K_1}$} & {$K_2$} & {$\sigma_{K_2}$} & {$\sigma_1$} & {$\sigma_2$}\\
\cline{10-17}
& {($^\circ$)} & {($^\circ$)} & & & {(d)} & {(d)} & {(HJD}& {(d)} & \multicolumn{8}{c}{(\kms)}\\
&                 &                &  &  &    &       & {$-$2\,400\,000)} \\
\hline
\endhead
\endlastfoot
\input{tab/combinedorbits}
\end{longtable}
\tablefoottext{a}{Component 1 refers to the cool star (the subgiant G star)} 
\tablefoottext{b}{The orbit listed corresponds to the F2 component, and includes HERMES measurements only.}
\tablefoottext{c}{The orbit listed on the first line is computed from HERMES RV only.}
\twocolumn
\normalsize

\subsection{Improving the reference SB9 orbits}\label{Sect:SB1}

We collect in Table~\ref{tab:revorbits_old_new} those systems for which an improved orbit could be derived by combining the SB9 data with the new HERMES observations, generally after applying a zero-point correction, as discussed in Sect.~\ref{sect:Observations}. 
We publish only orbits for which a substantial  gain in precision may be achieved (especially on the orbital period) from such a combination of recent with older measurements. HERMES-only orbits are not listed individually, since the combined SB9+HERMES orbit is usually much more accurate, with the exception of HIP~95176. The latter is a complex system for which the systemic velocity is difficult to evaluate; for that system, the HERMES-only orbit turned out to be valuable (see Appendix~\ref{sect:Star}, with all orbits displayed in Fig.~\ref{fig:revorbits1}).

\subsection{Combined spectroscopic and Hipparcos astrometric orbit}
\label{sect:Hipparcos}

When spectroscopic solutions are available, they may be combined with astrometric data to derive a combined spectro-astrometric orbit, with the advantage that astrometry data alone without the constraints provided by the spectroscopic orbital elements would not allow us to derive an astrometric orbit   \citep[][]{2005A&A...442..365J}. Table~\ref{tab:astrometry} collects such cases, obtained by combining Hipparcos Intermediate Astrometric Data with the new spectroscopic orbit (hence the importance of getting an accurate spectroscopic orbit). In that respect, Hipparcos \citep{Hipparcos} has not delivered its final word on binaries yet. We note that a similar approach is not yet possible with \Gaia data as the corresponding astrometric epoch data are not yet publicly available \citep[see however][]{Leclerc-2023}.

The combination of spectroscopic orbits with the Hipparcos astrometric data and its quality assessment have been extensively described elsewhere \citep{2001A&A...369L..22P,2001A&A...372..935P} and applied to various systems \citep[\emph{e.g.},][]{2000A&AS..145..161P,2005A&A...442..365J,Ren-2013}.  Among the spectroscopic orbits that were substantially revised in the present paper, only 13 yield spectro-astrometric orbits that successfully pass the quality checks\footnote{Namely, improvement of the orbital-model fit with respect to the single-star one (this criterion -- and all the others -- being based on a $F$-test with a 5\% probability threshold), the significance of the Thiele-Innes constants resulting from the fit, the consistency of the Thiele-Innes solution and the Campbell/spectroscopic one, and the likelihood of the face-on orbit.} of \citet{2003A&A...398.1163P}.  The relevant information is listed in Table~\ref{tab:astrometry}. Of special interest are HIP~50796, HIP~69929, HIP~84402, and HIP 90135, since neither the Hipparcos Catalogue nor Gaia DR3 did provide an astrometric orbit\footnote{More precisely, Hipparcos flagged those as accelerated solutions.} for those systems. 

The reasons why so few (\emph{i.e.}, 13) spectroscopic systems could deliver a spectro-astrometric orbit are multiple.  One reason relates to the object that was effectively observed by Hipparcos.  For instance, the two pairs of the quadruple system HIP~55505 (Sect.~\ref{sect:Star}) were observed by Hipparcos as a unique resolved pair with a fixed configuration.  Among the two systems of the other quadruple system HIP~115142, only the brighter system was observed, since  only component A was present in the Hipparcos Input Catalogue.  Therefore, component B (ADS 16676 B = SB9~2859) has no astrometry associated to it.  A second reason for the lack of astrometry is the small amplitude of the photocentric wobble in case of SB2 systems.  The more similar the brightness of the two components is, the smaller their combined orbital astrometric signature.  Finally, one should keep in mind that, although the amplitude of the radial-velocity variations is distance-independent, the astrometric signal is not.  It is therefore expected that most of the spectroscopic systems, being far away, will remain unreachable for astrometry.

Nine of the 13 spectro-astrometric orbits in Table~\ref{tab:astrometry} were originally processed in the Hipparcos Catalogue with an orbital model (DMSA/O) and four belong to the DMSA/G category (\emph{i.e.}, fitted with an acceleration model).  For the DMSA/O entries, only HIP~59750 and HIP~63742 had all their orbital parameters derived from scratch, without assuming any ground-based orbit (spectroscopic orbits as far as our systems are concerned).  Only for HIP~45527 and HIP~95066 is the reference orbit listed in Table~\ref{tab:ProgrammeStars} the same as the one adopted in the original Hipparcos reduction,  hence the necessity of reprocessing all the other systems with the new spectroscopic orbit.  For the remaining cases, the orbit we adopted as reference already supersedes the one used in the original Hipparcos reduction. For four stars (HIP 31205, HIP 39198, HIP 57791, HIP 91751), \Gaia DR3 \citep{Arenou2022} provides an astrometric orbit, as mentioned in Tables~\ref{tab:ProgrammeStars} and \ref{tab:astrometry}.

For those targets with no astrometric orbit provided by the Hipparcos catalogue,  we nevertheless list the parallax and proper motion computed by Hipparcos from a single-star model to evaluate the impact of the binary motion on these parameters. It is well known (and well visible on Table~\ref{tab:astrometry}) that the impact on the parallax is null unless the orbital period is close to 1 yr \citep[][]{2005A&A...442..365J}, contrary to the proper motion which can be largely offset as the orbital motion may add to the proper motion if the former is not properly taken into consideration through an orbital model \citep{Wielen-1999,Frankowski-2007:a,Makarov-2005:a}.

\section{Comparison with \Gaia DR3 binaries\protect\footnote{This section was added after the passing away of D. Pourbaix.}
}
\label{Sect:GaiaDR3}

Even though several papers \citep[\textit{e.g.,}][]{Bashi2022,Arenou2022,  Babusiaux2023, Halbwachs2023} have documented the selection processes of the Gaia DR3 binaries, the criteria to validate their orbital solutions, and the ensuing biases, we believe that it may still be of interest to illustrate them on an actual sample. As these criteria are complex, they will not be repeated in full detail here; only the most relevant will be mentioned. We will rather focus on illustrating their impact by exploring which binaries from the present subsample of the SB9 catalogue are found among the Gaia DR3 NSS (\emph{i.e.}, `non-single stars') sample, and why some others are not. A similar analysis for the full SB9 catalogue is in preparation. 

\begin{table*}
\scriptsize
\centering
\caption[]{\label{tab:astrometry}Astrometric parameters based upon the Hipparcos data \citep{Hipparcos} and the revised spectroscopic orbit (first line per entry), the Hipparcos DMSA/O or \citet{2005A&A...442..365J} solutions (second line), and the \Gaia DR3 astrometric orbit (third line) when available. \Gaia DR3 parallaxes and proper motions (but without orbital solution) are also provided when available.}
\begin{tabular}{@{}l*{12}{S}@{}}
\hline\hline
HIP & {$\varpi$} & {$\sigma_{\varpi}$} & {$\mu_{\alpha*}$} & {$\sigma_{\mu_{\alpha*}}$} & {$\mu_{\delta}$} & {$\sigma_{\mu_{\delta}}$} & {$a_0$} & {$\sigma_{a_0}$} & {$i$} & {$\sigma_i$} & {$\Omega$} & {$\sigma_{\Omega}$} \\
& {(mas)}  & {(mas)} & {(mas yr$^{-1}$)} & {(mas yr$^{-1}$)} & {(mas yr$^{-1}$)} & {(mas yr$^{-1}$)} & {(mas)} & {(mas)} &{($^\circ$)}  &{($^\circ$)} &{($^\circ$)}  &{($^\circ$)}\\  \hline
\input{tab/astromOrbits}
\end{tabular}
\end{table*}

\begin{table*}
\footnotesize
\centering
\caption[]{\label{Tab:astromSB9} Comparison of the DR3 astrometric and SB9 spectroscopic orbits for  HIP~ 31205, 39198, 57791, and 91751. The latter do not result from a simultaneous adjustment since the epoch DR3 astrometry was not yet made public by \Gaia DR3. 
To allow an easy comparison between the spectroscopic SB9 orbital solution and \Gaia DR3 astrometric solution, the value of $\omega$ in this Table corresponds to the spectroscopic value (component 1 with respect to the centre of mass). Hence all DR3 $\omega$ values have been shifted by $180^\circ$ with respect to the value given in the DR3 catalogue. The first line of $T_0$ values labelled SB9 corresponds to the original value shifted in time to match the DR3 epoch, whereas the second $T_0$ line labelled SB9 lists the original $T_0$ value from SB9.} 
\begin{tabular}{lrrrrrr}
  \hline\hline
                    HIP     & 31205 &  39198           & 57791 & 91751\\
  \hline         
$a$ (mas)                       &$2.87 \pm 0.05$ &  $5.7 \pm0.1$ & $6.6 \pm  0.2$ &  $6.39 \pm 0.15$\\
$i$ ($^\circ$)                  &$75.1\pm 1.8$ &. $123.2 \pm0.6$ & $73.7 \pm 0.6$ & $61.9\pm  0.9$\\
$\omega_1$ ($^\circ$) \hfill  DR3 &$-23.1\pm  56.1$ 
&$103.2 \pm2.6$ & $134.1 \pm 12.7$& $63.9 \pm8.5$ \\ 
                     \hfill SB9 &$73.7\pm32.1$ &
$108.1\pm3.2$ & $126.4\pm1.3$ & $75.50\pm0.16$\\
$\Omega_1$ ($^\circ$)\hfill  DR3  &$134.8\pm1.8$ & $246.8 \pm1.9$ & $108.2\pm0.5$ & $-77.5 \pm0.9$ \\
$e$   \hfill     DR3            &$0.052\pm	0.054$ 
&$0.48\pm	0.02$ & $0.21 \pm	0.02$ & $0.28\pm0.04$\\ 
\hfill       SB9                &$0.013\pm0.008$ 
&$0.52\pm	0.02$ & $0.272\pm	0.006$ & $0.2019\pm	0.0006$\\
$P$ (d) \hfill      DR3                  &$464.8\pm	5.4$ 
& $364.7\pm0.6$ & $488.2 \pm	2.2$ & $481.6\pm2.1$\\
\hfill   SB9                    &$457.40 \pm 0.14$ 
&$365.4\pm	0.6$ & $490.77\pm0.09$ & $485.21\pm0.02$\\ 
$T_0$ (JD$-$2\,400\,000) DR3&$57374.4\pm80.6$ 
&$57482.3\pm 1.5$ & $57564 \pm16$ & $57463.6\pm9.1$\\
\hfill SB9&\multicolumn{1}{l}{$ 57282.9\pm43.6$}&\multicolumn{1}{l}{$57480.9\pm7.3$}
&\multicolumn{1}{l}{$57556.9\pm2.7$}
&\multicolumn{1}{l}{$57480.7\pm9.2$}\\
\hfill SB9&$47677.5\pm40.7$ 
&$53826.9\pm	1.3$ & $51667.7\pm1.6$ & $54084.27	\pm0.19$\\
\\
$M_1$ (M$_\odot$) \hfill(FLAME)& &&&  1.42 ${+0.10}\atop{-0.06}$\\
$M_2$ (M$_\odot$)  & &&&  0.64 ${+0.03}\atop{-0.02}$\\

\hline\\
\end{tabular}
\end{table*}

\begin{figure}[t]
\vspace*{-4.5cm}
\resizebox{1.15\hsize}{!}{\includegraphics{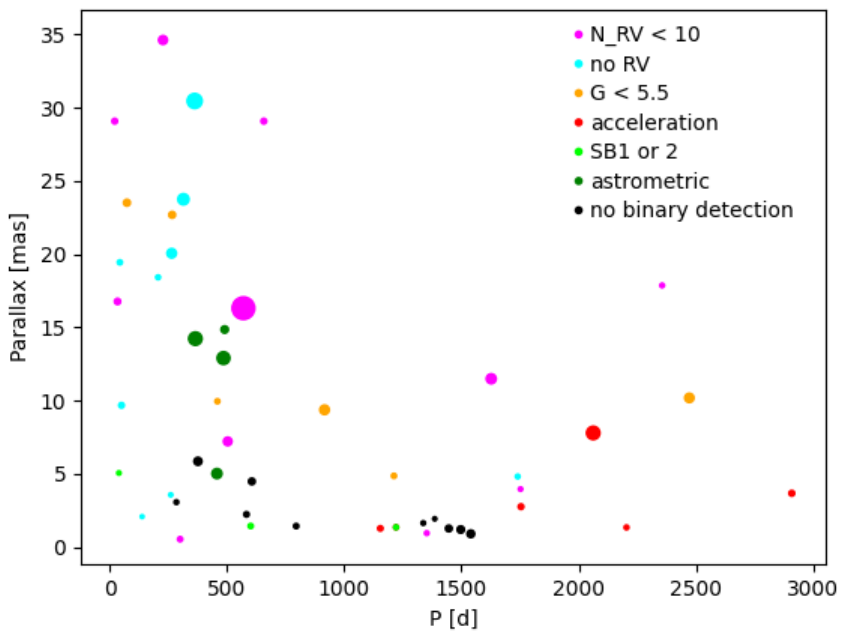}}
\vspace*{-4.5cm}
\caption[]{\label{fig:DR3_SB9}
SB9 systems from the present study with a DR3 entry in the period -- parallax plane. The symbol size is proportional to the RUWE indicator \citep{2021A&A...649A...2L}. The color codes are as follows: cyan: no RV available from \Gaia DR3, magenta: $N_{\rm RV} < 10$; orange: $G_{\rm RVS} < 5.5$, dark green: astrometric binary, red: astrometric acceleration solution, green: SB1 or SB2, black: no binary detection in the NSS DR3 catalogue (even though the RV $\chi^2$ $p-$value is lower than $10^{-4}$).  
}
\end{figure}

\subsection{Selection biases for DR3 spectroscopic binaries}

The \Gaia DR3 NSS  catalogues \citep{Arenou2022} contains the largest number of astrometric, spectroscopic and eclipsing binaries collected so far.
Even though the completeness of this huge catalogue has been discussed in some details by the papers listed in the introduction to this section, the present subsample of SB9 systems offers an interesting illustration of the complex selection biases at work. We believe that it is of interest to clarify on a case-by-case basis the reasons why a large fraction of the subsample of SB9 systems studied in the present paper is not yet present in the Gaia DR3 NSS sample. These reasons involve selection criteria based on the RV properties and others related to the $G$ magnitude and to the effective temperature \citep{2022gdr3.reptE...7P,Katz2023}, to the stellar environment \citep[\emph{i.e.}, the presence of close visual companions,][]{2021A&A...653A.160S}, and to the existence of period aliases \citep[][their Figure~18]{Holl2023}.

To this aim, Fig.~\ref{fig:DR3_SB9} presents the SB9 systems from the present paper in a period -- parallax diagram, and the different situations encountered relative to Gaia DR3 NSS were colour-coded as summarized in the figure legend. We provide below a more extensive discussion of these possible cases:
\begin{enumerate} 
    \item no orbit computation was attempted by the Gaia NSS  consortium for systems with less than 10 RV observations \citep[`$N_{RV} < 10$';][Sect. 7.4.2]{2022gdr3.reptE...7P};
    \item some SB9 systems have no Gaia DR3 RV (`no RV') because they are either too warm or too cold for the available RV templates (namely they lie outside the temperature range 3875~K to 8125~K -- \citealt{2022gdr3.reptE...7P}, Sect. 7.4.2 --) or because of some other disturbance -- like a resolved close visual companion, etc.;
    \item systems with the Gaia RVS magnitude ($G_{\rm RVS}$) outside the range [5.5\footnote{This bright-magnitude threshold is in fact an indirect consequence of the criterion set on the \texttt{rv\_renormalised\_gof} parameter, which is not defined for such bright objects (Gosset et al., in prep. and priv. comm.).} -- 12] have no RV either (`$G_{\rm RVS} < 5$') hence preventing any SB detection.
\end{enumerate}
Furthermore, for obvious reasons, the following cases appearing in Table~\ref{tab:ProgrammeStars} were not plotted in Fig.~\ref{fig:DR3_SB9}:
\begin{enumerate}
\setcounter{enumi}{3}
    \item two SB9 entries that our HERMES study flagged as single;
    \item systems with a multiple-star catalogue (MSC) entry \citep[][Sect. 7.4.6]{2018ApJS..235....6T,2022gdr3.reptE...7P};
    \item systems with no DR3 entry.
\end{enumerate}

The remaining targets should then show binary signatures (either as acceleration solutions, or as eclipsing, astrometric or spectroscopic binaries -- or a combination of these), unless their orbital period is short (almost no astrometric binaries with $P < 100$~d are present in Gaia DR3 -- \citealt{Arenou2022} -- more details in Table~1 and Fig.~3 of  \citealt{Halbwachs2023}) or much longer than the time span of Gaia DR3 (about 1000 d). It appears that, among  the 11 systems with $P < 1000$~d (for which the DR3 time coverage is thus long enough to have covered a full orbital period) not belonging to any of the non-detection criteria 1--6 listed above, only 4 are flagged as astrometric binaries (dark green dots in Fig.~\ref{fig:DR3_SB9}; see Sect.~\ref{Sect:astrometric}), and only  2 have been flagged as SB (light green dots) in the Gaia DR3 NSS catalogue,  thus leaving 5 SB9 targets with no \Gaia DR3 SB detection whatsoever. This could appear even more surprising given that the $p$-value for RV constancy based on a $\chi^2$ criterion\footnote{The $p$-value for RV constancy \citep{2022gdr3.reptE...7P} is defined as the probability that the $\chi^2$ value of the RV time series (with $k = N_{\rm RV} - 1$ degrees of freedom) will, under the null hypothesis of RV constancy, exceed the observed value $\chi^2_{\rm obs}$, as a result of random fluctuations. In other words, $p = \int_{\chi^2_{\rm obs}}^\infty {\rm d}P_{\chi^2}/{\rm d}{\chi^2}\; {\rm d}{\chi^2}$, where ${\rm d}P_{\chi^2}/{\rm d}\chi^2$ is the $\chi^2$ probability density function. Thus $p$  ranges from 0 for very low probability of RV constancy to 1 for very strong probability of RV constancy. If one selects for example all sources having $p < 0.05$ from a sample, this will yield 5\% of RV constant sources that may still contaminate the selected subset of RV variable sources.} reveals RV variability  ($p < 10^{-4}$) for all these stars. Furthermore, the semi-amplitude $K$ (several km~s$^{-1}$), as revealed by the SB9 catalogue, is large enough for Gaia DR3 to be able to detect these as binary systems. Why then were these 5 SB9 systems with $P < 1000$~d (black dots in Fig.~\ref{fig:DR3_SB9}) not detected as SB by \Gaia DR3? The reason thereof is their $F2 > 3$ (possibly caused by underestimated error bars), although these filtered-out, unpublished orbits have periods and eccentricities in good agreement with their SB9 counterparts (Damerdji \& Gosset, priv. comm.). A posteriori, this selection criterion $F2 > 3$, albeit seemingly justified on statistical grounds, appears to have been too conservative.

Overall, for the sample of relatively bright binary stars (\emph{i.e.}, $V \la 11.5$) considered by our analysis, only  6 (\emph{i.e.}, 4 astrometric and  2 SB) were flagged as binaries by the Gaia DR3 NSS catalogue among the 31 SB9 systems with a DR3 parallax available and with $P_{\rm SB9} < 1000$~d.

\subsection{Comparison DR3 - SB9 orbits for astrometric and spectroscopic binaries}
\label{Sect:astrometric}

Even though the SB9 binaries studied in the present paper ought not to appear systematically as astrometric binaries, we nevertheless identify in Table~\ref{tab:ProgrammeStars} and plot in Fig.~\ref{fig:DR3_SB9} the DR3 astrometric binaries (HIP 31205, 39198, 57791, and 91751) and acceleration binaries (HIP~25912, 32467, 38217, 53717, 73007, and 104785) present in our SB9 subsample.

The four astrometric orbits are compared to the SB9 spectroscopic solutions in Table~\ref{Tab:astromSB9}, and the agreement is good. We note that a simultaneous spectroscopic and astrometric solution (as done for visual or Hipparcos astrometry; see Tables~\ref{Tab:VBSBorbits} and \ref{tab:astrometry}) cannot yet be computed in these cases, since the epoch astrometry was not yet provided by \Gaia DR3. 

 In the context of \Gaia DR3, the Final Luminosity Age Mass Estimator, FLAME, aiming to produce the stellar mass and evolutionary parameters for each \Gaia source (see Sect.~11.3.6 in the \Gaia DR3 documentation), has provided a mass estimate for HIP~91751: $M_1 = 1.42$~M$_\odot$ (with an uncertainty range 1.36 -- 1.52~M$_\odot$). This mass can be attributed to the primary component, since we will find that the companion is significantly less massive than the primary, hence much
fainter than the primary. Combining the primary mass with the mass function $f(M) = 0.0420\pm0.00009$~M$_\odot$ and the inclination $i = 61.9^\circ\pm0.9^\circ$ (Table~\ref{tab:astrometry}), we find $M_2 = 0.64$~M$_\odot$ (with the uncertainty range 0.62 -- 0.67~M$_\odot$). 

The comparison between the DR3, old SB9 and revised  spectroscopic orbits for HIP~22000, HIP~32467, and HIP~37041 may be found in Table~\ref{tab:revorbits_old_new}. The agreement between the SB9 and DR3 solutions is excellent in all cases, considering the large errors on $\omega_1$ for HIP~31205 caused by the orbit being compatible with a circle.  For the SB2 system HIP 22000,  the longitude of periastron $\omega$ differs by $180^\circ$ between the SB9 catalogue and \Gaia DR3, and we found no explanation for this discrepancy.

\section{Conclusions}

This paper has provided 51 orbits that have been improved with respect to the solution listed in the SB9 catalogue, out of the
58 SB9 systems studied. Two systems (HIP~22701 and 26563) appear to be single stars.
Five SB (3 SB2 and 2 SB1) belong to five triple stellar systems (namely HIP 22000, 31205, 45527, 69974 and 115126), and 7 SB (four SB1 and three SB2) belong to four quadruple stellar systems (namely HIP 55505, 73182, 92726 and 115142).  
HIP~115142~A is the only star which we detected as a new SB2 system. The B component of the visual binary HIP~92726 has now been found to be a spectroscopic system as well, which makes HIP 92726 a newly discovered quadruple system. The high HERMES resolution allowed us moreover to better isolate the signature of the secondary component of HIP~12390, HIP~73182 and HIP~111170. More accurate masses have thus been derived for them. Finally, by comparing the NSS DR3 diagnostic for the 58 SB9 systems studied here, we conclude that the recovery rate is not yet very large, but that situation should improve with the coming Gaia data releases.

\begin{acknowledgements}
We thank E. Gosset and Y. Damerdji for useful discussions on the SB processing within \Gaia DPAC, as well as the anonymous referee for useful comments. 
T.M. was supported by a grant from the Fondation ULB and is now granted by the BELSPO Belgian federal research program FED-tWIN under the research profile Prf-2020-033\_BISTRO.
This work was partly supported by the Belgian PRODEX grant C4000119826 \textit{Gaia mission Belgian consolidation} and  
by the FNRS-F.R.S. grant PDR T.0115.23.
Based on observations made with the Mercator Telescope, operated on the island of La Palma by the Flemish Community, at the Spanish Observatorio del Roque de los Muchachos of the Instituto de Astrof\'\i sica de Canarias. Based on observations obtained with the HERMES spectrograph, which is supported by the Research Foundation - Flanders (FWO), Belgium, the Research Council of KU Leuven, Belgium, the Fonds National de la Recherche Scientifique (F.R.S.-FNRS), Belgium, the Royal Observatory of Belgium, the Observatoire de Gen\`eve, Switzerland and the Th\"uringer Landessternwarte Tautenburg, Germany.
This research has made use of the SIMBAD database, operated at CDS, Strasbourg, France \citep{2000A&AS..143....9W}.
This research has made use of the Washington Double Star Catalog maintained at the U.S. Naval Observatory \citep{2001AJ....122.3466M} available at \url{https://crf.usno.navy.mil/wds}.
\end{acknowledgements}

\bibliographystyle{aa}
\bibliography{sb9hermes}

\begin{appendix}

\section{Comments on individual systems and revised orbits}
\label{sect:Star}
\begin{figure*}[ht]
\resizebox{0.33\hsize}{!}{\includegraphics{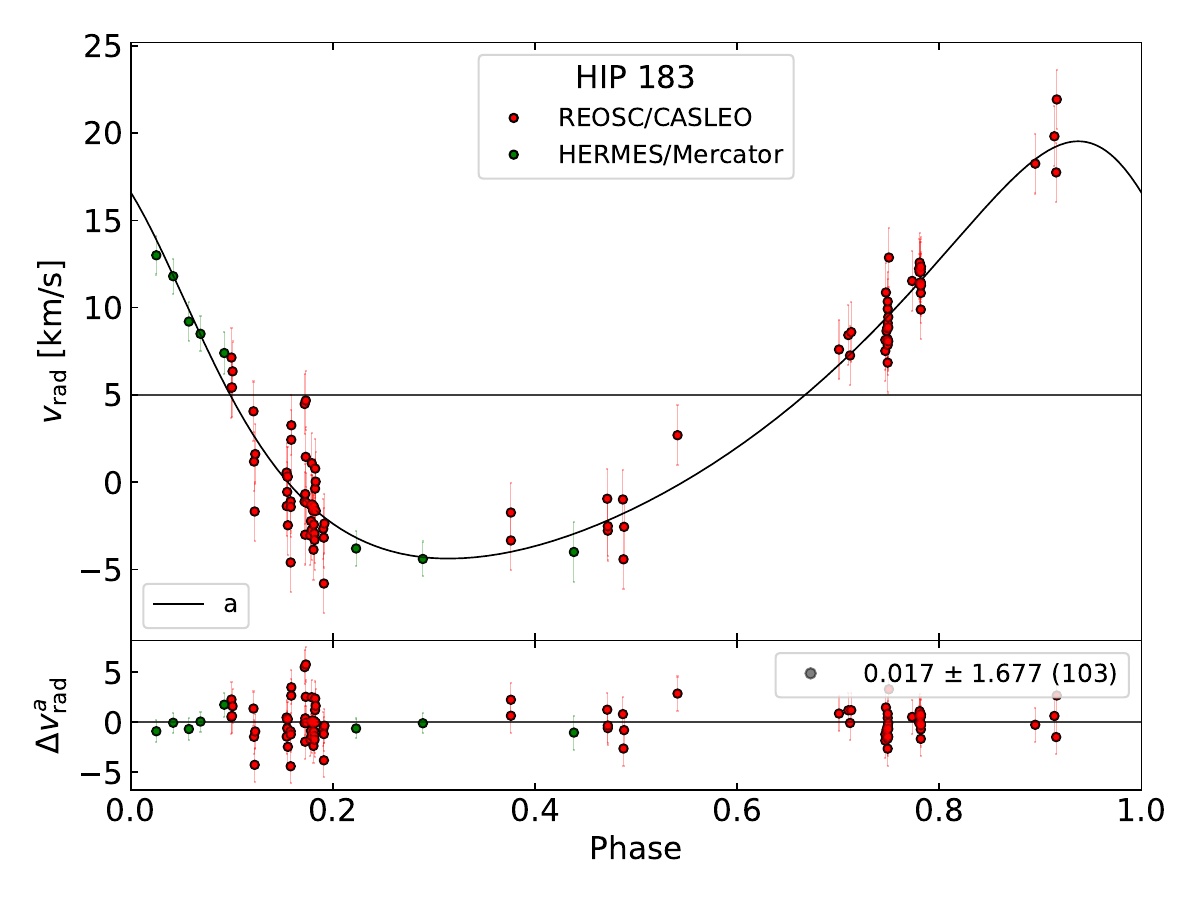}}
\resizebox{0.33\hsize}{!}{\includegraphics{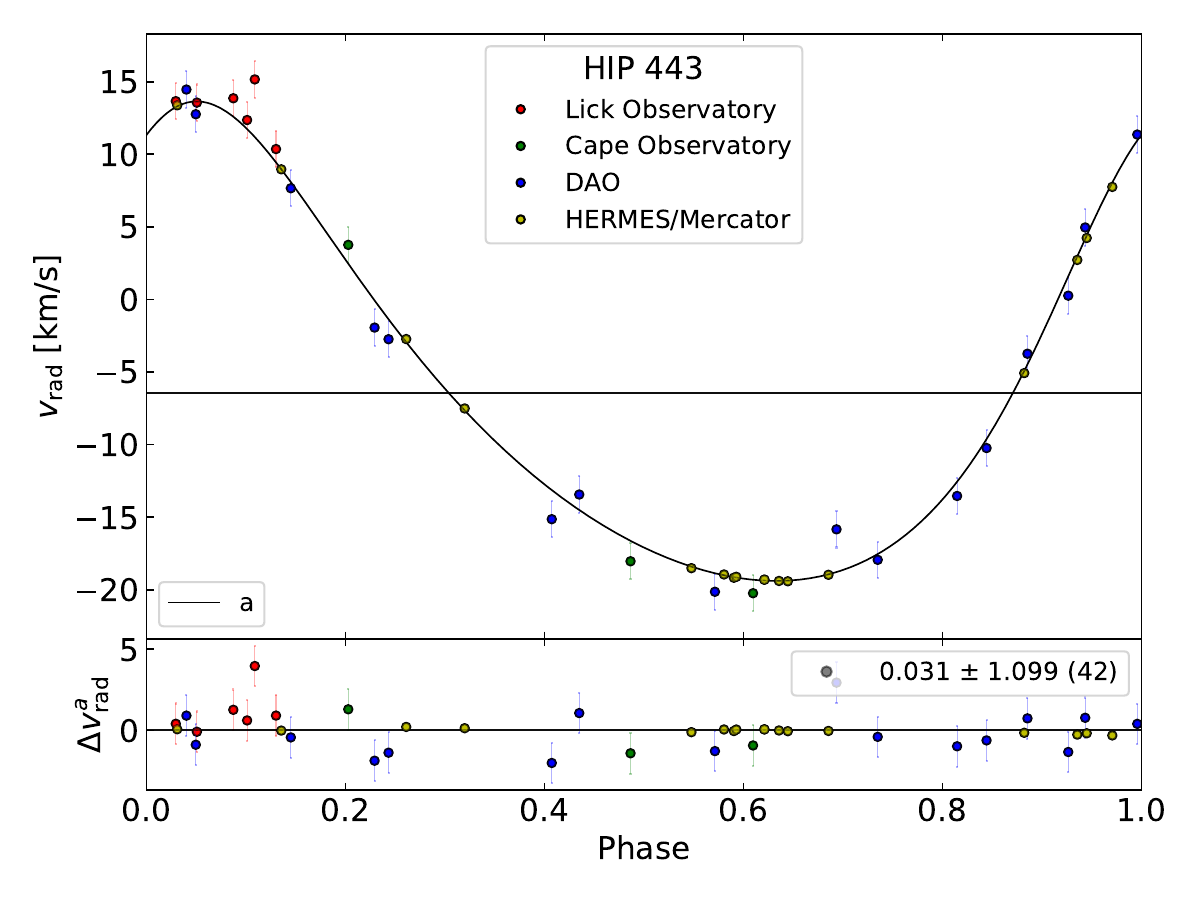}}
\resizebox{0.33\hsize}{!}{\includegraphics{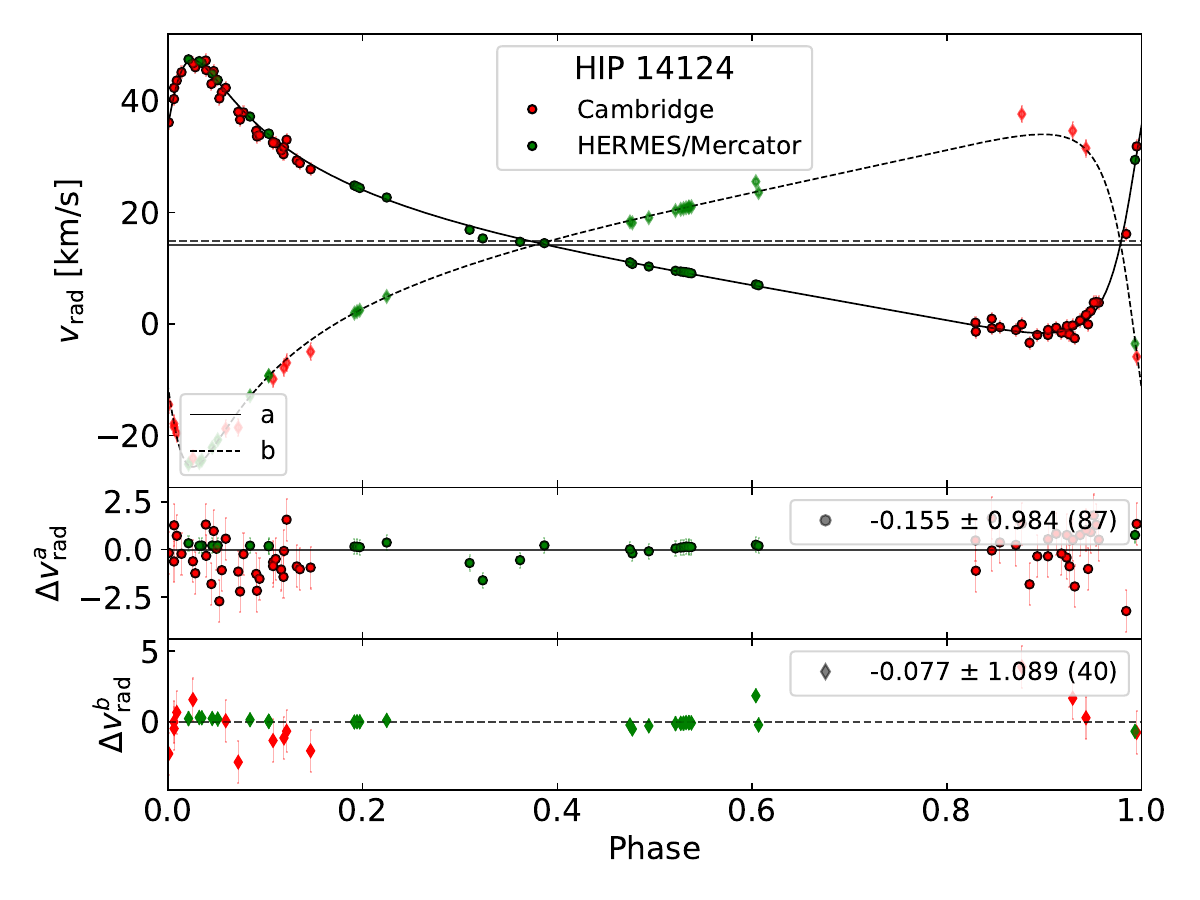}}\\
\resizebox{0.33\hsize}{!}{\includegraphics{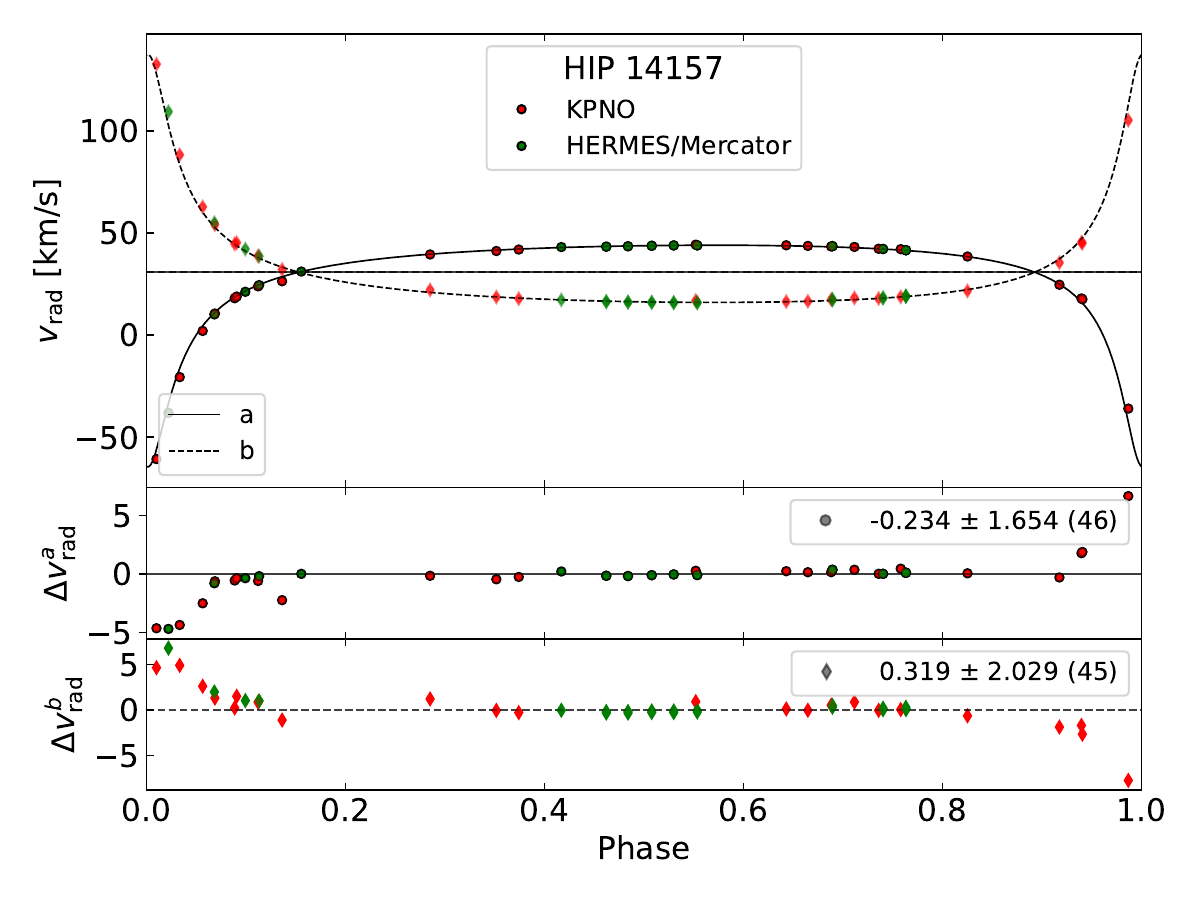}}
\resizebox{0.33\hsize}{!}{\includegraphics{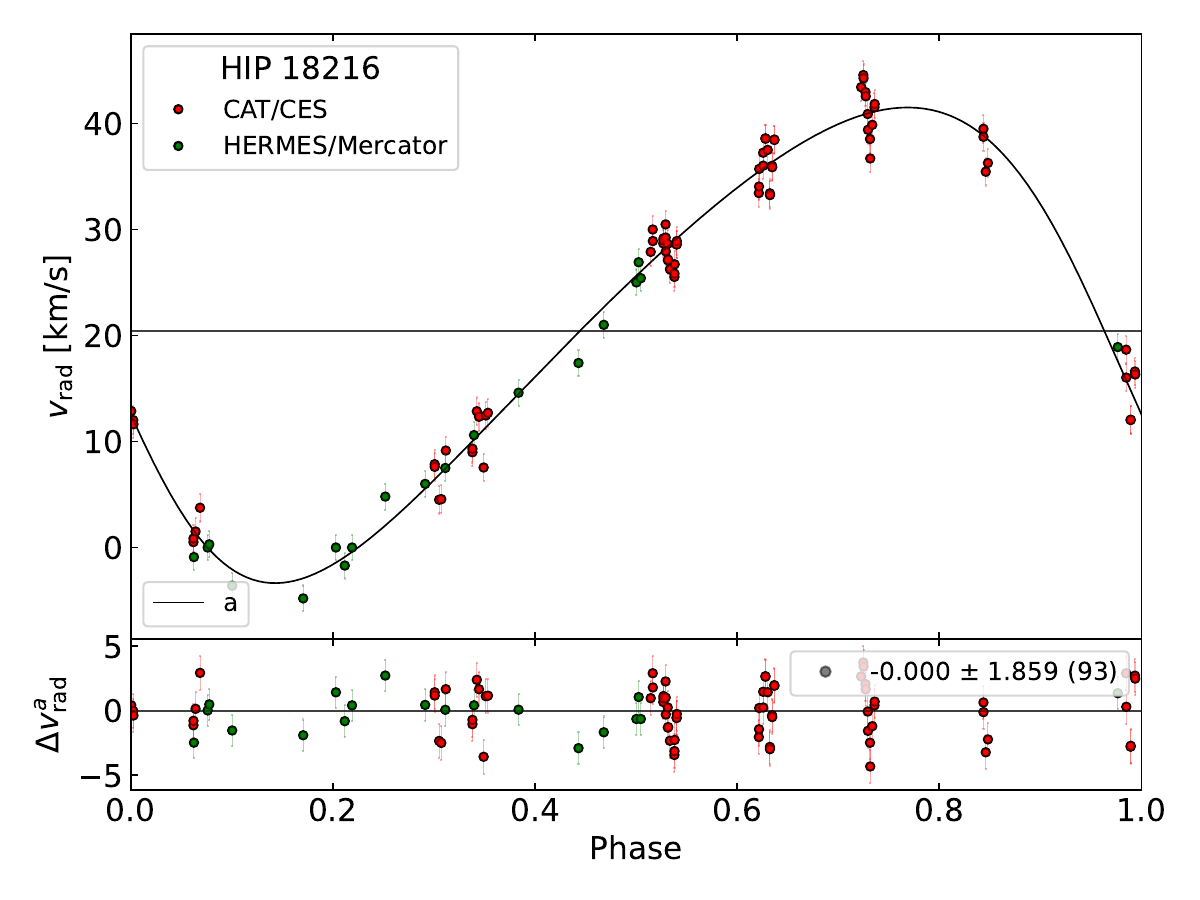}}
\resizebox{0.33\hsize}{!}{\includegraphics{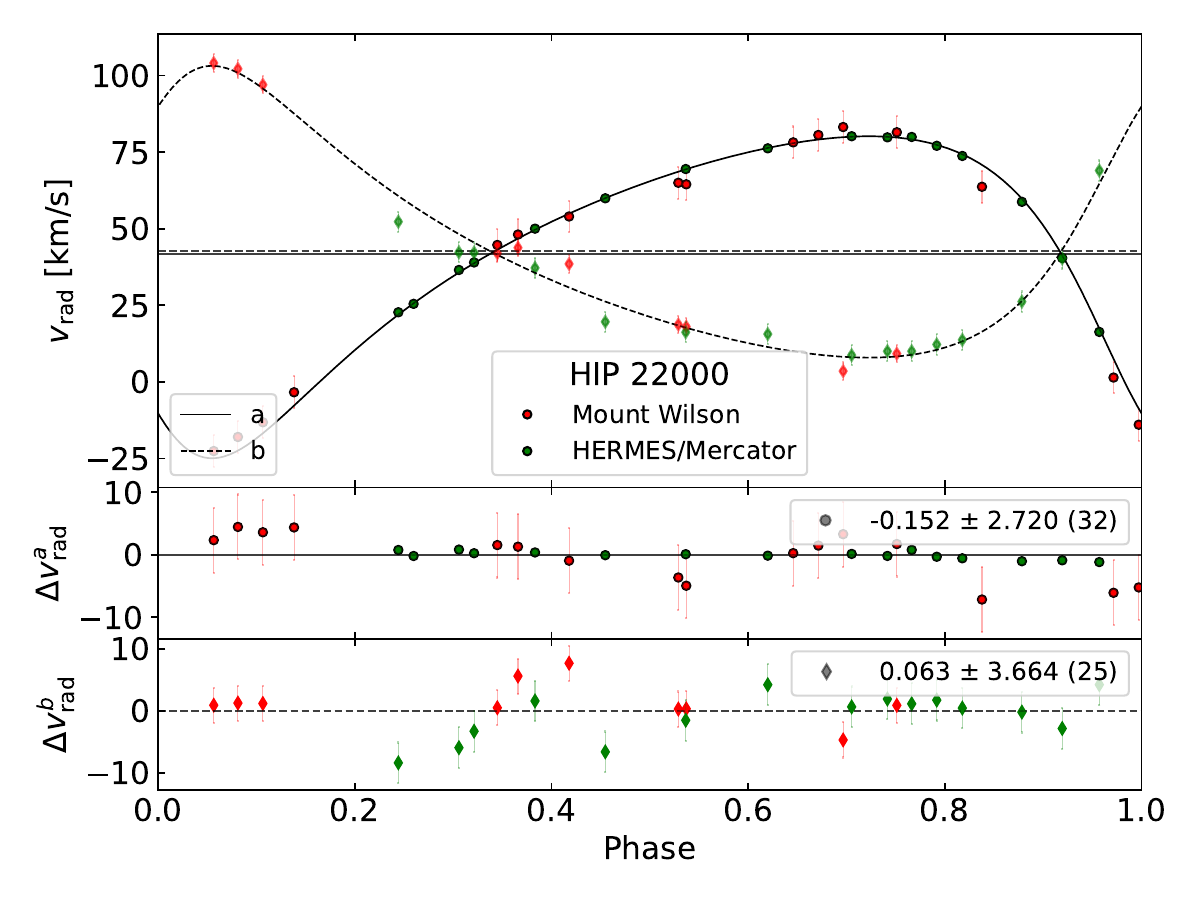}}\\
\resizebox{0.33\hsize}{!}{\includegraphics{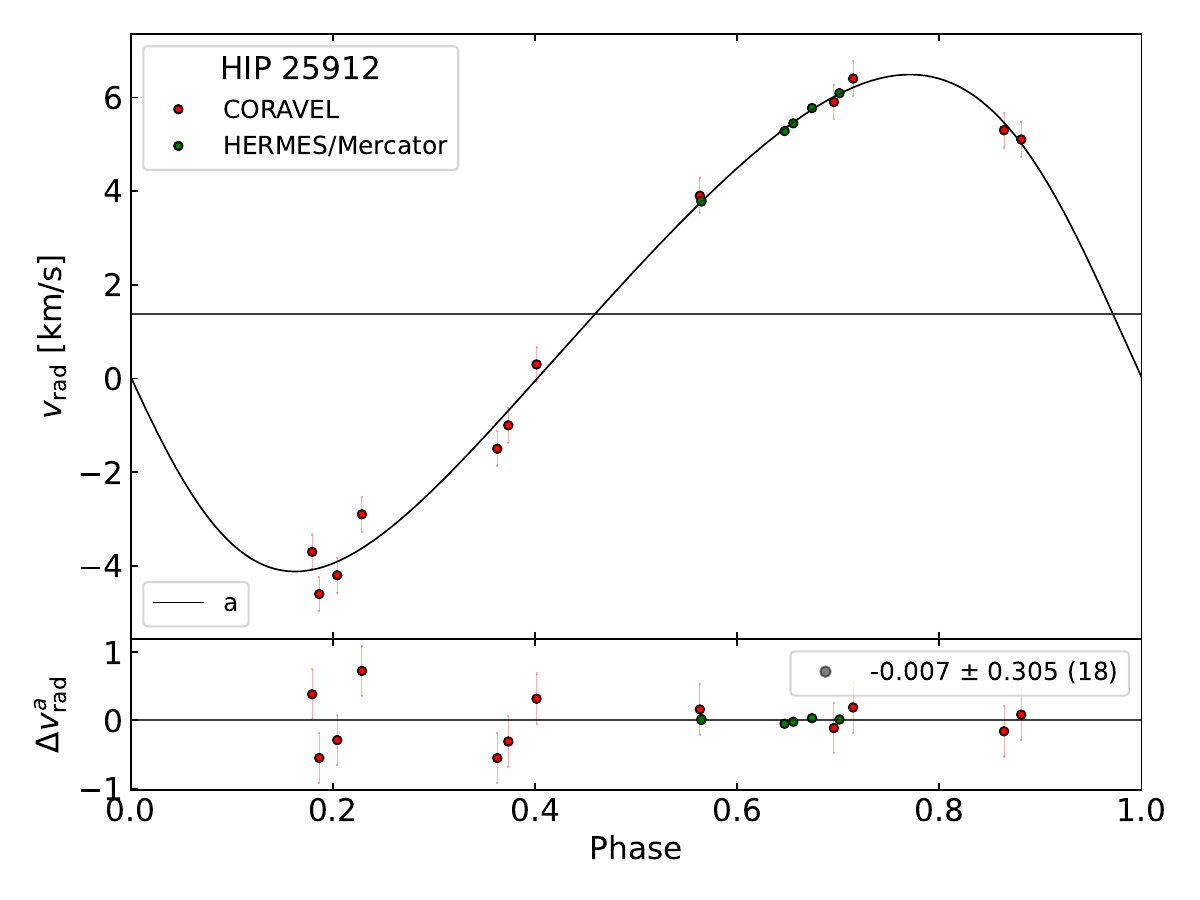}}
\resizebox{0.33\hsize}{!}{\includegraphics{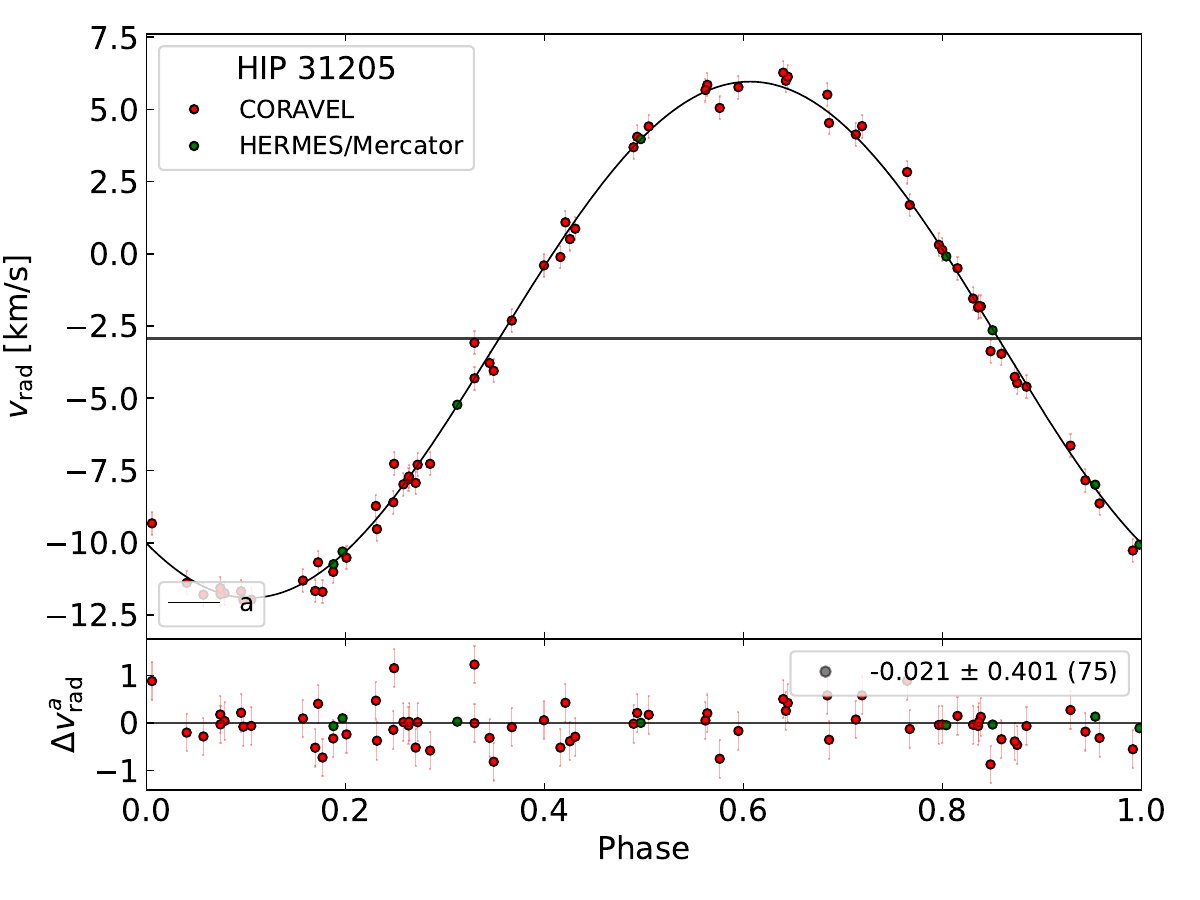}}
\resizebox{0.33\hsize}{!}{\includegraphics{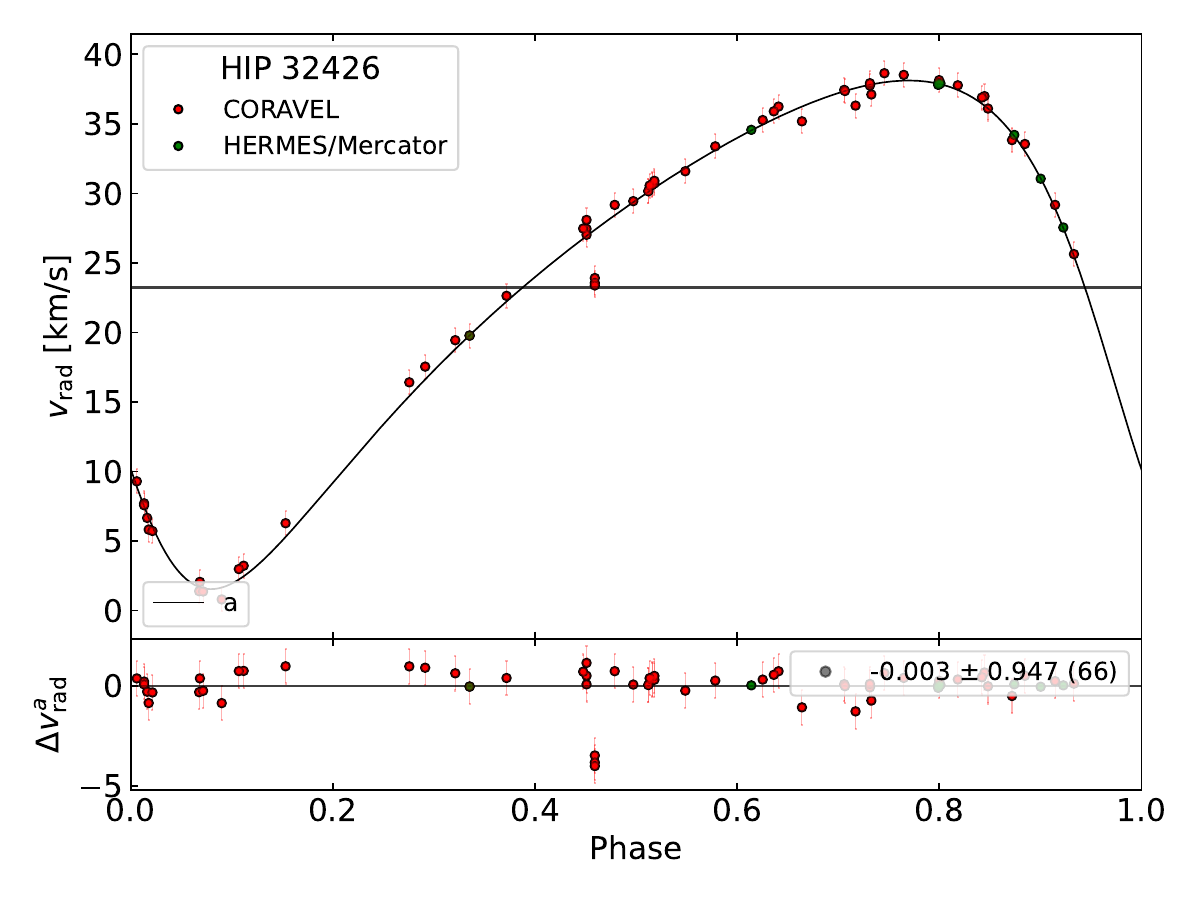}}\\
\resizebox{0.33\hsize}{!}{\includegraphics{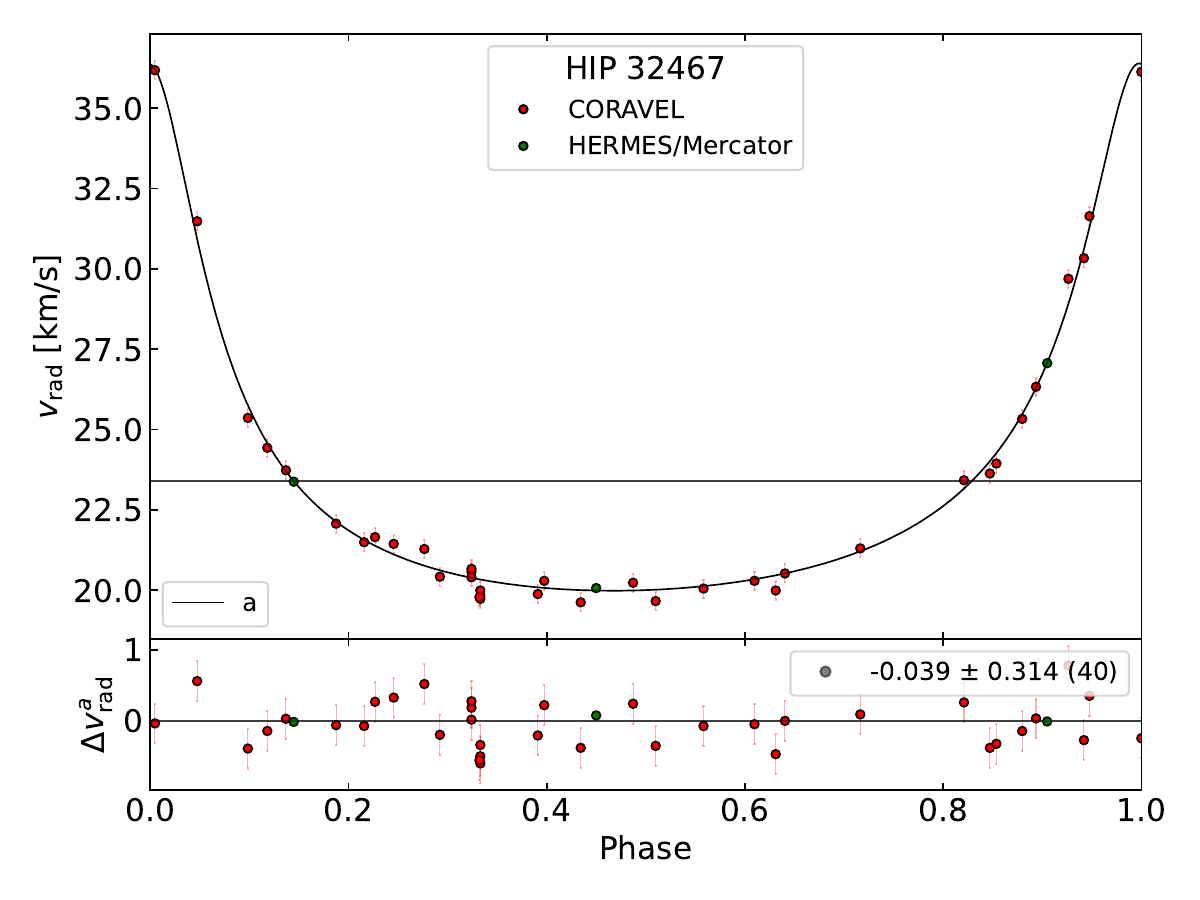}}
\resizebox{0.33\hsize}{!}{\includegraphics{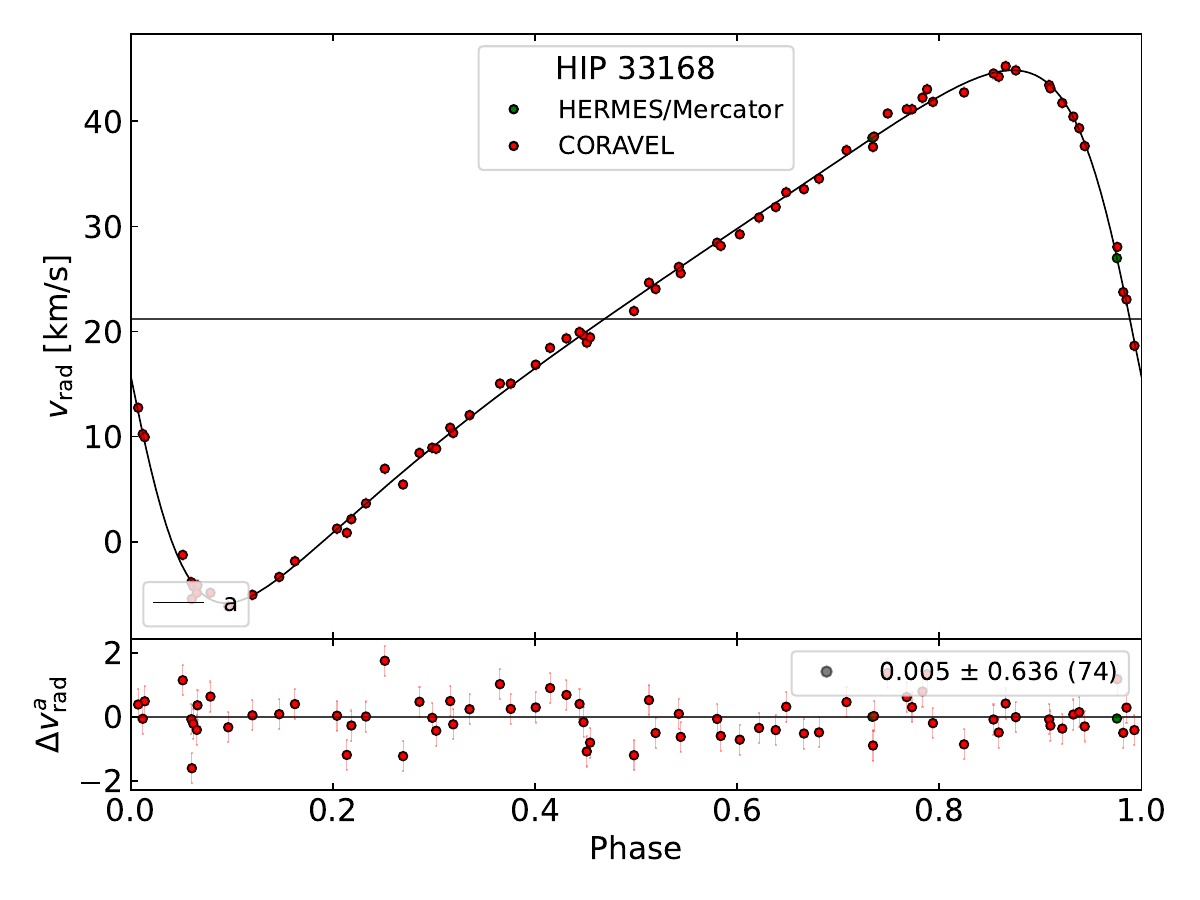}}
\resizebox{0.33\hsize}{!}{\includegraphics{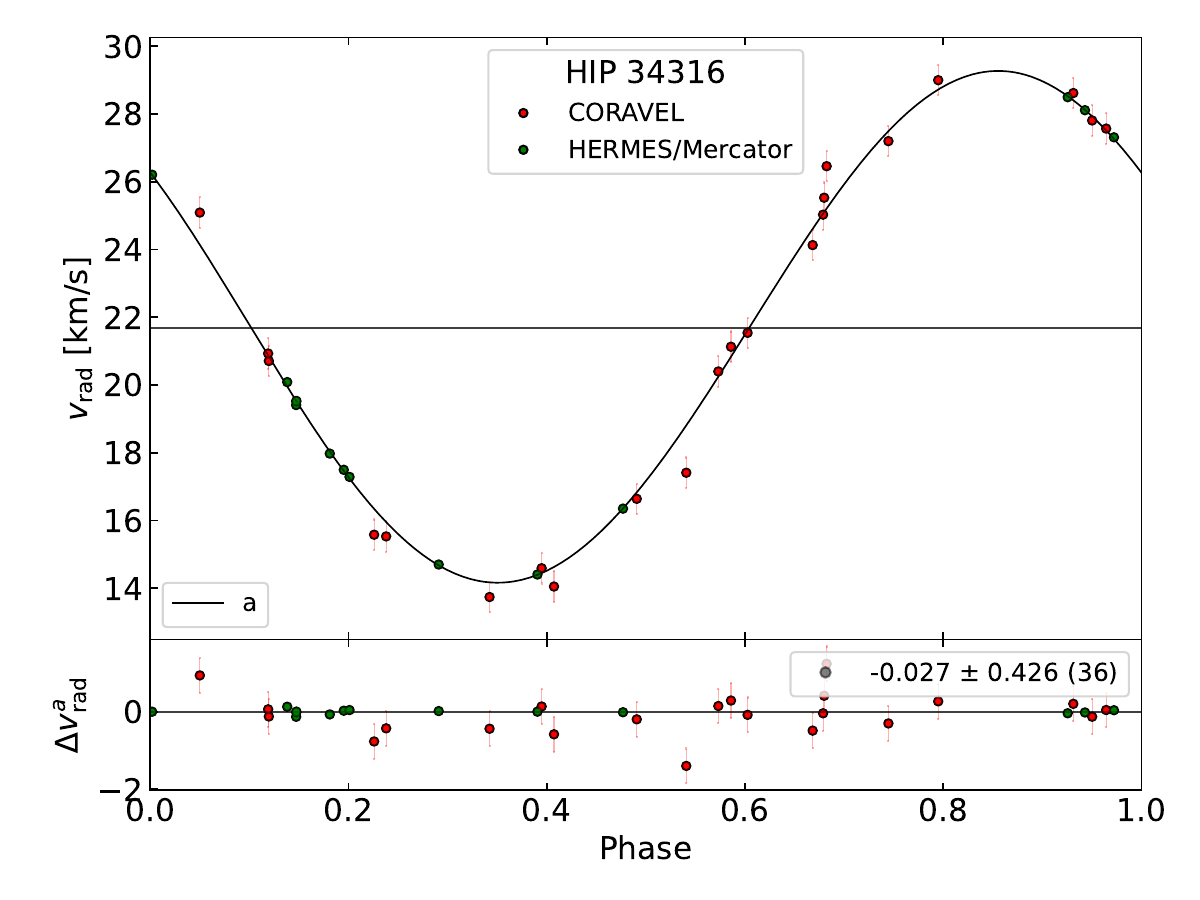}}\\
\resizebox{0.33\hsize}{!}{\includegraphics{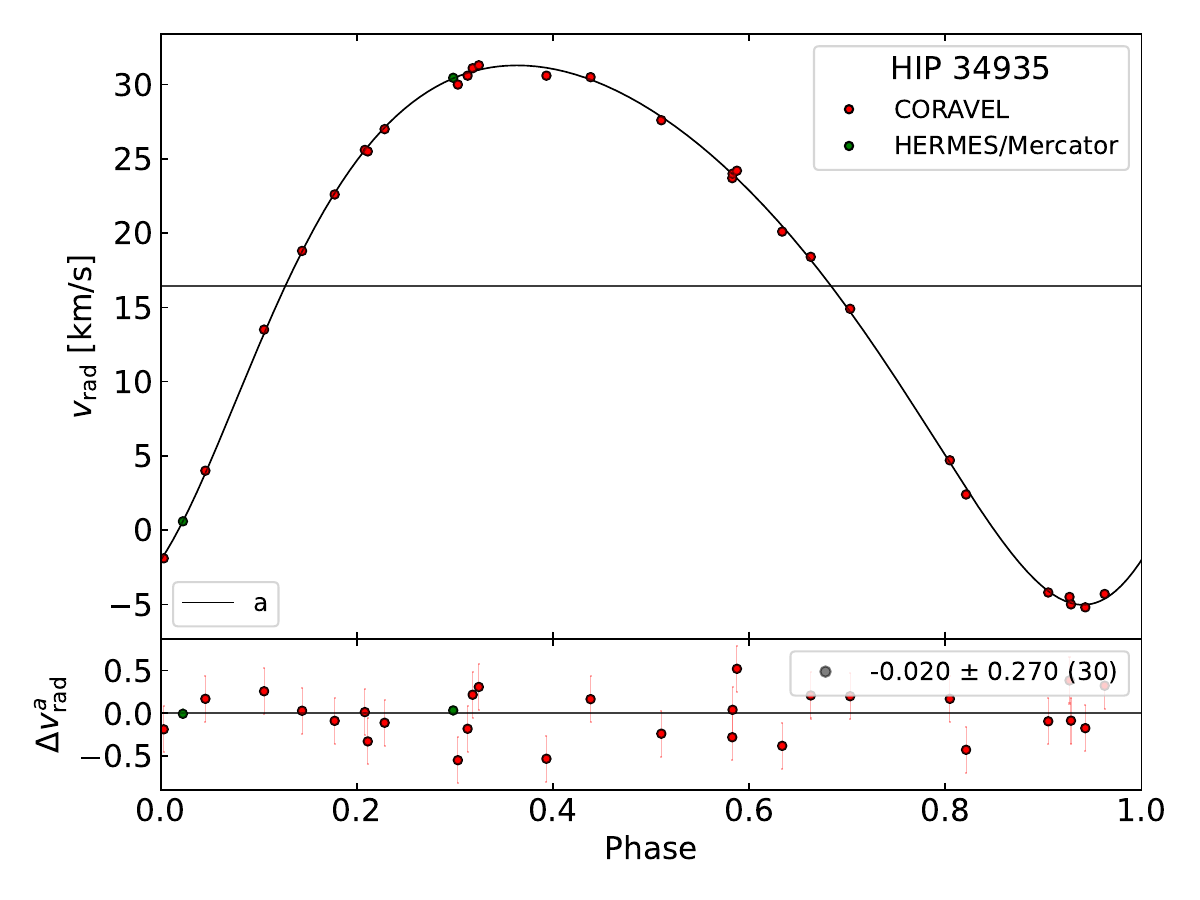}}
\resizebox{0.33\hsize}{!}{\includegraphics{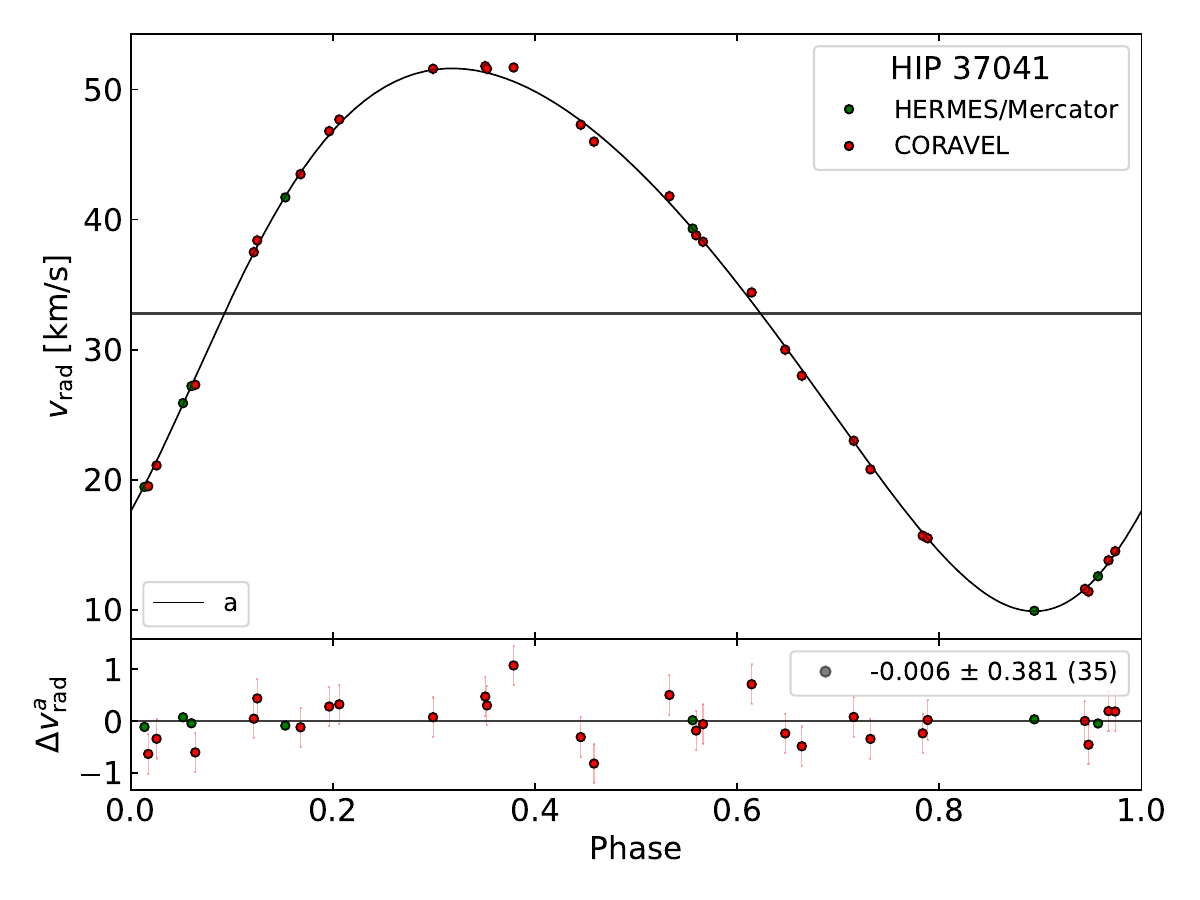}}
\resizebox{0.33\hsize}{!}{\includegraphics{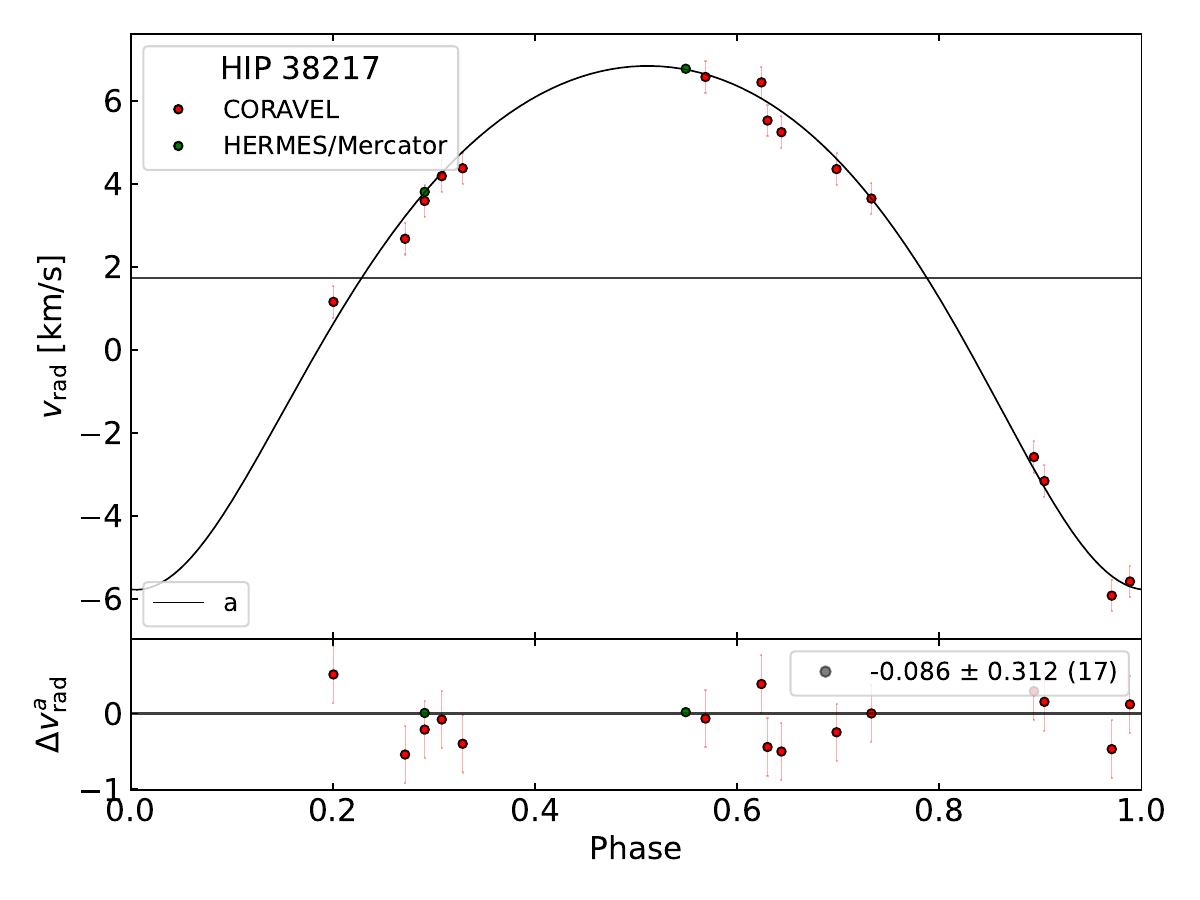}}\\
\caption[]{\label{fig:revorbits1} Revised spectroscopic orbits except for HIP 12390, 28816, 73182 and 111170 for which a simultaneous adjustment of the visual and spectroscopic orbits are displayed in Figs.~\ref{Fig:12390_HERMES}, \ref{fig:HIP073182Comb}, \ref{fig:HIP111170Comb} and \ref{fig:HIP028816Comb}.  The bottom panels display the residuals and the insert list their average, standard deviation, and number of data points (between parentheses).}
\end{figure*}

All revised orbits are listed in Table~\ref{tab:revorbits_old_new} with reference to the old SB9 orbit given in Table~\ref{tab:ProgrammeStars}. The radial velocity curves of the new orbital solutions are displayed in Fig.~\ref{fig:revorbits1}, with the system zero-point offsets included. 
In the remainder of this Appendix, we give comments on individual systems.\medskip\\

\addtocounter{figure}{-1}
\begin{figure*}[ht]

\resizebox{0.33\hsize}{!}{\includegraphics{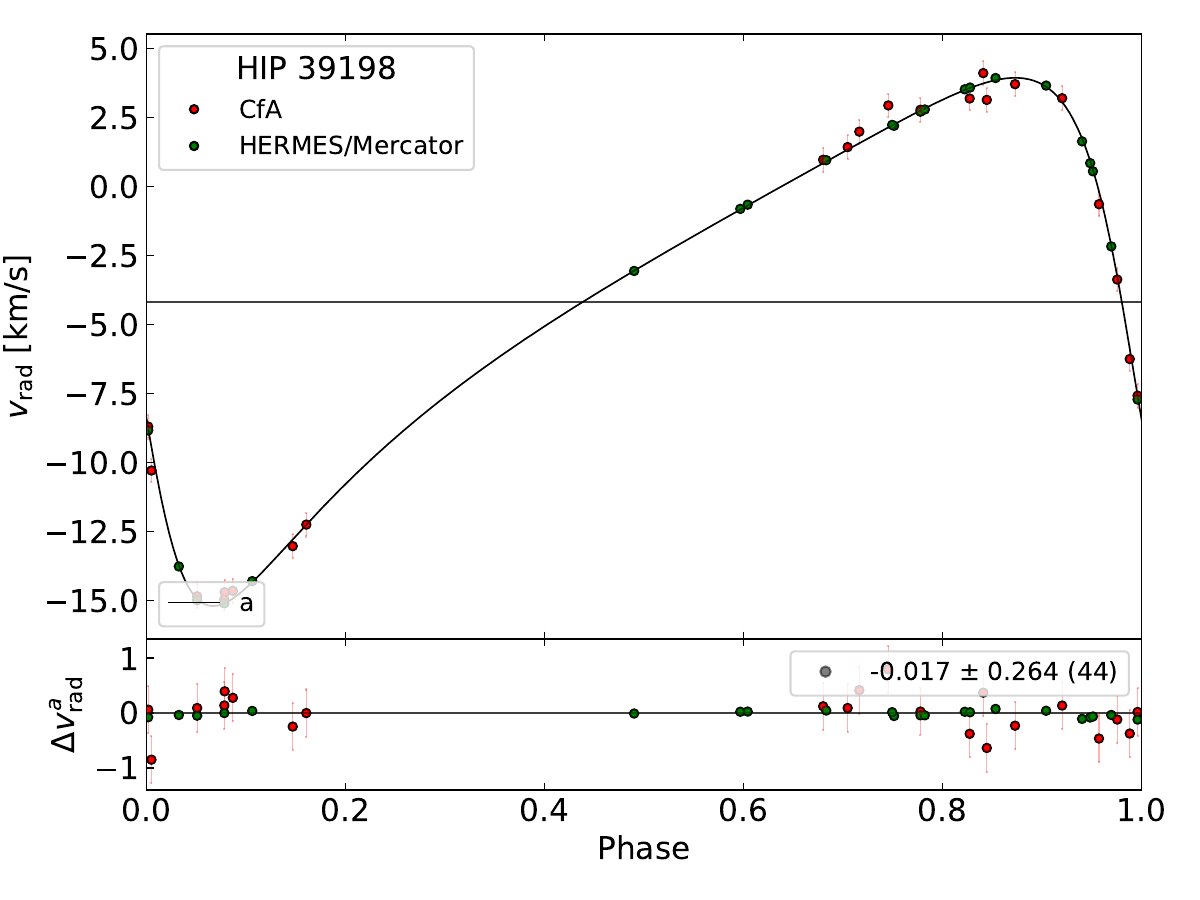}}
\resizebox{0.33\hsize}{!}{\includegraphics{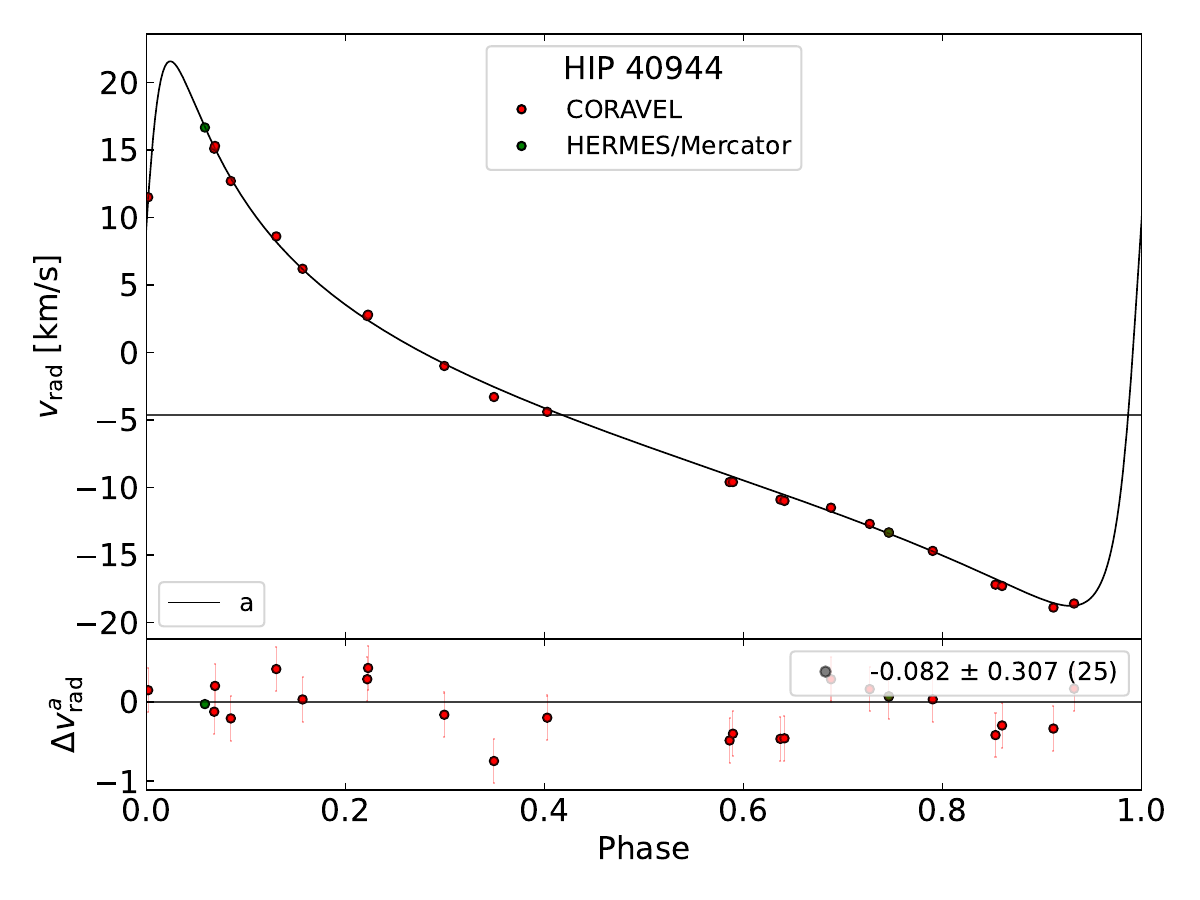}}
\resizebox{0.33\hsize}{!}{\includegraphics{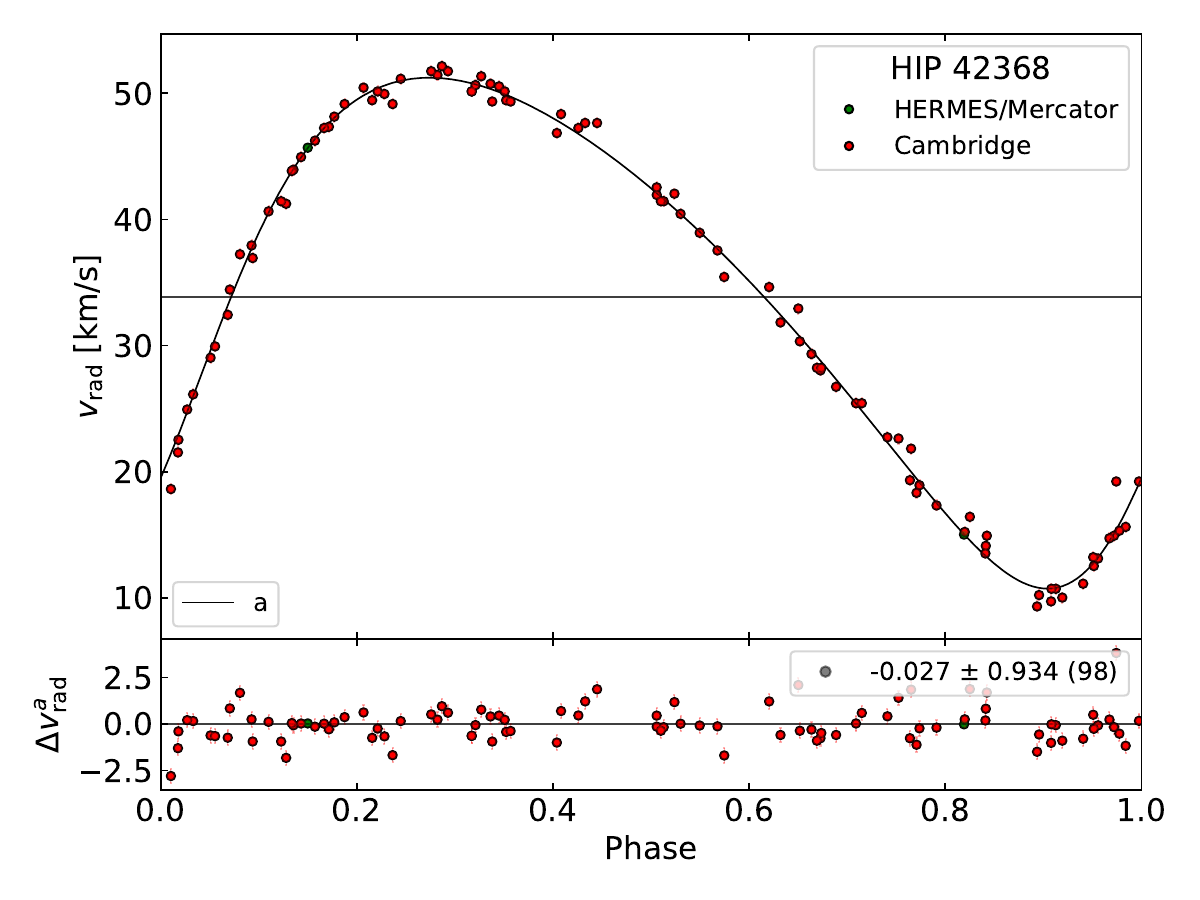}}\\
\resizebox{0.33\hsize}{!}{\includegraphics{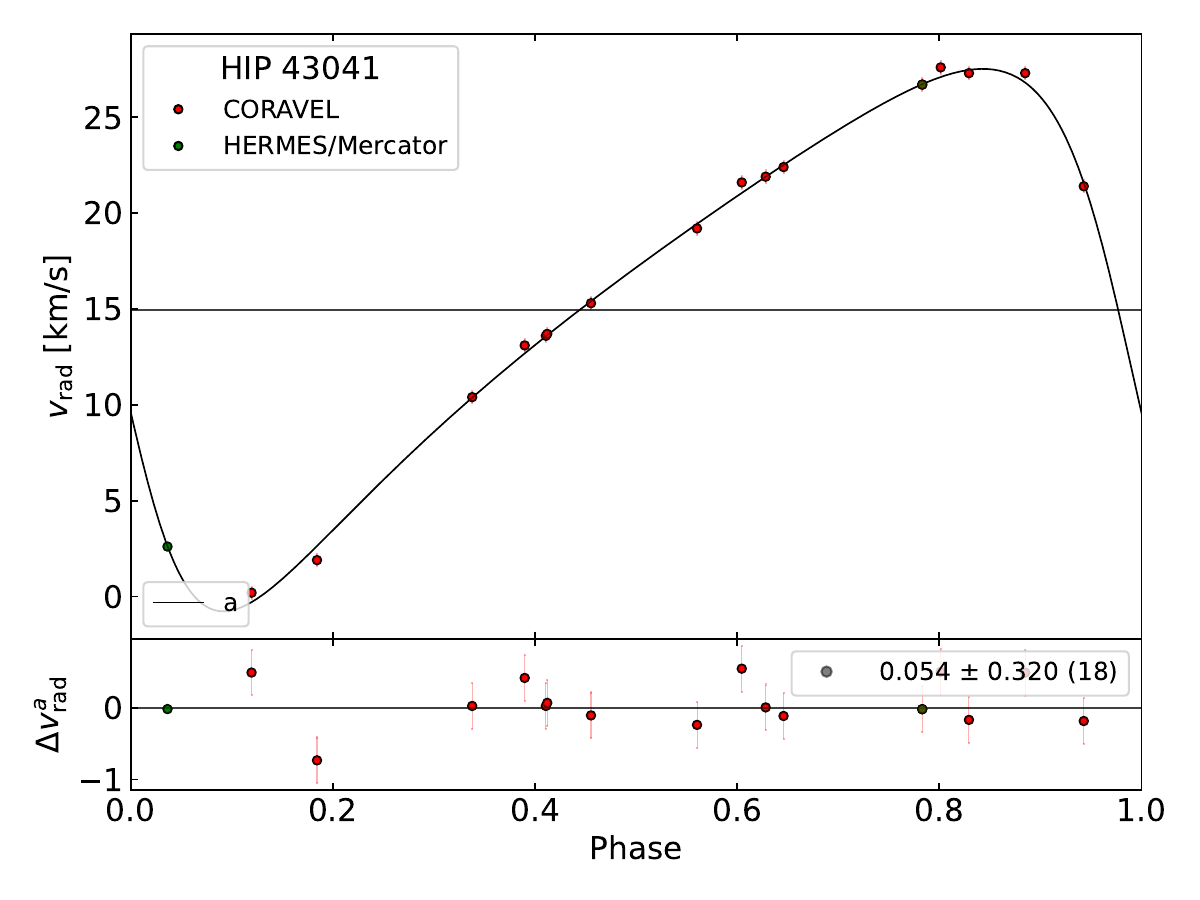}}
\resizebox{0.33\hsize}{!}{\includegraphics{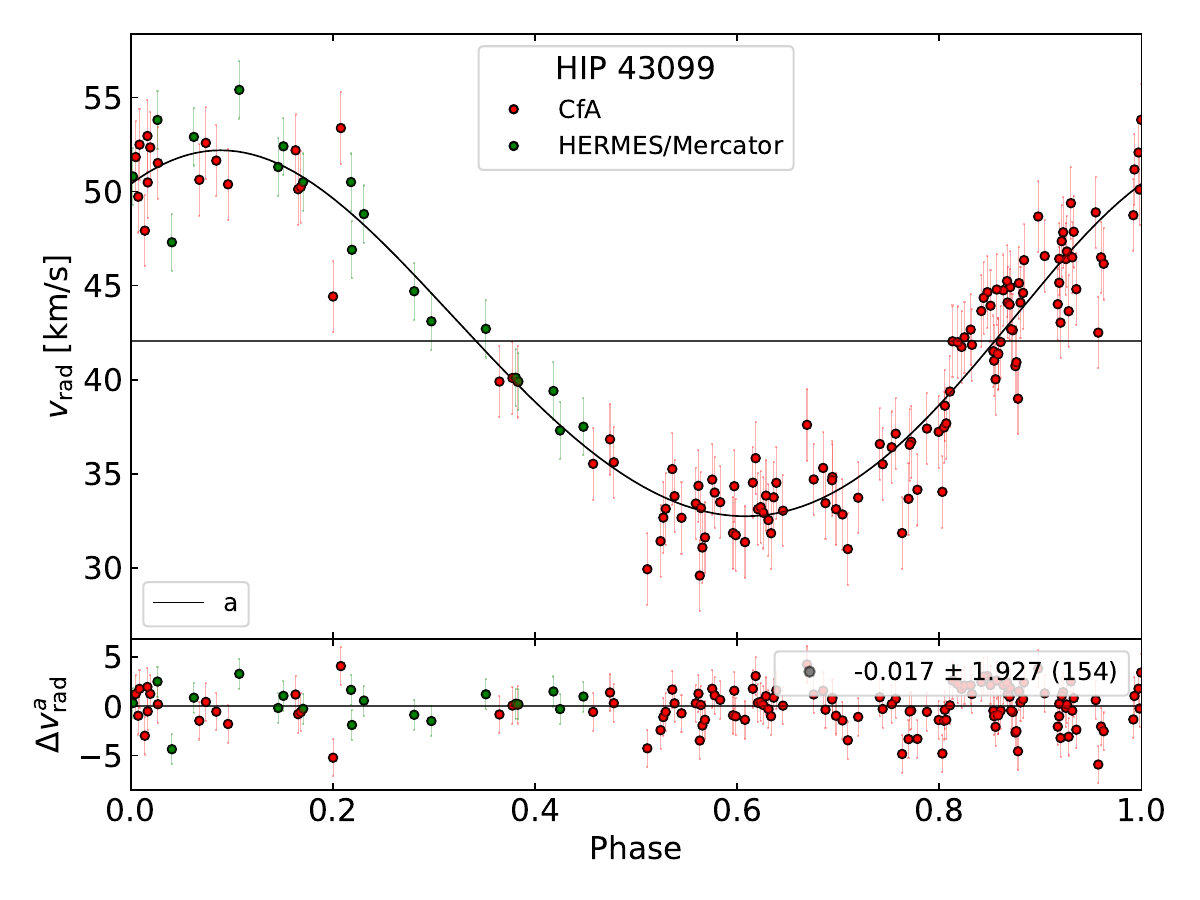}}
\resizebox{0.33\hsize}{!}{\includegraphics{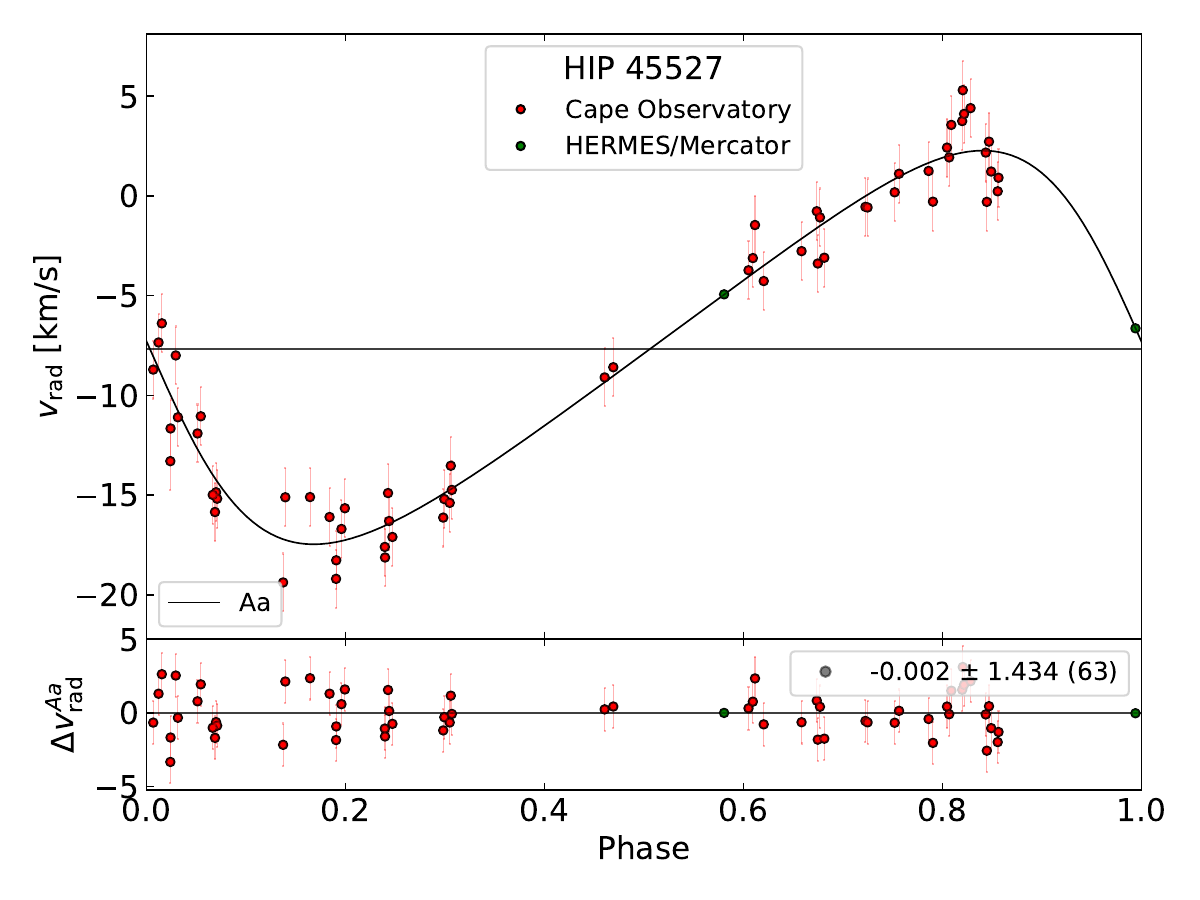}}\\
\resizebox{0.33\hsize}{!}{\includegraphics{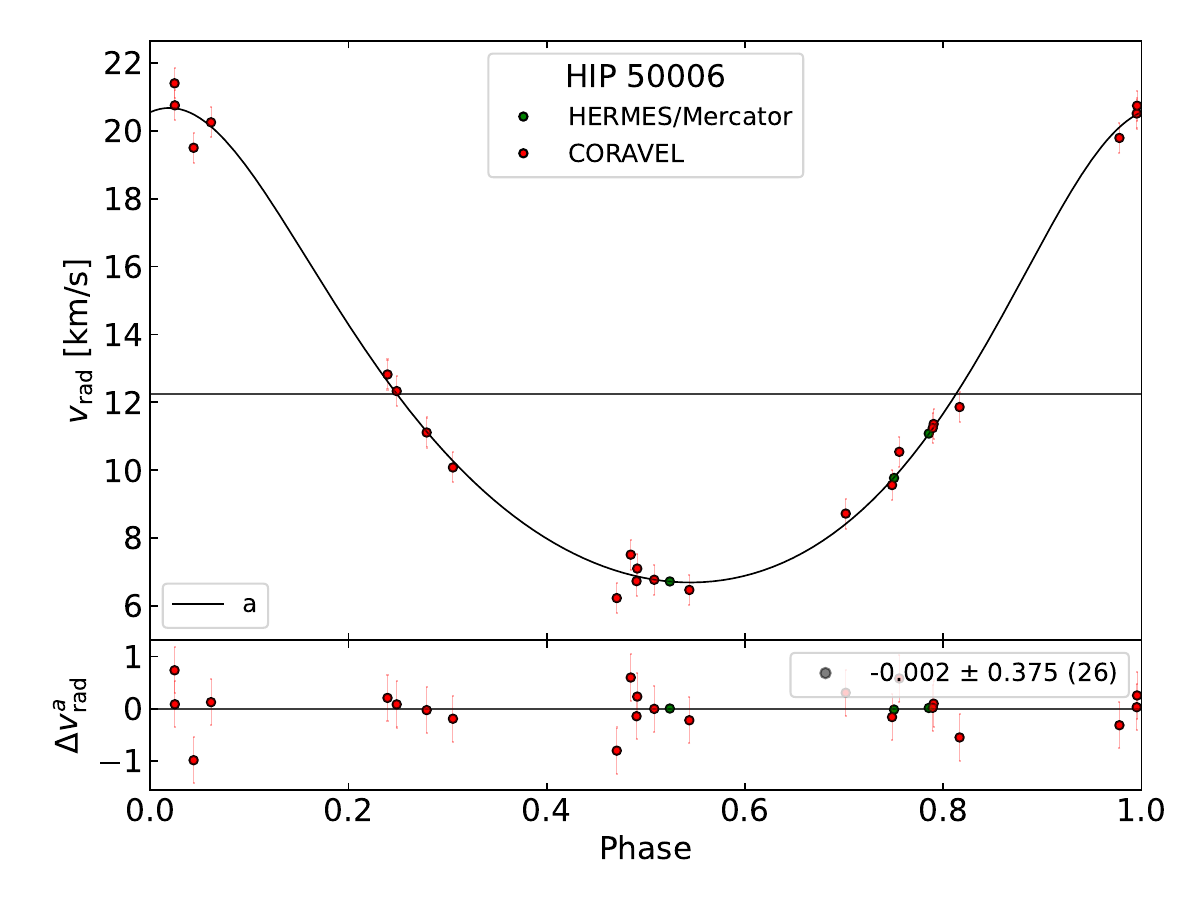}}
\resizebox{0.33\hsize}{!}{\includegraphics{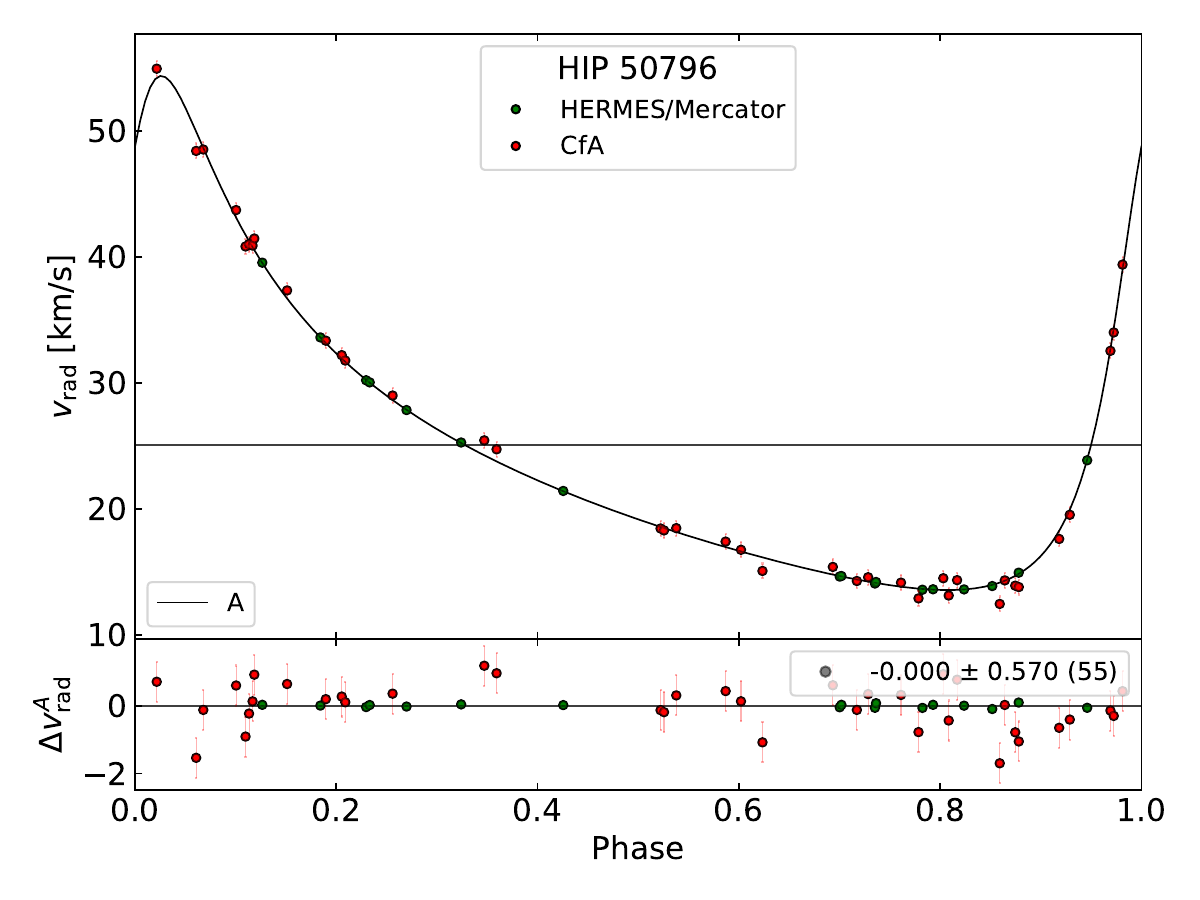}}
\resizebox{0.33\hsize}{!}{\includegraphics{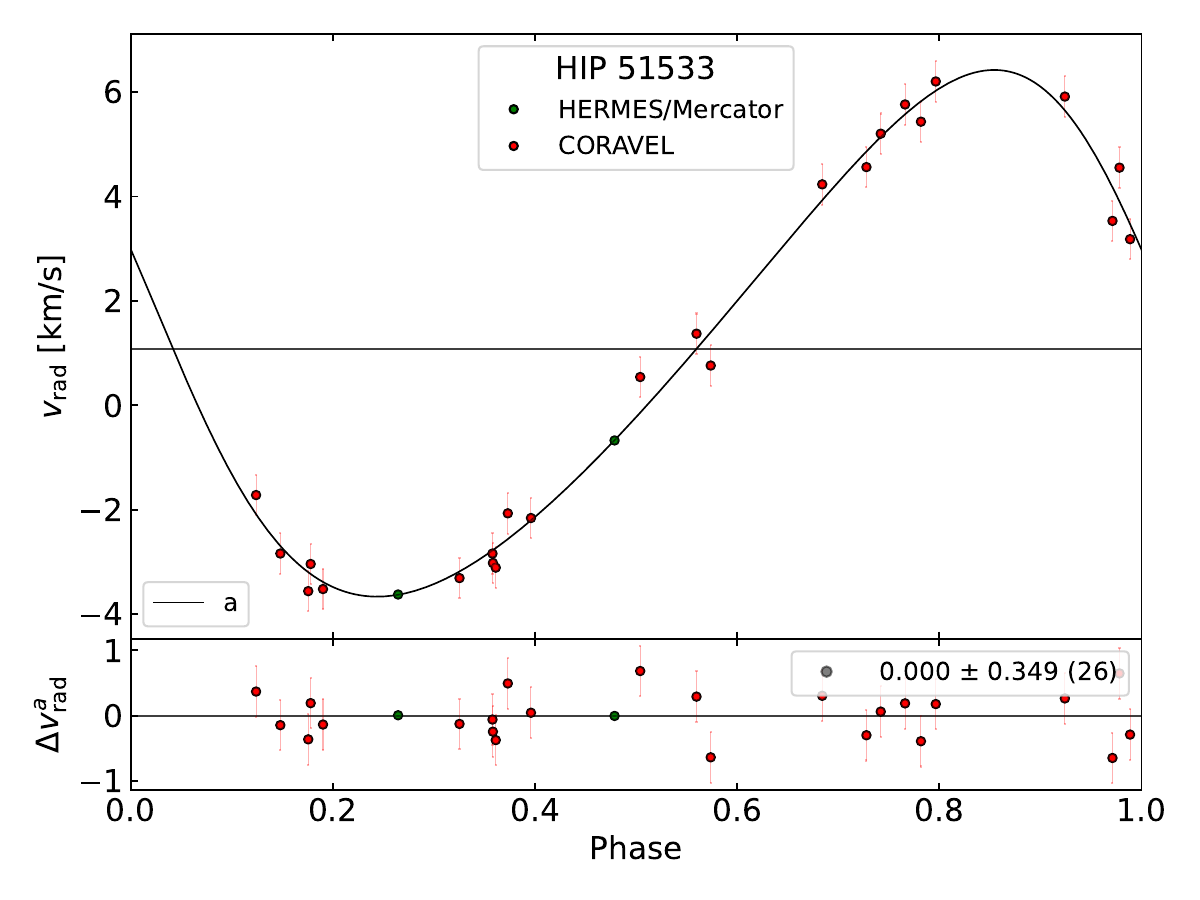}}\\
\resizebox{0.33\hsize}{!}{\includegraphics{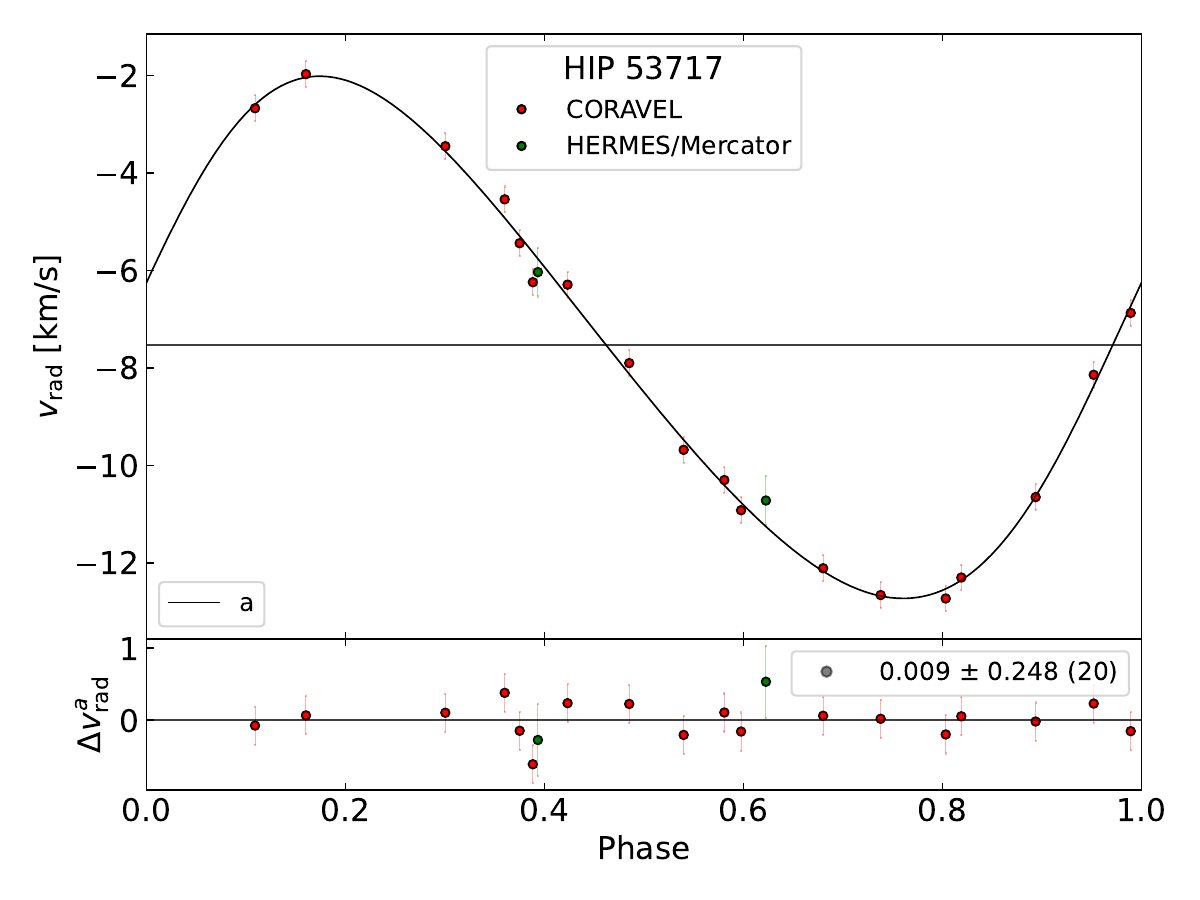}}
\resizebox{0.33\hsize}{!}{\includegraphics{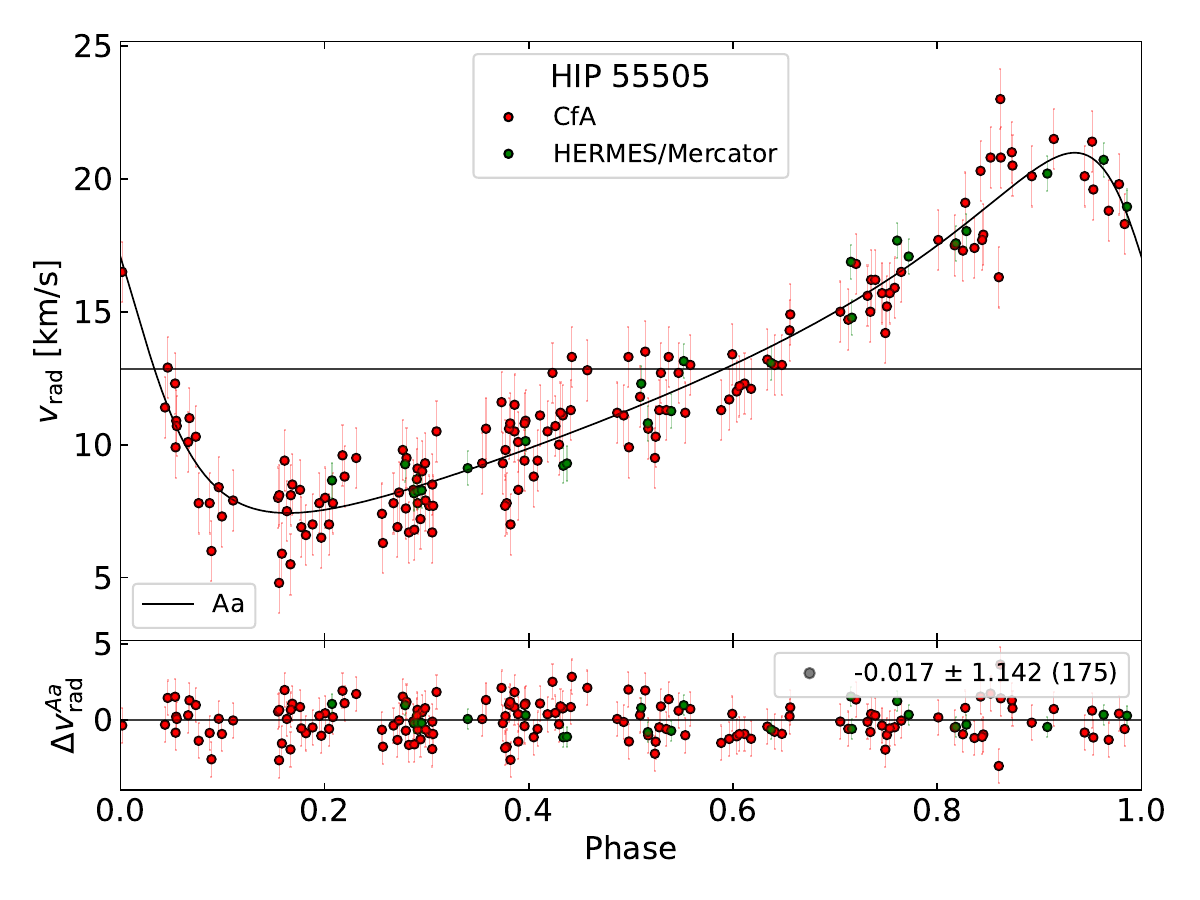}}
\resizebox{0.33\hsize}{!}{\includegraphics{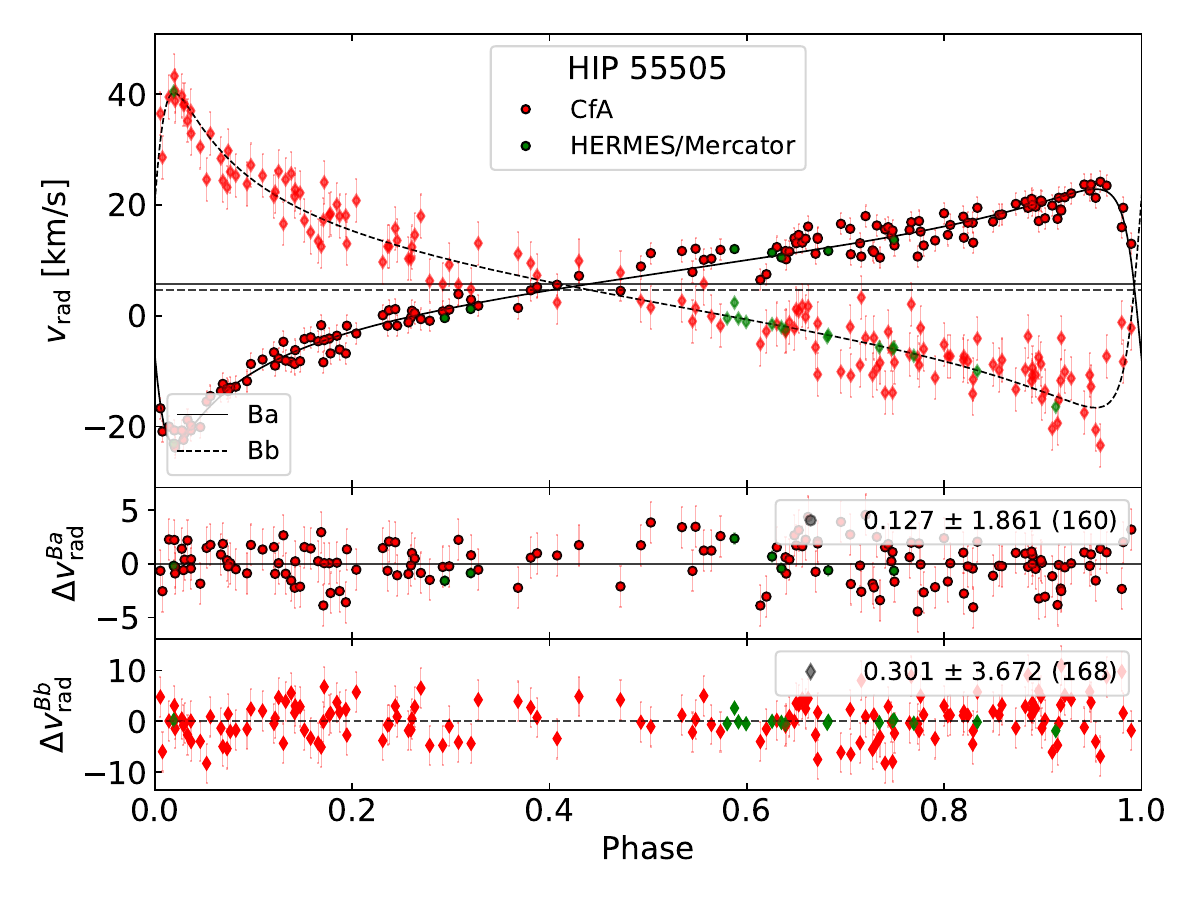}}\\
\resizebox{0.33\hsize}{!}{\includegraphics{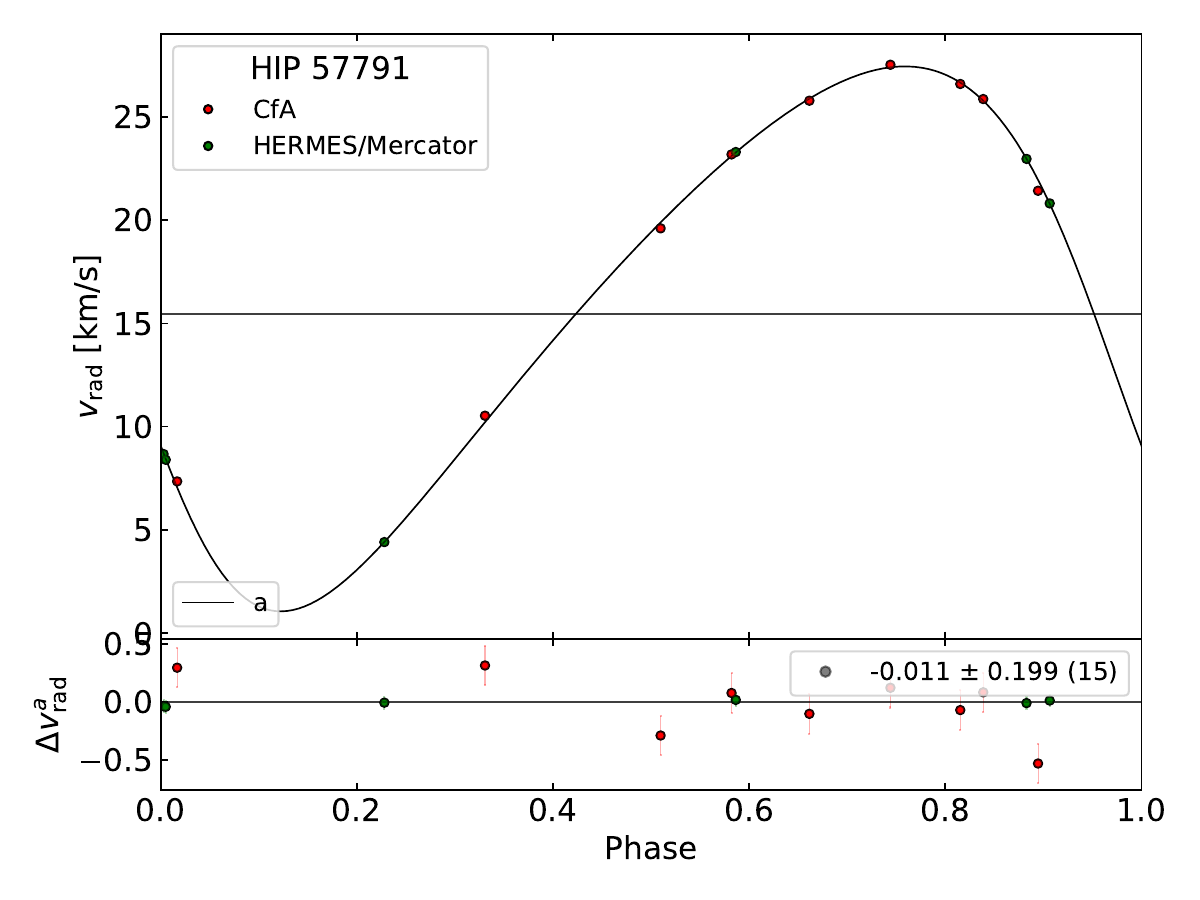}}
\resizebox{0.33\hsize}{!}{\includegraphics{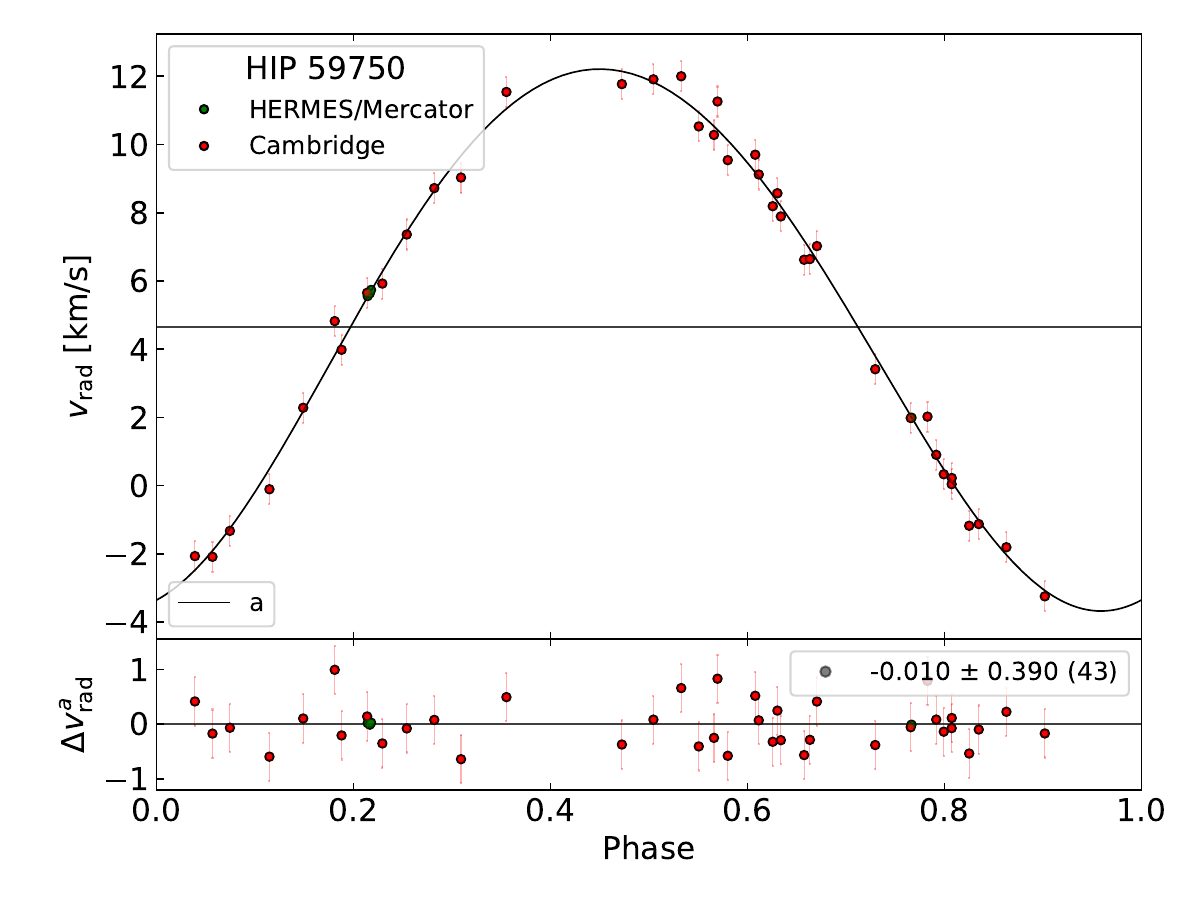}}
\resizebox{0.33\hsize}{!}{\includegraphics{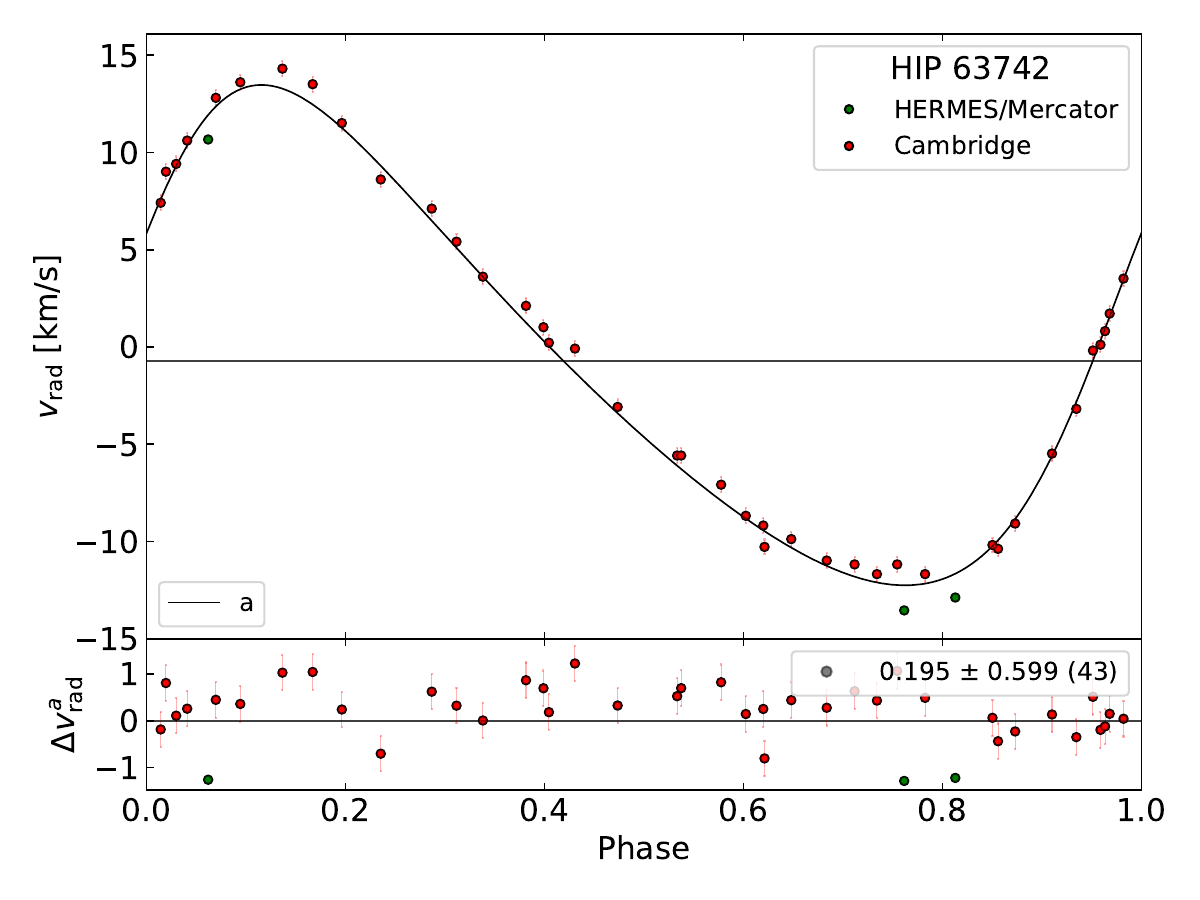}}\\
\caption[]{\label{fig:revorbits2} Continued. The two plots displayed for HIP 55505 correspond to component A and component B.}
\end{figure*}

\addtocounter{figure}{-1}
\begin{figure*}[ht]
\resizebox{0.33\hsize}{!}{\includegraphics{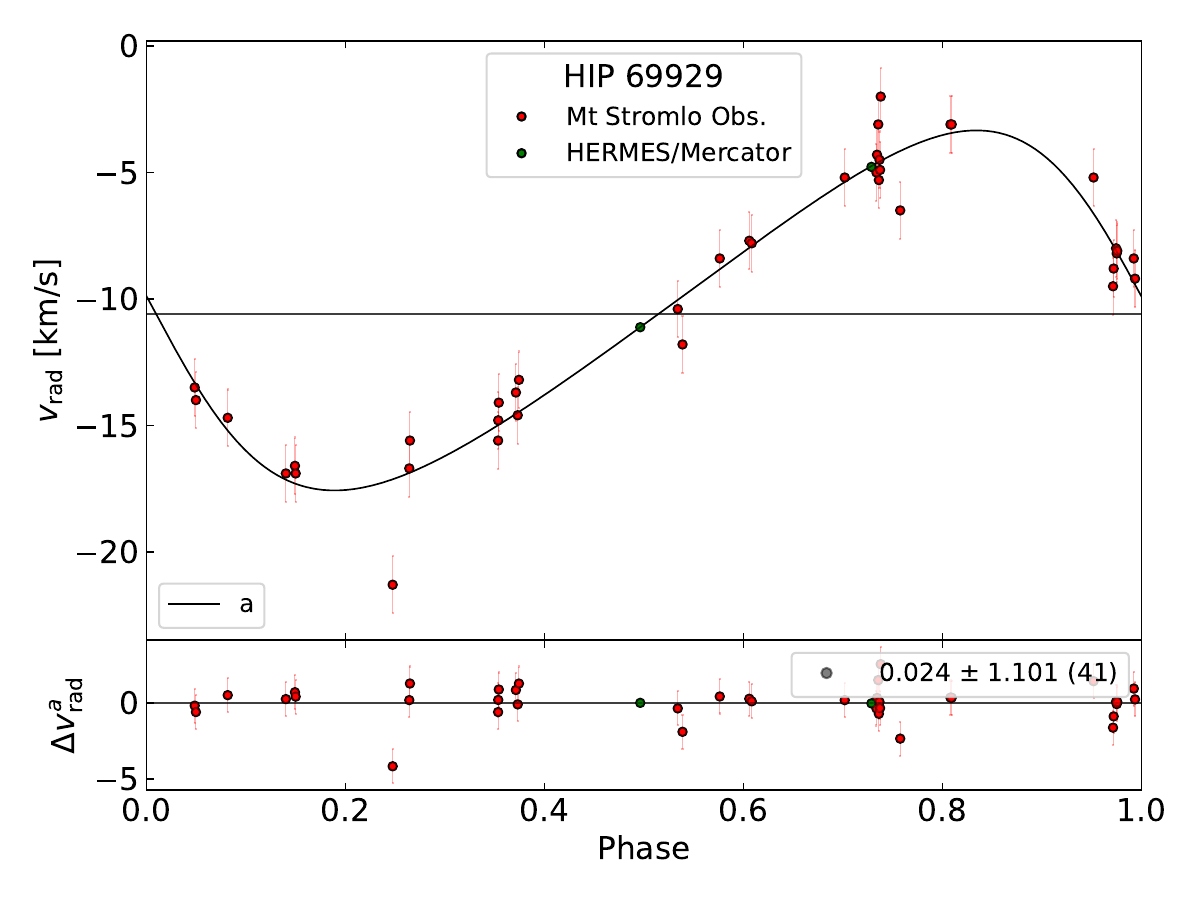}}
\resizebox{0.33\hsize}{!}{\includegraphics{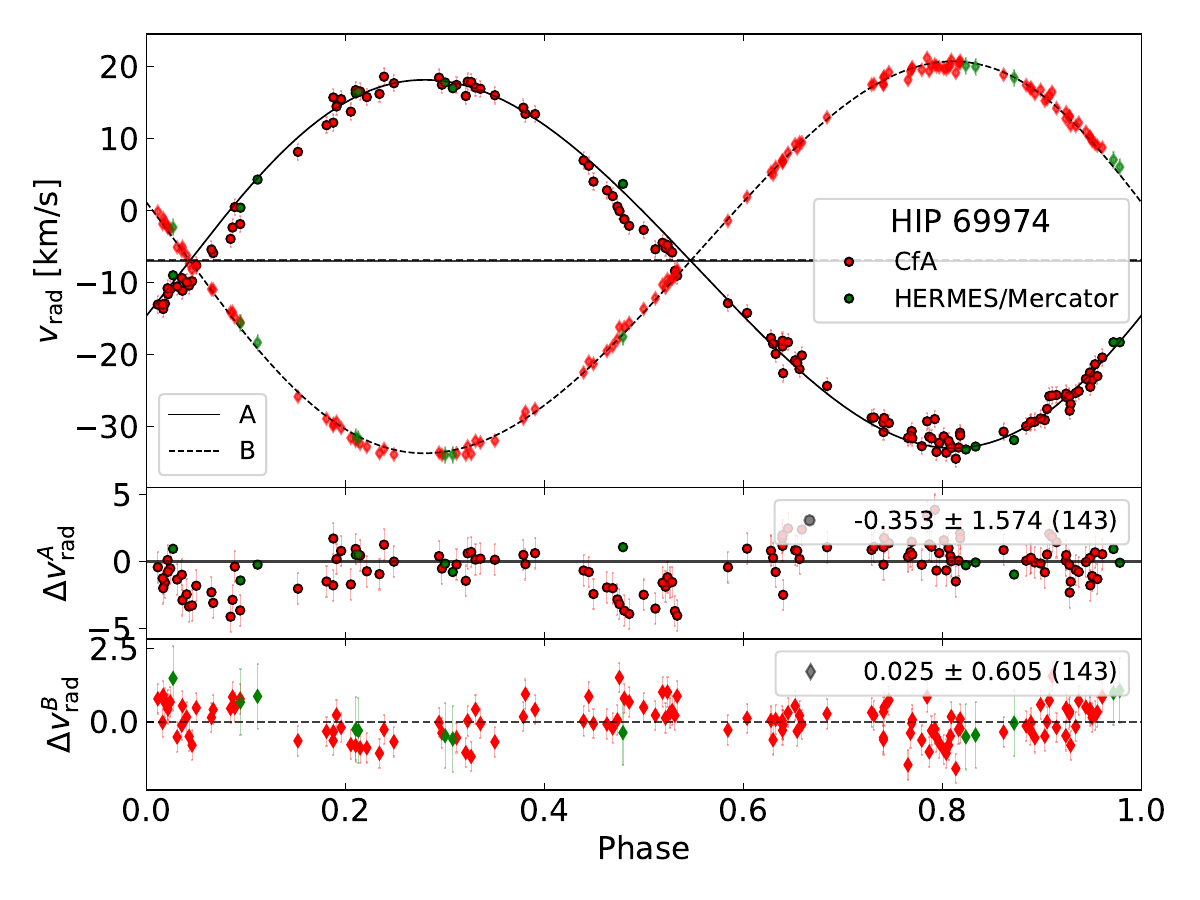}}
\resizebox{0.33\hsize}{!}{\includegraphics{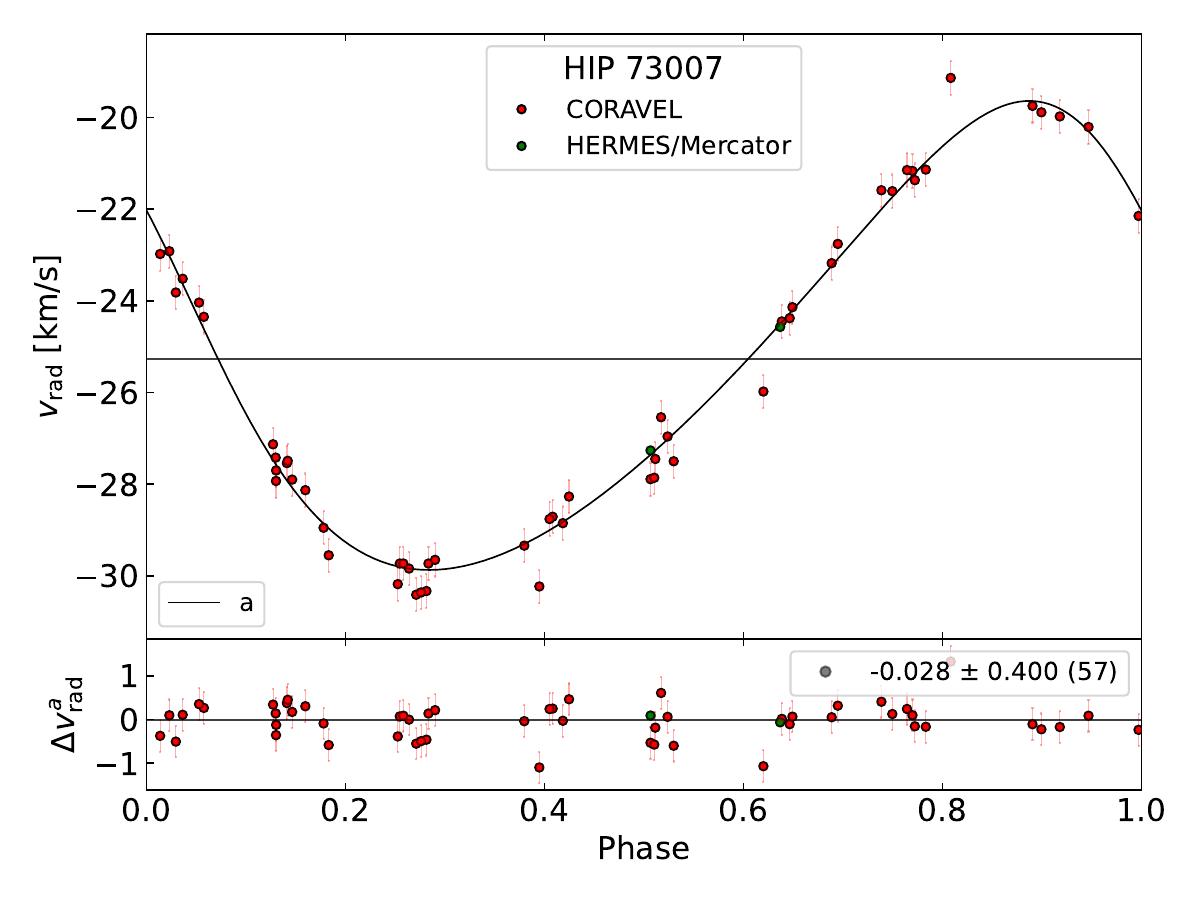}}\\
\resizebox{0.33\hsize}{!}{\includegraphics{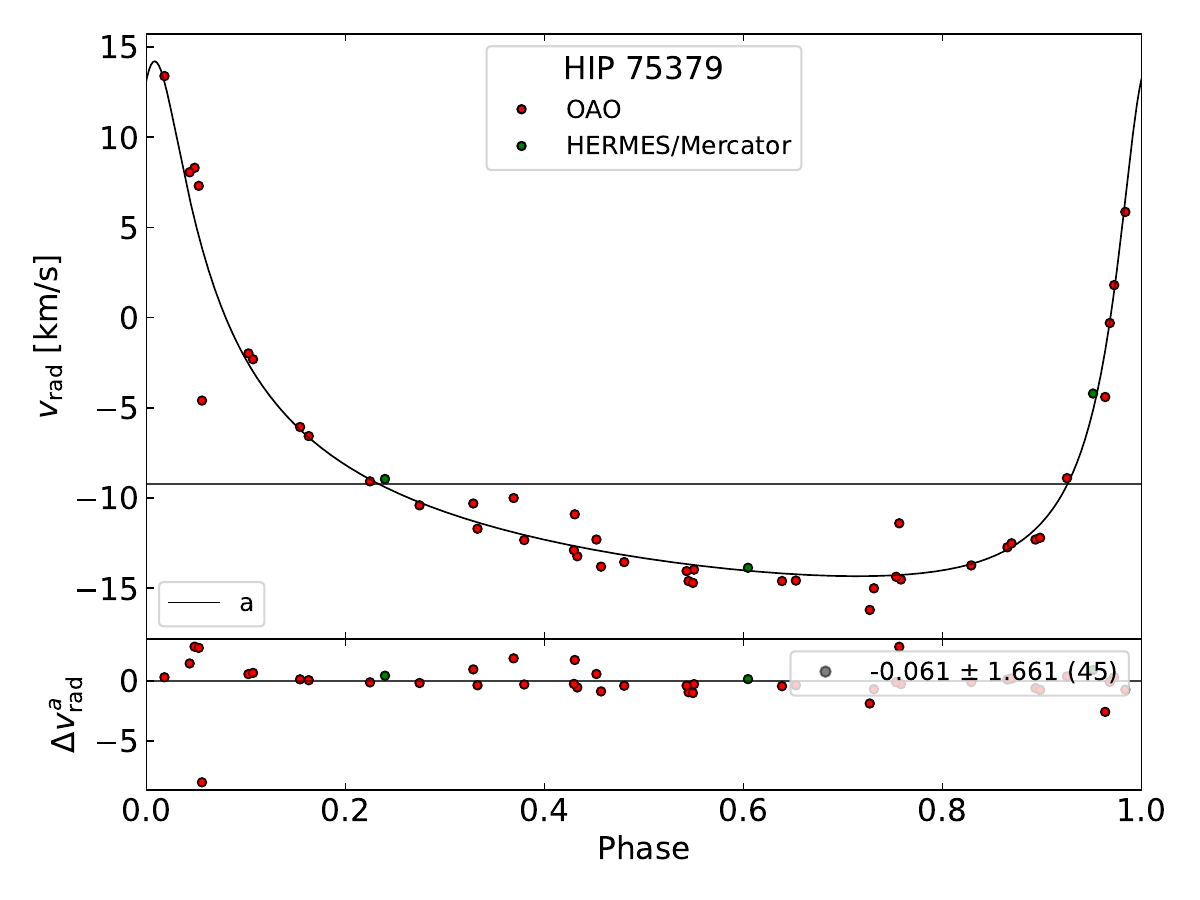}}
\resizebox{0.33\hsize}{!}{\includegraphics{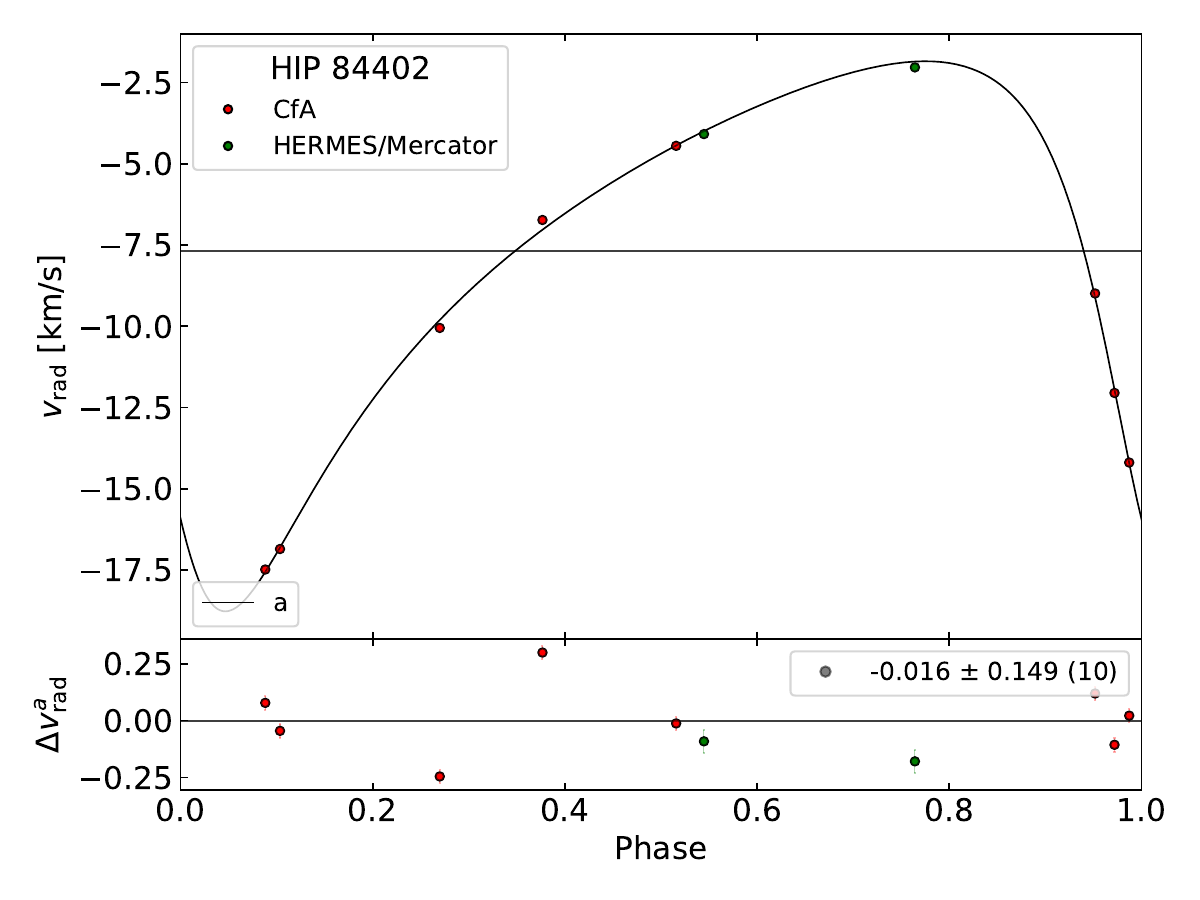}}
\resizebox{0.33\hsize}{!}{\includegraphics{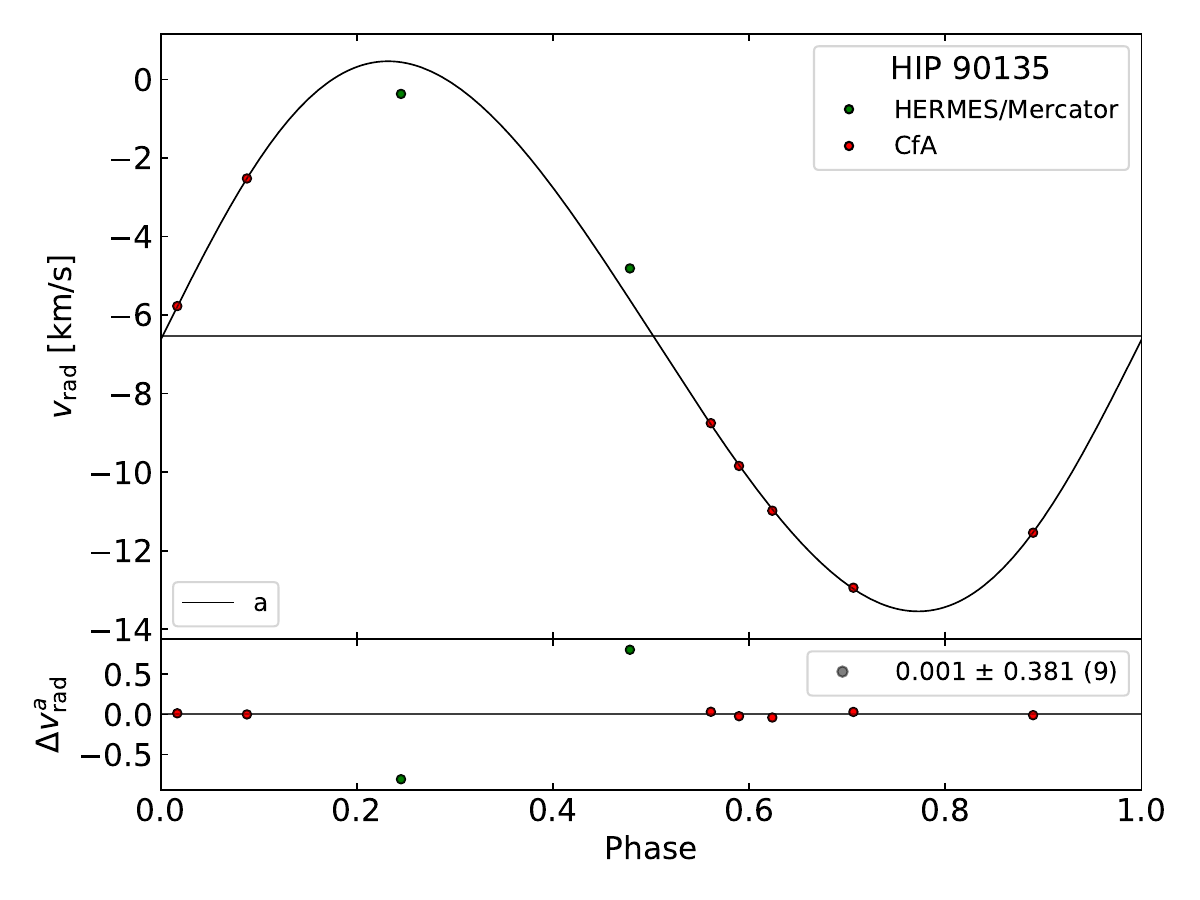}}\\
\resizebox{0.33\hsize}{!}{\includegraphics{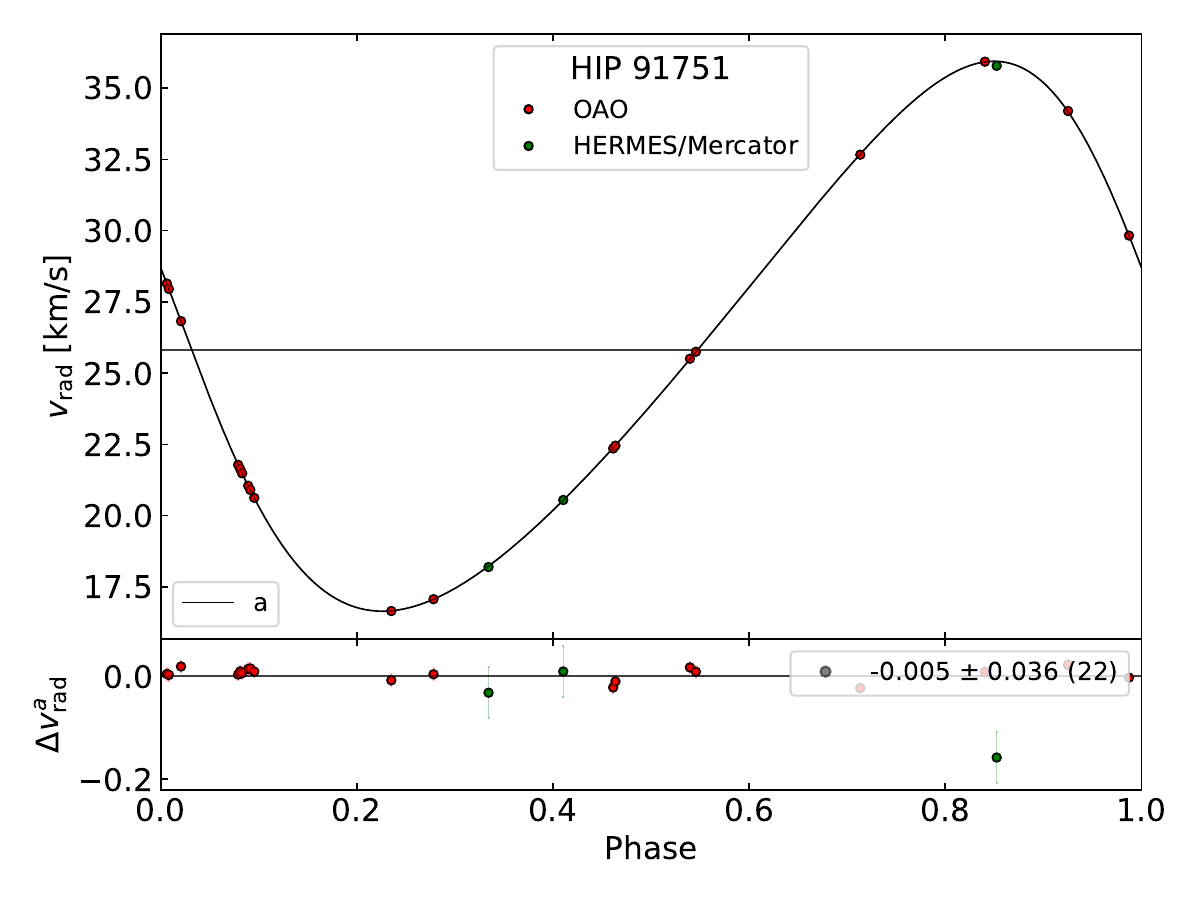}}
\resizebox{0.33\hsize}{!}{\includegraphics{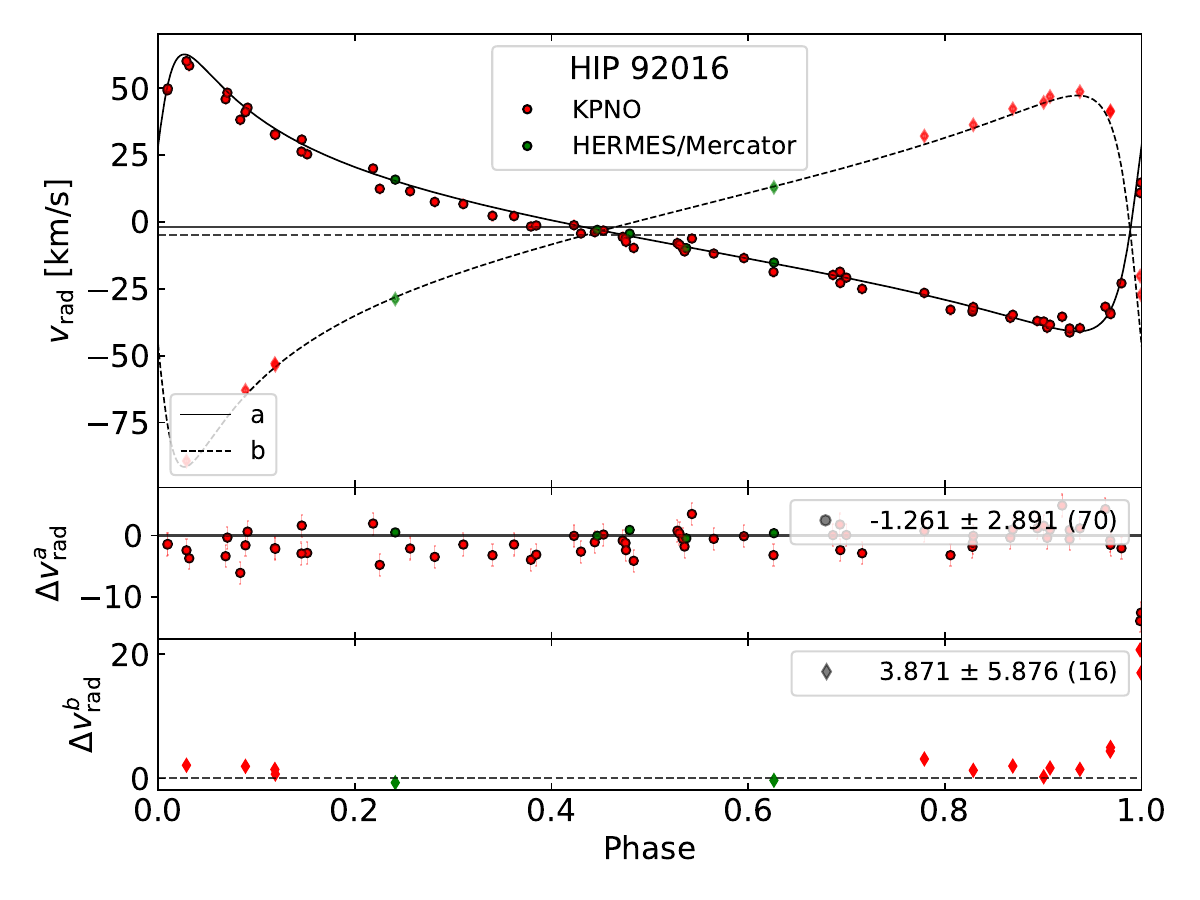}}
\resizebox{0.33\hsize}{!}{\includegraphics{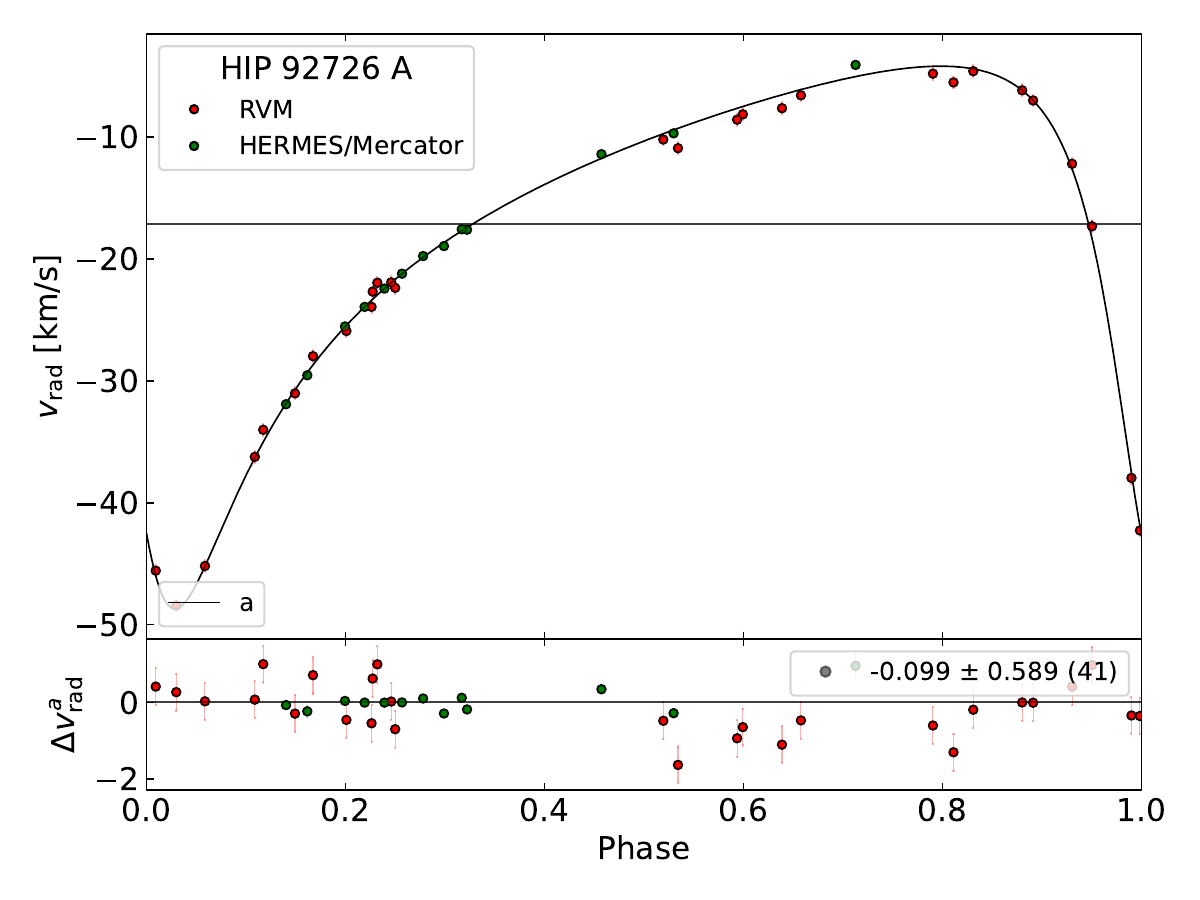}}\\
\resizebox{0.33\hsize}{!}{\includegraphics{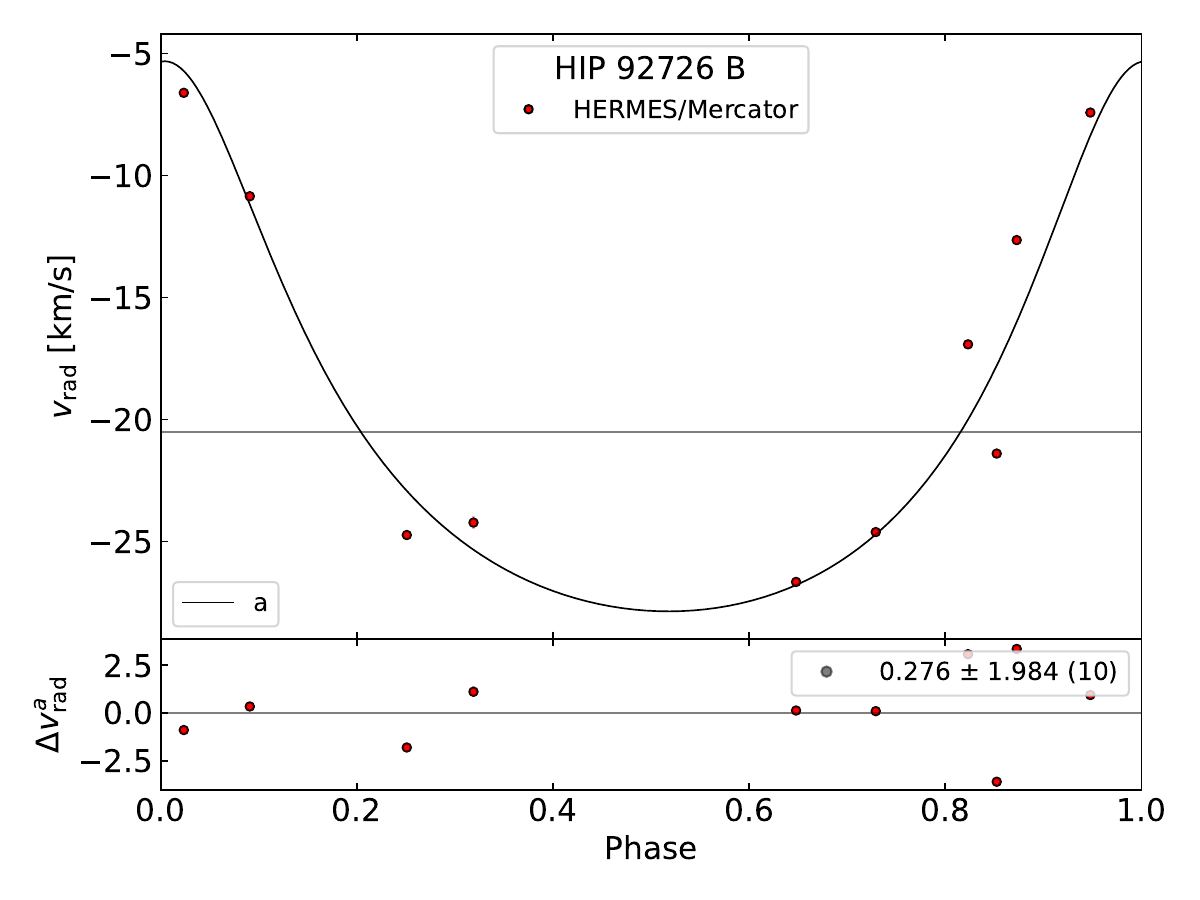}}
\resizebox{0.33\hsize}{!}{\includegraphics{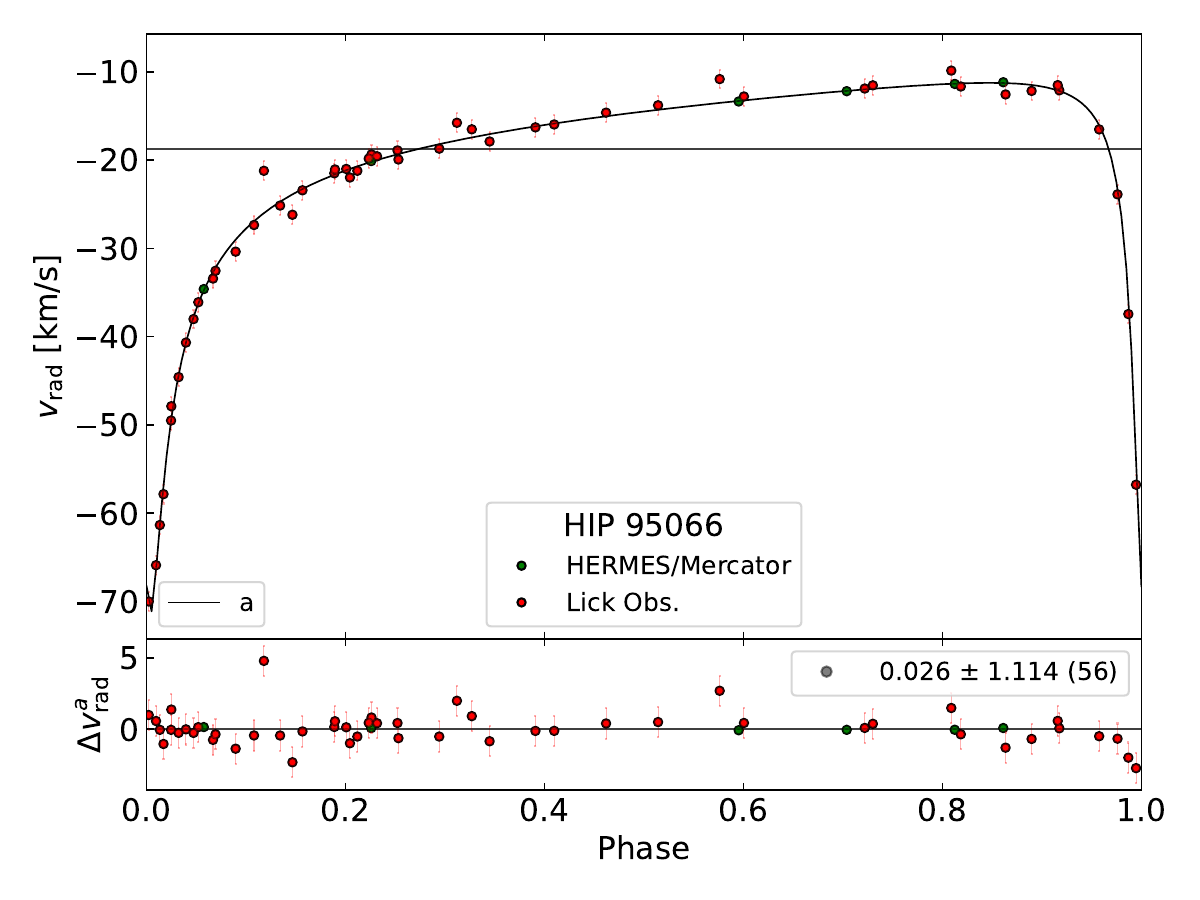}}
\resizebox{0.33\hsize}{!}{\includegraphics{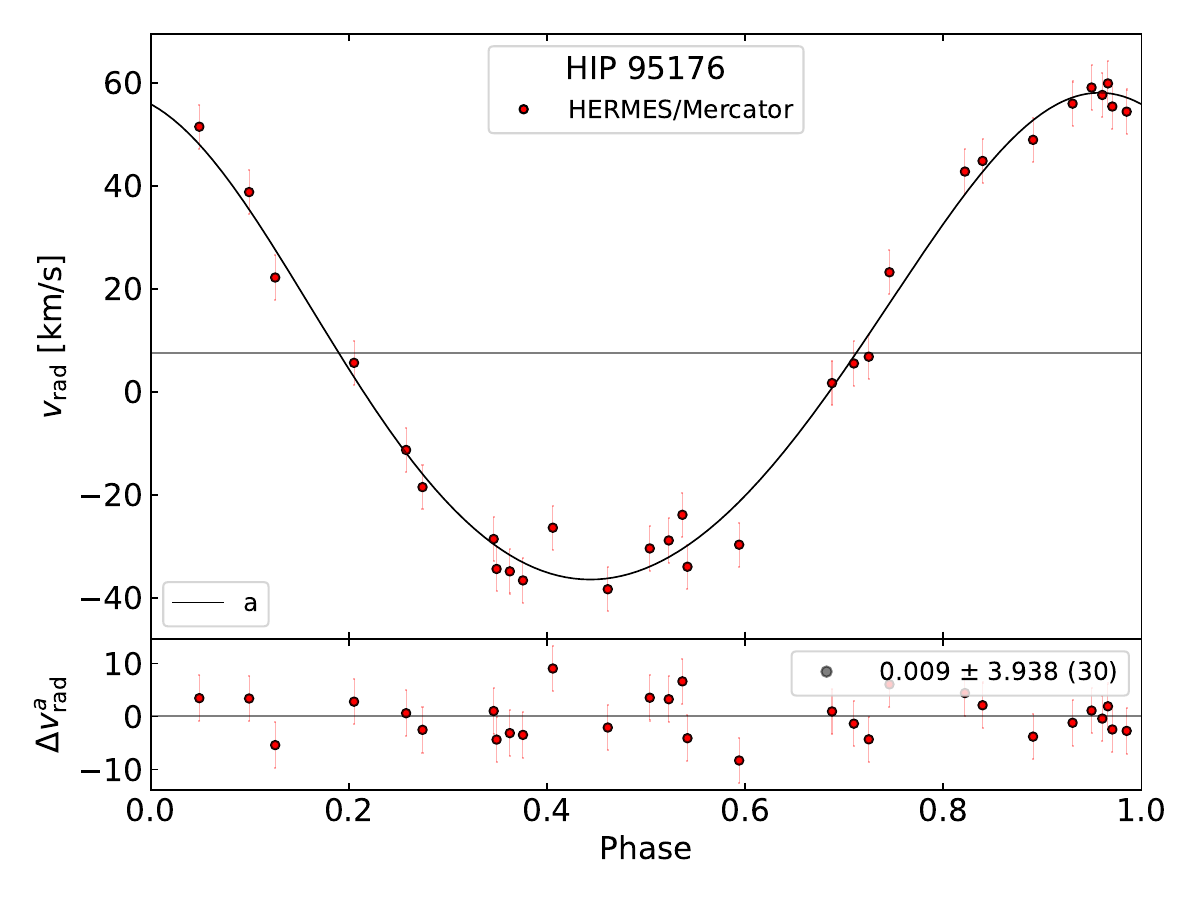}}\\
\resizebox{0.33\hsize}{!}{\includegraphics{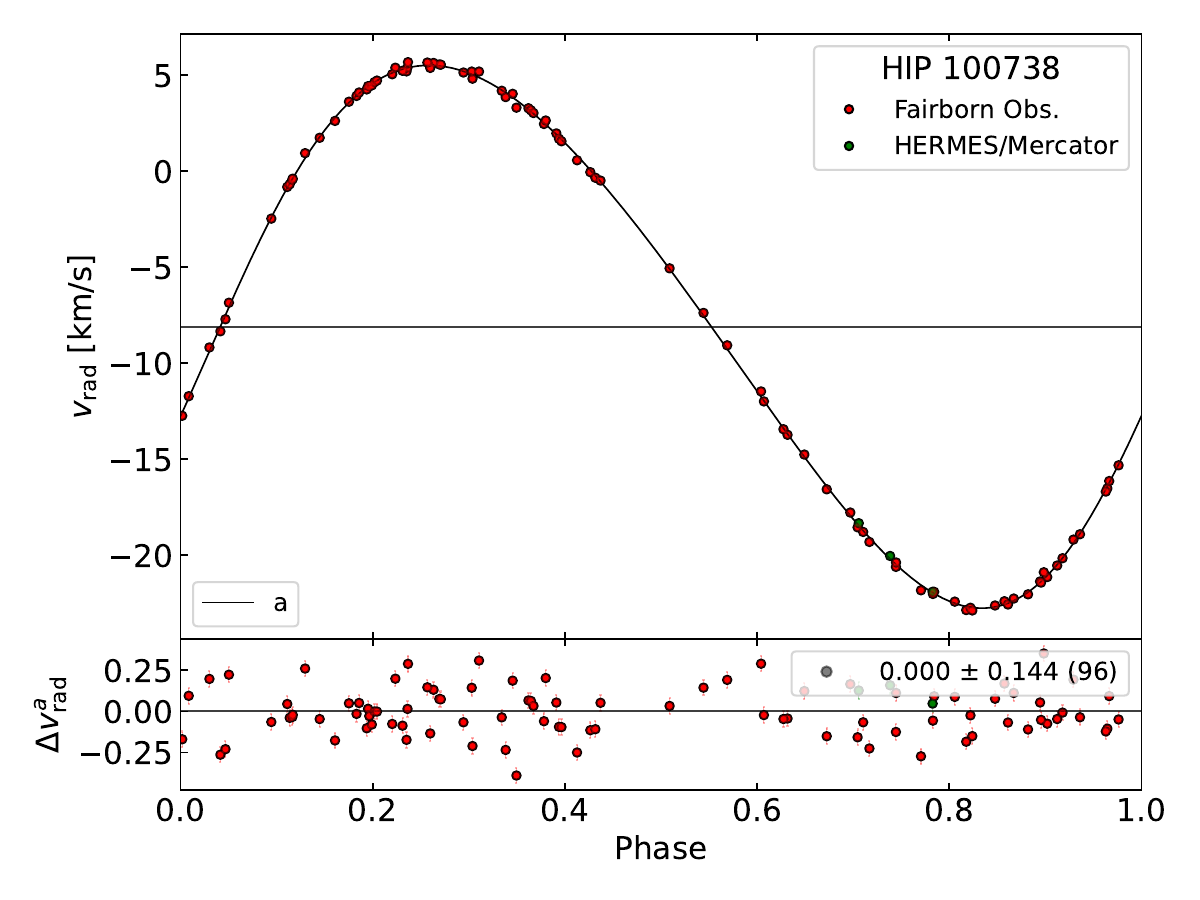}}
\resizebox{0.33\hsize}{!}{\includegraphics{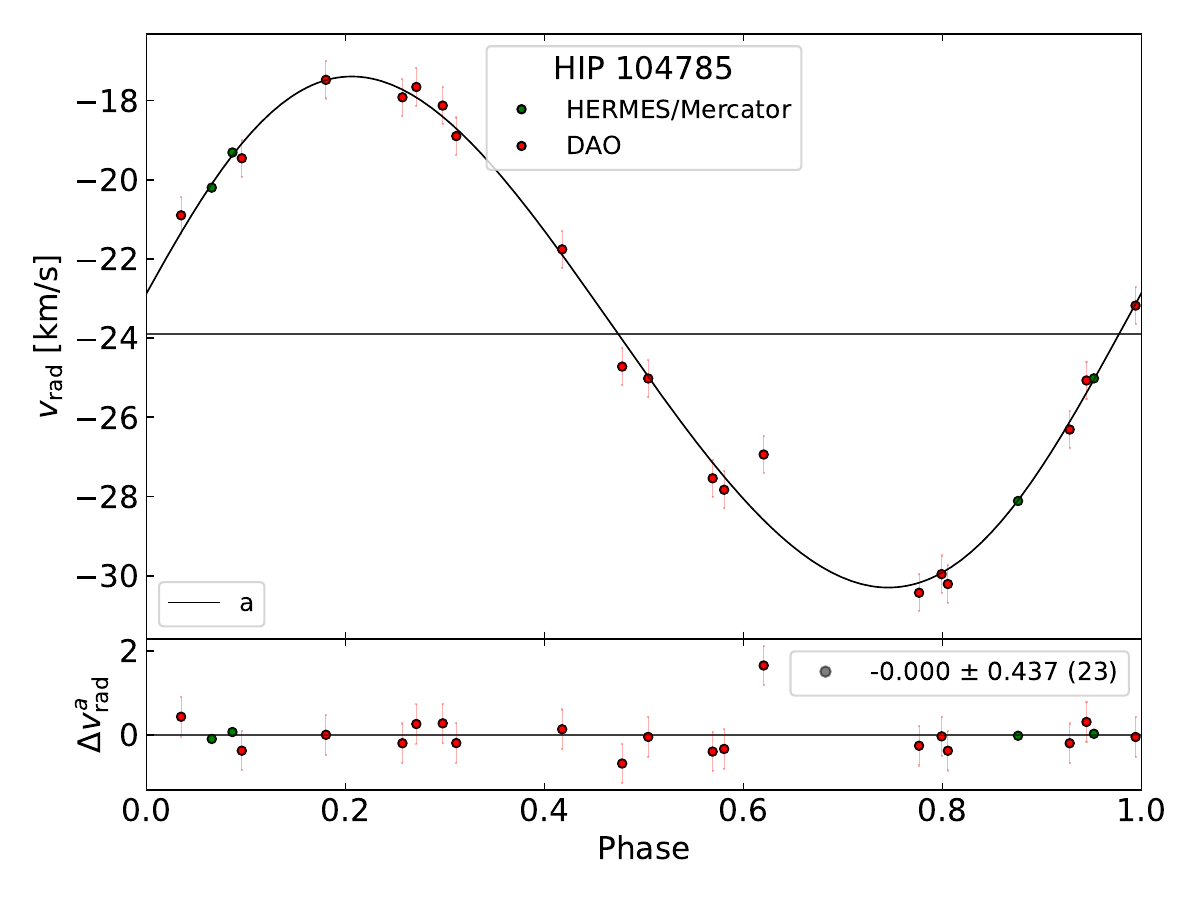}}
\resizebox{0.33\hsize}{!}{\includegraphics{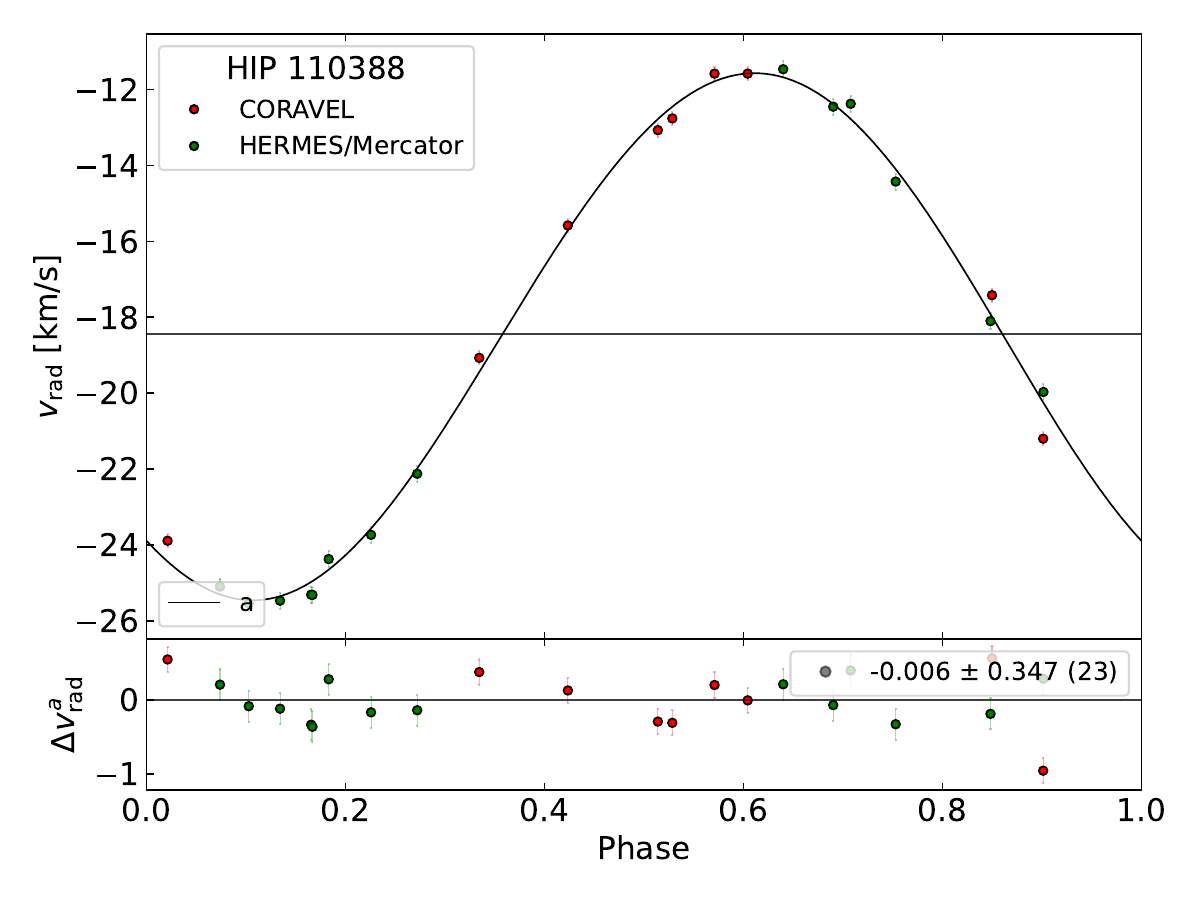}}\\
\caption{\label{fig:revorbits3} Continued.}
\end{figure*}

\addtocounter{figure}{-1}
\begin{figure*}[ht]

\resizebox{0.33\hsize}{!}{\includegraphics{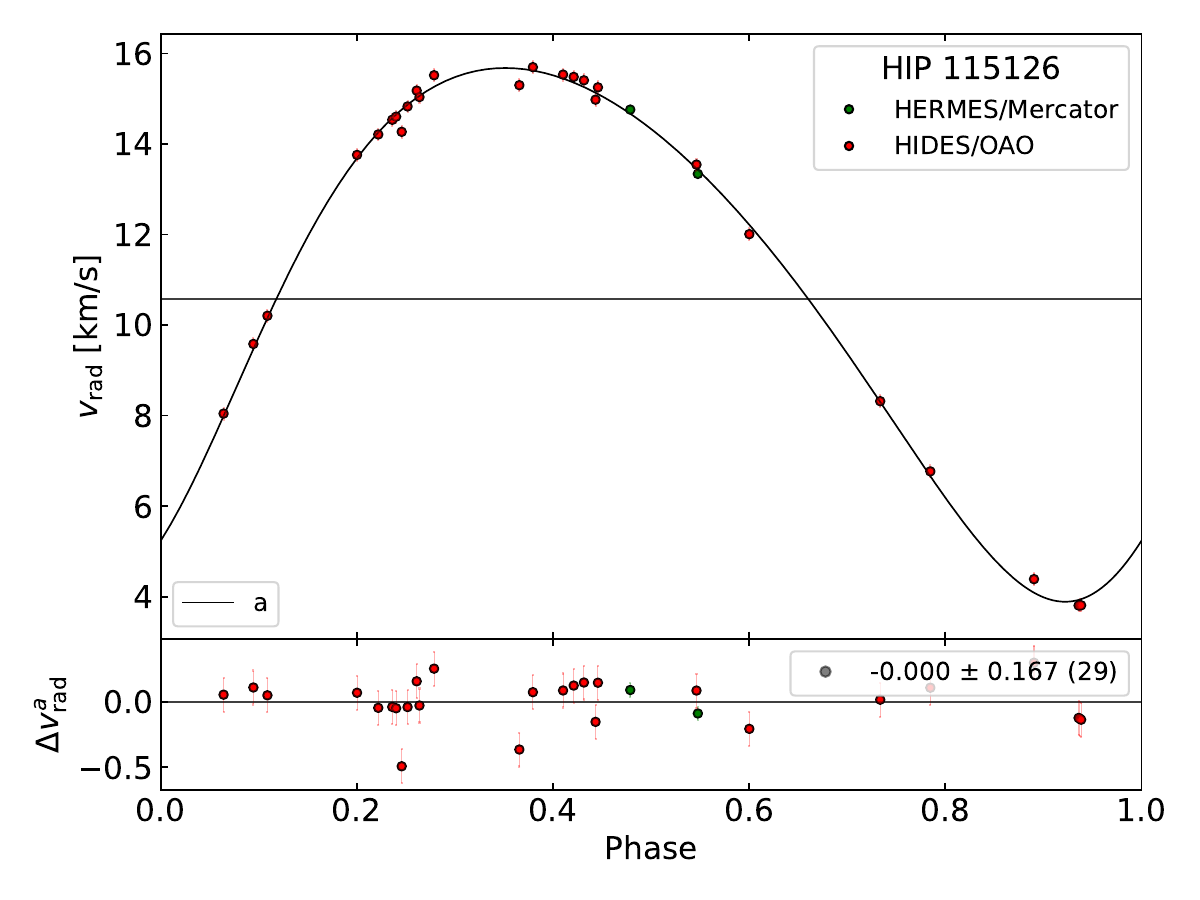}}
\resizebox{0.33\hsize}{!}{\includegraphics{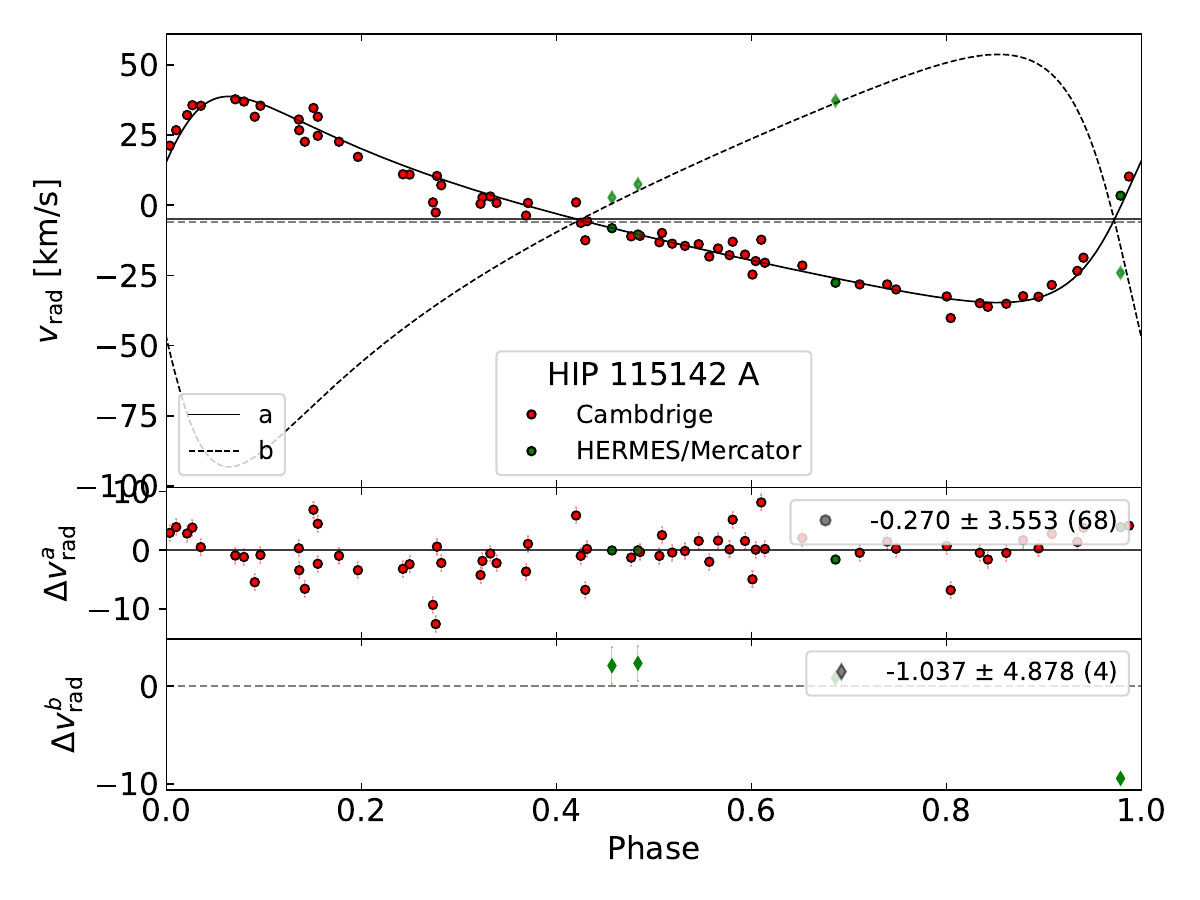}}
\resizebox{0.33\hsize}{!}{\includegraphics{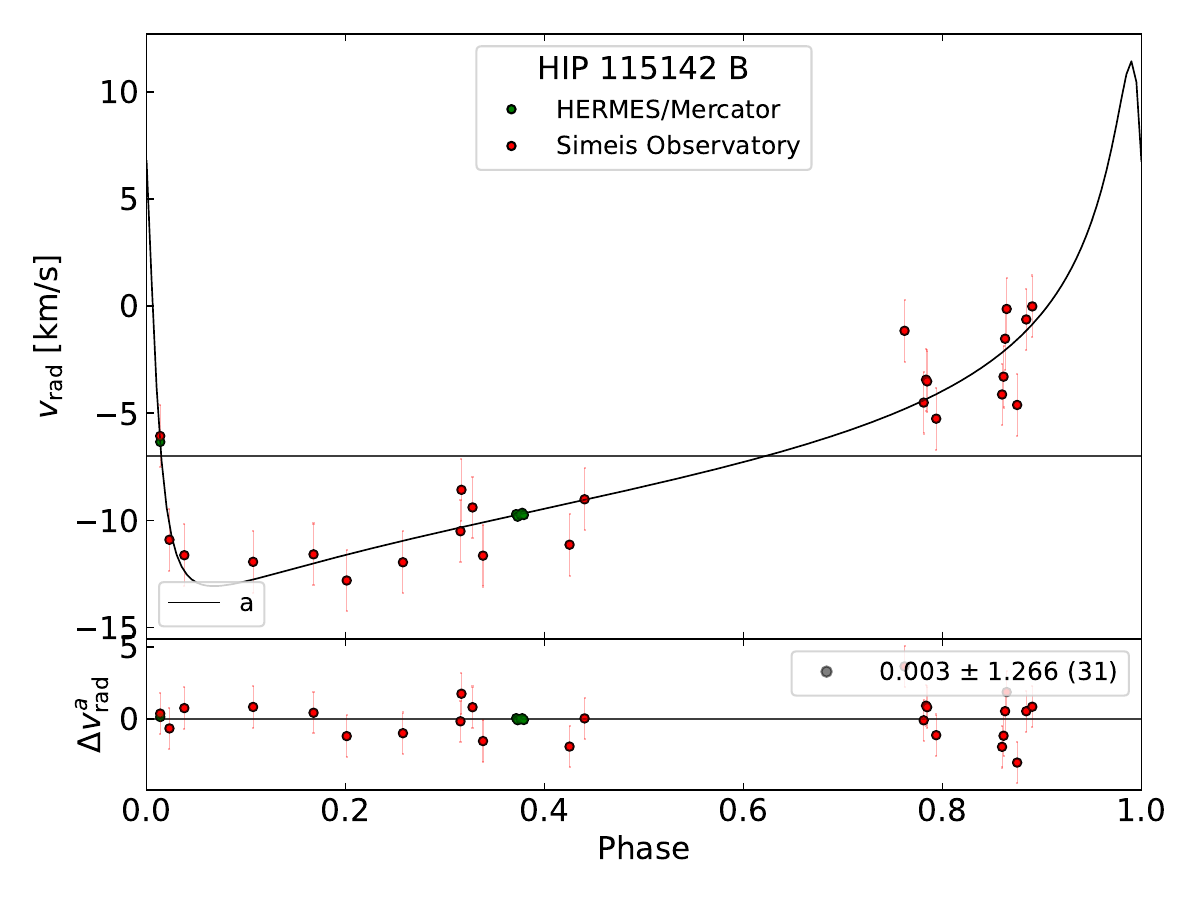}}\\
\caption{\label{fig:revorbits4} Continued.}
\end{figure*}

\noindent {\bf  HIP 183} (HD 224990, $\zeta$ Scl).   
The star is a member of the open cluster Blanco 1. The new combined SB9/HERMES spectroscopic orbit has more accurate orbital elements than the SB9 orbit, and has therefore been listed in Table~\ref{tab:revorbits_old_new}. The HERMES radial velocities of this B3/5V star were measured from the Balmer H$\alpha$ line. 
\medskip\\ 
\noindent {\bf  HIP 443} (HD 28, BC Psc).
Only some elements of the spectroscopic orbit \citep{Harper-1926} were adopted in the Hipparcos DMSA/O solution.  For that solution, the periastron time was left free and fitted.  The result does not match our refined spectroscopic value.  Even though a combined spectroscopic-astrometric solution passes all the statistical criteria, the large uncertainties on the semi-major axis, inclination and longitude of the ascending node would make that orbit useless. A spectroscopic orbit for this system has recently been published by \citet{Wang-2023}.
\medskip\\
\noindent {\bf  HIP 14124} (HD 18894).
The reference SB9 orbit from \citet{Griffin-1980} has a period very close to 1~yr (363~d), and its phase coverage has been largely improved when combined with the new HERMES measurements, resulting in reduced uncertainties for the combined orbit.
\medskip\\ 
\noindent {\bf  HIP 14157} (HD 18955, IR Eri).
A revised orbit including HERMES measurements has recently been published by \citet{Halbwachs-2016}. This star was also re-analyzed by \citet{2022AJ....163..220V}. Residuals are large ($\pm5$~km~s$^{-1}$) close to the extrema of the RV curves which   are not well constrained due to the high eccentricity of the orbit ($e = 0.76$). This system is actually one of the most eccentric of the sample with quite a short orbital period (43~d).
\medskip\\ 
\noindent {\bf  HIP 18216} (HD 24587, $\tau^8$ Eri).  This star is a slowly pulsating B star \citep{2000A&A...355.1015D}. With the Balmer line used here, the RV data points are less scattered and do not reveal the pulsation as seen in Fig.~4a of \citet{2000A&A...355.1015D}.  The orbit has been improved.
\medskip\\  
\noindent {\bf  HIP 22000} (HD~30050, RZ~Eri). This is a RS~CVn eclipsing system with a 39.3~d SB2 orbit from \citet{1988AJ.....96.1040P} in SB9. Extended studies of this system have been performed by \citet{1945ApJ...101..370C}, \citet{1988AJ.....96.1040P}, \citet{1992A&A...256..463B}, and \citet{1994JApA...15..165V}.

The primary star is a rapidly rotating A8-F0IV star and the secondary a G8-K2 IV-III star \citep{1992A&A...256..463B}. It is the secondary which shows signatures of stellar activity, in the form of photometric variations caused by spots \citep{1992A&A...256..463B} and Ca~II H \& K lines with an emission component \citep{1988AJ.....96.1040P}. The HERMES CCF indeed appears distorted by spots.  \citet{2008AJ....136.1736G} compiled radii and rotational velocities for the two components of the RZ~Eri system from previous studies \citep{1992A&A...256..463B,1993AJ....106.1200E,1994JApA...15..165V}, and suggest the following values: $R_A = 2.84$~R$_{\odot}$,  $V_{\rm rot,A} \sin i = 69 \pm 8$ ~\kms\  and $R_B = 6.94$~R$_{\odot}$,  $V_{\rm rot,B} \sin i = 12.3 \pm 1.2$~\kms.  The fast velocity of the A component is as well confirmed by the broad HERMES CCF. \citet{1992A&A...256..463B} and \citet{1994AcA....44...33S} showed that these velocities correspond to a faster than synchronous rotation for A, and a slower than synchronous rotation for B (at least under the assumption of aligned spin and orbital axes). 

Since \citet{1988AJ.....96.1040P} mentions that lines from the hot (A) component are well visible in the 380 -- 420~nm region, we used the HERMES violet orders to derive the velocities of that component.  The A star is moreover eclipsed at primary minimum (around spectroscopic phase 0.3) by the subgiant G8 star \citep{1945ApJ...101..370C,1992A&A...256..463B}. According to the latter authors, the primary minimum has a depth of 0.817~mag and occurs at spectroscopic phase 0.32 whereas the secondary minimum has a depth of 0.073~mag around phase 0.67. On one occasion (HJD 2\,457\,008.43) indeed, the broad CCF component is not visible, and the remaining CCF becomes 12\% deeper than usual (contrast of 20\% instead of 8\% in the visible band). This is consistent with an eclipse of the warm, rapidly-rotating component.

The HERMES orbit for HIP~22000 is more precise than the one obtained by  \citet{1988AJ.....96.1040P}. In particular, the semi-amplitudes $K_A$ and $K_B$ are more dissimilar than they were in Popper's orbit, leading to a mass ratio not so close to unity; more precisely $q = M_{A:F0IV} / M_{B:G8 IV-III} = K_B / K_A = 1.10$ (as compared to 1.04 for Popper's orbit). This system thus fulfils the Algol criterion: the most evolved component (B: G8 IV-III) is also the least massive. The presence of circumstellar matter through significant IR excess \citep{1990MmSAI..61...77B,2012MNRAS.427..343M} and reddening \citep{1992A&A...256..463B}, along with the rapid rotation of the more massive AV component, hint at a significant mass transfer  in recent times.  RZ~Eri could thus be considered an Algol with an IR excess \citep{2012MNRAS.427..343M}, a class seeked by \citet{2015A&A...577A..55D}.

RZ~Eri is moreover a visual binary (WDS~04438-1041) with a 0.6~arcsec separation, increasing to 0.7 arcsec from 1991 to 2014, according to {\it The Washington Visual Double Star Catalogue} \citep{2001AJ....122.3466M} and the Hipparcos Catalogue \citep{Hipparcos}, where HIP~22000 appears with flag C ('component solution') in the Double and Multiple Star Annex. A third peak could indeed be seen in the CCF, remaining stationary at 44~\kms, and  which very likely corresponds to the visual component.  Since that velocity value is close to the centre-of-mass velocity of the spectroscopic system, it is very likely that the visual component forms a physical triple with the SB2 system. The tertiary component, with a magnitude of $V=11.6$, has an estimated period of about 643 y from the {\it Multiple Star Catalogue}\footnote{\url{http://www.ctio.noirlab.edu/~atokovin/stars/}} \citep[MSC,][]{2018ApJS..235....6T}. \Gaia DR3 might have detected this companion since the image was seen as non-single in 17\% of the transits (\emph{i.e.}, \texttt{ipd\_frac\_multi\_peak}=0.17).
\medskip\\ 
\noindent {\bf  HIP 22701} (HD 31109, $\omega$~Eri). This system was declared to be a long-period ($P = 3057$~d) SB1 system by \citet{Abt-1965:a}, but the HERMES data neither confirm Abt's orbit, nor do they hint at orbital variations. The velocity measurements based on the Balmer H$\alpha$ line have a standard dispersion of 0.9~\kms\ (for a time span of 383~d, or 0.125 in Abt's orbital phase), as compared to the average uncertainty on one measurement of 0.33~\kms, just at the 3$\sigma$ level thus. The phase span covered by the HERMES observations would correspond to a maximum velocity variation of  $4\times 0.125 \times 18.1 = 9$~\kms\ if optimally located on the velocity curve of semi-amplitude 18.1~\kms, according to Abt's orbit.  Velocity measurements based on metallic lines are even more difficult since these lines appear to be severely rotationally-broadened, with $V_{\rm rot} \sin i$ of the order of 200~\kms\ (Royer et al. 2007 find 186~\kms)\nocite{2007A&A...463..671R}.  We thus conclude that the SB1 nature of HIP~22701 must be seriously questioned.
\medskip\\ 
\noindent {\bf  HIP 25912} (HD 36562). 
The new HERMES data points allow us to substantially improve the precision of the orbital elements, given the long orbital period involved (2204~d) which is now better covered by the available observations.
\medskip\\ 
\noindent {\bf  HIP 26563} (HD 37507 = d Ori). This is a special case. In just fifty years, that object went from a SB2 classification to likely single.  The orbit used to compute the residuals in Table~\ref{tab:ProgrammeStars} \citep[from][]{Abt-1965:a} was rejected ten years later \citep{Abt-1974}.  The HERMES data confirm that the \citeyear{Abt-1965:a} orbit is wrong.  The HERMES radial velocity scatter [$\sigma (O-C) = 0.73$~\kms\ as compared to $<\epsilon> = 0.50$~\kms; this value of the average uncertainty is much larger than the value mentioned earlier, because RV for this A star were measured from the Balmer H$\alpha$ line] is consistent with HIP~26563 being a single star. The \Gaia DR3 RV are also consistent with the star being single, since the \Gaia $p$-value of the RV $\chi^2$ is 25\%, far too large for a binary system.  
An orbital solution was nevertheless adopted both in the original and revised processing of the Hipparcos data.  Yet, the object fails at all the binary tests \citep{2003A&A...398.1163P} and the single-star fit is excellent, yielding $\varpi=20.80\pm1.00$~mas, $\mu_{\alpha*}=-14.64\pm0.66$~mas\,yr$^{-1}$, and $\mu_{\delta}=-49.80\pm0.53$ mas\,yr$^{-1}$. 

 These are essentially the same values as with the orbital model but with an improved precision. 
Nevetheless, the \Gaia DR3 parallax is significantly different ($\varpi=22.96\pm0.11$~mas, but with a very large value -- 1336~mas -- of the excess-noise standard deviation) and this star shows a proper motion anomaly \citep{2019A&A...623A..72K} and astrometric acceleration \citep{2021ApJS..254...42B}. It is worth noting that this star is a fast A4V rotator with $v\sin{i}\approx190$~\kms\ \citep{2007A&A...463..671R} which is high compared to the semi-amplitude of 28.6~\kms\ of \citet{Abt-1965:a} rejected orbit.
\medskip\\ 
\medskip\\
\noindent {\bf  HIP 31205} (HD 46407, HR CMa). This is a prototypical barium star \citep{1957ApJ...126..357B}. The uncertainties on all orbital parameters have been much reduced. Our combined astrometric-spectroscopic orbit (Table~\ref{tab:astrometry}) recovers the Hipparcos DMSA/O solution \citep[][note the negative value for the semi-major axis $a_0$  in the Hipparcos solution, associated with a 180$^{\circ}$ offset in the longitude of the node ${\Omega}$]{Hipparcos}.
\citet{1991Msngr..66...53J} claimed that this system might be eclipsing. However, for this to be the case, the inclination should be close to 90$^{\circ}$, whereas the astrometric solution yields $73^\circ\pm12^{\circ}$  (\Gaia DR3 NSS provides $i = 75.1^\circ\pm1.8^\circ$) so that the claim that this star is an eclipsing binary seems strange. 
In addition it appears to be a triple system according to MSC \citep{2018ApJS..235....6T}. The tertiary component, located at 0.161~arcsec, has a period of about 50~y. 
\Gaia DR3 adds an excess noise with a standard deviation of 2001~mas to the astrometric data of this star, which however does not seem related to the tertiary companion since \texttt{ipd\_frac\_multi\_peak}, an indicator that the \Gaia image has multiple components, is null.
\medskip\\ 
\noindent {\bf HIP 32426} (HD 49126). This star belongs to the open cluster M~41 (NGC 2287).
It is a complex system whose spectral assignments are uncertain. It 
is a visual double star (ADS 5437 AB) separated by about 1.2~arcsec and with a magnitude difference of 4 mag, one component at least being itself a spectroscopic binary. 
In the first place, \citet{1979PASP...91..636L} detected a composite spectrum (F8IV-V + B9.5V). Then \citet{2004AJ....127.2915P} 
has suggested K0II for Aa and B9 for Ab and no spectral type for B. We cannot decide whether the F8IV assignment from \citet{1979PASP...91..636L} 
holds for component B or whether it is another spectral-type determination for component Aa.
The Michigan catalogue \citep{1988mcts.book.....H} gives a spectral type B8III for the system as a whole, not being able to resolve the visual double.
\citet{2007A&A...473..829M} obtained a SB1 orbit for the late-type star (K0 II or F8 IV?), and this is the orbit listed in the SB9 catalogue (which quotes however the wrong spectral type B8III) which has been improved by the new HERMES measurements. 

With an angular size of 2.5 arcsec for the HERMES optical fibre on the sky, wider thus than the AB angular separation of 1.2~arcsec, the HERMES spectrograph thus potentially catches both components in the fiber. Their appearance is F8+B9 (or K0+B9) composite, without any possibility however to decide whether they correspond to the Aa-Ab or Aa-B pair.
Nevertheless we were only able to derive accurately the RV from the  late-type component because the cross-correlation function computed from a B5V template is very broad (FWHM of about 200~km/s) and weakly contrasted, and therefore forbids us from deriving accurate enough velocities to compute a spectroscopic orbit for the companion. 
\medskip\\
\noindent{\bf HIP 32467} (HD 49212). This star also belongs to M~41 and is a red giant \citep{2007A&A...473..829M}. The orbital elements (especially the period) have somewhat improved thanks to the addition of the HERMES measurements.
\medskip\\
\noindent{\bf HIP 33168} (HD 50730/50731). This star has been classified as composite (K1III + B) by the HD catalogue (K0 + A3) and subsequently by the Michigan Spectral Survey \citep{1999MSS}. 
\cite{2005Obs...125...81G} obtained a very densely sampled orbit for the binary system, using spectral lines from the cool component as sampled by the CORAVEL spectral templates. Our HERMES spectra were meant at detecting the hot component, but were unsuccessful at doing so. Interestingly,  
Griffin's orbit further points at a very large mass function of $1.82\pm0.03$~\Msun\ for the system. Based on simple arguments relating to the mass function, the spectral classes and their associated luminosities, Griffin concludes that it is very unlikely that the companion could be a single B star. A close pair of main-sequence stars (one possibly of B-type) appears much more likely. Griffin further notes that the cool component appears to be rapidly rotating ($V_{\rm rot} \sin i \sim 12$~\kms). The HERMES spectra yield  a more modest 5.9~\kms\ for the standard deviation of the Gaussian profile fitted to the CCF, after quadratically subtracting the 3~\kms\ intrinsic width of the spectrograph, but this value is nevertheless larger than average for K giants.
The orbital period has become more precise thanks to the addition of the recent HERMES measurements.
\medskip\\
\noindent{\bf HIP 34316} (BD-09$^\circ$1851). This system belongs to the open cluster NGC 2335. The new measurements improved the orbital parameters.\medskip\\
\noindent {\bf  HIP 34935} (HD 55684). 
This is a composite system with spectra K3 II + B7.5 (III) \citep{2003MNRAS.342.1271C}. The wide Balmer lines H$\delta$, H$\epsilon$, H$\zeta$, H7, and H8 shortwards of 410~nm visible in the HERMES spectra may  clearly be assigned to the B component. Unfortunately, the accuracy of the velocities derived from the broad Balmer lines from the B star is not good enough to derive a satisfactory spectroscopic orbit for the companion. 
\medskip\\
\noindent {\bf  HIP 36419} (HD 59435, V827 Mon). The new HERMES observations confirm the existing orbital solution for this SB2 system. The reduction of the uncertainty on the orbital period is not large enough to warrant the publication of a new orbital solution though. 
\medskip\\ 
\noindent {\bf  HIP 36652} (HD 60092). The new HERMES observations do not improve the existing orbital solution. Component  B ($V = 10.1$ at 0.4 arcsec) is not seen in the CCF. It has a period of about 144~y. \Gaia has seen component~B since \texttt{ipd\_frac\_multi\_peak}~=~29\%.
\medskip\\ 
\noindent {\bf  HIP 37041} (HD 60966). The 7 new HERMES measurements reduce the uncertainties on all orbital elements.
\medskip\\ 
\noindent {\bf  HIP 38217} (HD 63733). Two new HERMES measurements of this S-type star have reduced the uncertainty on the orbital period.
\medskip\\ 
\noindent {\bf  HIP 39198} (HD 65938). The HERMES velocities double the number of available measurements for this system with a period very close to 1~yr, and considerably reduce the uncertainties on the orbital elements. 
\medskip\\ 
\noindent {\bf  HIP 40944} (HD 70442). The 2 new HERMES observations reduce the uncertainty on the orbital period. There is no clear evidence for the presence of the AV companion from the violet Balmer lines of the HERMES spectrum.
\medskip\\
\noindent {\bf  HIP 42368} (HD 73451). The 2 new HERMES observations do not improve substantially the existing orbital solution, which is nevertheless listed in Table~\ref{tab:revorbits_old_new}.  The wide Balmer lines H$\delta$, H$\epsilon$, H$\zeta$, H7, and H8 shortwards of 410~nm visible in the HERMES spectra may  clearly be assigned to the A1 component. Unfortunately, the accuracy of the velocities derived from the broad Balmer lines from the A star is not good enough to derive a satisfactory spectroscopic orbit for the companion.
\medskip\\ 
\noindent {\bf  HIP 43041} (HD 74946). The 2 new HERMES observations reduce the uncertainty on the orbital period. The Balmer H$\epsilon$ line of the B component in this composite spectrum (K3III + B/A) is well visible in the HERMES spectra, but its FWHM of $\sim 7.5$~\AA\ makes it difficult to derive a precise RV.
\medskip\\
\noindent {\bf  HIP 43099} (BD$-$12$^\circ$2669). This star is a field blue straggler \citep{2001AJ....122.3419C}. The new HERMES measurements (based on the Balmer H$\alpha$ line) reduce the uncertainties on the orbital period.  We have no explanation for the large residuals along the orbit 
other than the fact that this seems common among low-metallicity binary stars \citep{2003AJ....125..293C,2016A&A...586A.158J}.
\medskip\\ 
\noindent {\bf  HIP 45527} (HD 79910, 23 Hya).
Two new HERMES measurements allow us to confirm the validity of this old orbital determination \citep{1928AnCap..10....8S}. However, the absence of any zero point offset cannot be ascertained, except for the fact that the recent HERMES measurements do not lead to a substantial revision of the orbital period or eccentricity, as could be expected in the presence of a zero-point offset.  Our combined astrometric/spectroscopic solution  (Table~\ref{tab:astrometry}) essentially reproduces the DMSA/O solution \citep{Hipparcos}. It forms with component B located at 1.4 arcsec a triple system according to MSC \citep{2018ApJS..235....6T}.
\medskip\\ 
\noindent {\bf  HIP 50006} (HD 88562). This is a barium star.
Three new HERMES measurements have reduced the uncertainty on the orbital period.
\medskip\\
\noindent {\bf  HIP 50796} (BD$-$09$^\circ$3055).
The 17 new HERMES observations have reduced the uncertainty on the orbital period.  The original Hipparcos reduction adopted an acceleration model even though combining astrometry and spectroscopy yields a valid astrometric orbit (Table~\ref{tab:astrometry}). A detailed analysis of this system (including estimates of the component masses) is provided by \citet{2006AJ....131.1022T} who concluded that the unseen companion, more massive than the primary, probably consists of two M-type stars, making this system a stellar triple.
\medskip\\ 
\noindent {\bf  HIP 51533, HIP 53717} (HD 91208, HD 95193). These are two barium stars.
The 2 new HERMES observations have reduced the uncertainty on the orbital period.  
\medskip\\ 
\noindent {\bf  HIP 55505} (HD 98800, TV Crt).  This is a pre-main-sequence quadruple 2+2 system (Aab - Bab), with B being a 10th magnitude star located 1.1 arcsec away from A. Aab and Bab are SB1 and SB2 systems, respectively.  The new HERMES observations of A have reduced the uncertainty on the orbital period.
The orbit of Bab has been much improved thanks to the higher-resolution HERMES observations: it is now much less noisy than the reference orbit but there is an offset of $-1.1\pm0.3$~\kms\ between the centre-of-mass velocities derived from Ba and Bb.  \cite{2019NatAs...3..230K} reported the existence of a protoplanetary/circumbinary disc in a polar configuration around Bab, and \cite{2021A&A...655A..15Z} recently provided a joint astrometric (PIONIER/VLTI) and spectroscopic orbit yielding the individual masses. 
\medskip\\ 
\noindent {\bf  HIP 57341} (HD 102171).
The two new HERMES observations confirm the reference orbit without improving its accuracy; it is therefore not listed in Table~\ref{tab:revorbits_old_new}. The wide Balmer lines H$\delta$, H$\epsilon$, H$\zeta$, H7, and
H8 shortwards of 410 nm visible in the HERMES spectra may
clearly be assigned to the A component, but again, the accuracy on the velocities is not good enough to derive a spectroscopic orbit for the companion.
\medskip\\
\noindent {\bf  HIP 57791} (HD 102928).
For HIP~57791, SB9 presently contains three orbits \citep{Ginestet-1985,Imbert-2005,Massarotti-2008} which, despite their similarities, are not consistent with each other within uncertainties.  The original Hipparcos processing to derive an astrometric orbit as well as the reprocessing by \citet{2005A&A...442..365J} were based on the spectroscopic solution from \citet{Ginestet-1985}.  
Our revised spectroscopic orbit (Table~\ref{tab:revorbits_old_new}) has been used to generate a new combined astrometric/spectroscopic orbit (Table~\ref{tab:astrometry}), 
These astrometric solutions have $\Omega$ differing however by 180$^{\circ}$ from the recent value obtained by \citet{Ren-2013}. The similarity between our astrometric solution and the Hipparcos one casts some doubts on the accuracy of the \citeauthor*{Ren-2013} solution.
\medskip\\ 
\noindent {\bf  HIP 59750} (HD 106516).
The combined astrometric-spectroscopic solution (Table~\ref{tab:astrometry}) reproduces the original DMSA/O solution derived from scratch (except for the 180$^{\circ}$ ambiguity on $\omega$, which could not be lifted in the absence of radial velocities).  A combined astrometric-spectroscopic solution was also provided by \citet{Ren-2013}, but with $\Omega$ differing by about 90$^{\circ}$ from our solution. The orbit is almost circular so $\omega$ is poorly constrained whereas $\Omega$ remains well defined.  Once again, our value is consistent with the Hipparcos result, pointing towards a problem in the  \citeauthor*{Ren-2013} solution.
\medskip\\ 
\noindent {\bf  HIP 63742} (HD 113449, PX Vir).
HIP~63742 is one of the rare Hipparcos objects for which the astrometric orbital solution was fitted from scratch (no reference spectroscopic orbit was used).  The resulting period was $231.2\pm2.0$~d which is substantially different from the 216~d period derived from the radial velocities.  When the astrometric fit takes advantage of the spectroscopic orbit, the resulting proper motion is in much better agreement with the Tycho-2 \citep{Hog-2000:a} value.
\medskip\\ 
\noindent {\bf  HIP 69929} (HD 125248, CS Vir). This star is a magnetic Ap star with CCFs distorted by spots.
Two new HERMES measurements allow us to improve the determination of the orbital period (although no zero-point offset was applied). The astrometric situation of HIP~69929 is a bit puzzling.  In the original Hipparcos reduction, an acceleration model was adopted, yielding a rather good fit.  In the revised reduction \citep{Hip2}, a single-star solution was adopted despite the bad fit (according to the statistical indicator which is part of the published solution).  Despite the different models used for computing these two solutions, the proper motion remains essentially the same.  This value is now carved in stone in UCAC-4 \citep{Zacharias-2013:a} even though it is based upon a questionable model.  
The \Gaia DR3 single-star solution has been listed in Table~\ref{tab:astrometry}  for comparison. This solution is very different from the Hipparcos single-star solution, so that the astrometric data for this system must definitely be processed with an orbital model. Although the \Gaia data definitely reveal extra noise ($\sigma_\varpi = 0.536$~mas, ruwe = 7.38, \emph{i.e.}, `reduced unit-weight error', and the standard deviation of the excess noise is 7037~mas), \Gaia DR3 was not yet processed with an orbital model for this star. 
\medskip\\ 
\noindent {\bf  HIP 69974} (HD 125337, $\lambda$ Vir).  A combined SB2 + IOTA orbit is reported by \citet{2007ApJ...659..626Z}, whereas none was seen in the Hipparcos Intermediate Astrometric Data  \citep{2005A&A...442..365J} because the two components have almost equal magnitudes. \citeauthor{2007ApJ...659..626Z} provide masses to better than 1.5\% for the two components.
The new HERMES observations confirm the reference orbit without improving its accuracy. HIP~69974 forms a triple system with component A being itself a binary. Components Aa and Ab are separated by less than 1 mas. The SB2 + IOTA orbit is for the AB pair. 
\medskip\\ 
\noindent {\bf  HIP 73007} (HD 131670).
This is a barium star studied by \citet{1990ApJ...352..709M}, with an orbit rederived by \citet{1998A&AS..131...25U}. Due to the large time span elapsed between HERMES and the latter CORAVEL measurements, the orbital period has become more precise.
\medskip\\ 
\noindent {\bf  HIP 75379} (HD 137052, $\epsilon$ Lib). 
The new HERMES measurements do not improve significantly the  \citet{2013AJ....145...41K} orbit, except that they provide the systemic velocity (not available from Kato's data obtained with an iodine cell).  The combined astrometric-spectroscopic orbit  \citep[already investigated by][]{2005A&A...442..365J} changes the longitude of the ascending node.  
\medskip\\
\noindent {\bf  HIP 75718} (HD 137763, BD$-$08$^\circ$3981).
The new HERMES measurements do not improve the  reference orbit.
The original Hipparcos accelerated solution was converted into an orbital one by \citet{2005A&A...442..365J}, and  we confirm that result.
The SB2 nature of this system \citep[as announced by][]{1992A&A...254L..13D} was not seen in the available HERMES spectra. 
It is the most eccentric system of the sample with $e=0.9733\pm0.0006$. \cite{2022AJ....163..118A} refined the orbit  and derived the masses of the components.
\medskip\\ 
\noindent {\bf  HIP 84402} (HD 155970, BD-14$^\circ$4585). 
The two HERMES measurements  allow us to improve the orbital elements.
The Hipparcos data of this system was originally processed with an acceleration model.  The combined astrometric/spectroscopic orbit yields a substantial revision of the proper motion  (Table~\ref{tab:astrometry}).
The comparison of the \Gaia DR3 astrometric parameters with the Hipparcos ones, as listed in Table~\ref{tab:astrometry}, reveals a great variability of the proper motion, indicative of an orbital motion.
\begin{figure}[t]
\vspace*{-4cm}
\includegraphics[width=\linewidth]{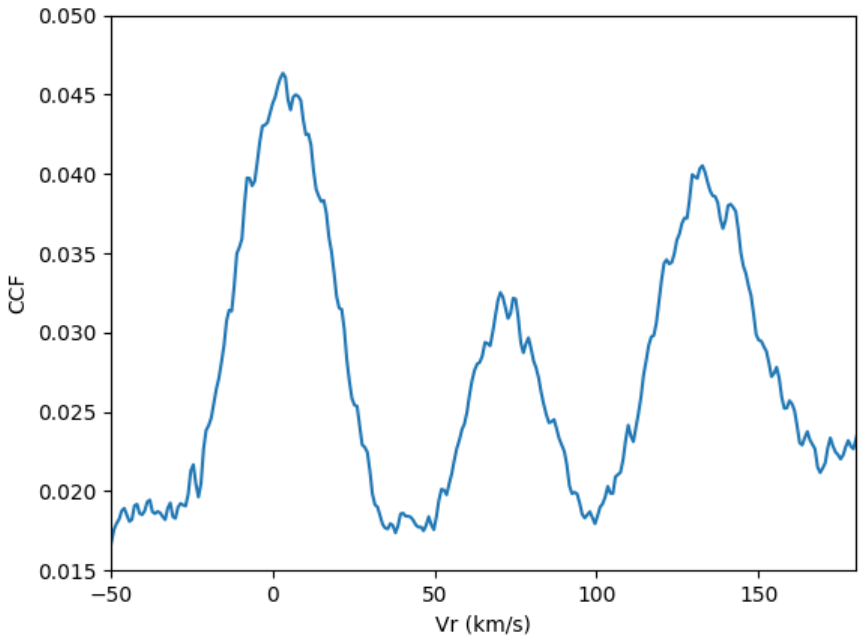}
\vspace*{-4cm}
\caption[]{\label{Fig:95176_CCF}
The CCF of  HIP~95176 on HJD 2\,456\,429.7 (phase 0.72 with the convention adopted on Fig.~\ref{Fig:stream}) in spectral order 85 ($\lambda\lambda 4205 - 4277$~\AA) with mask F0, showing the stellar F2 component (around velocity 0~\kms) and the two absorption features (around 70 and 130~\kms). 
}
\end{figure}
\medskip\\ 
\noindent {\bf  HIP 90135} (HD 169156, $\zeta$ Sct).  
Thanks to the 12 new HERMES data points, the spectroscopic orbit has been substantially improved with respect to the SB9 reference orbit. 
The combined solution provided in Table~\ref{Tab:astromSB9} was computed with HERMES measurements prior to 2016 only.
\medskip\\
\noindent {\bf  HIP 90692} (HD 170547, BD-05$^\circ$4675).
The 6 extra HERMES velocities bring no improvement to the existing orbit based on 67 measurements from \citet{1977Obs....97..173R}, especially since this G8 giant suffers from a significant velocity jitter which makes the derivation of the offset between the two RV systems inaccurate. Hence publishing a revised orbit for this star does not seem appropriate.
\medskip\\ 
\noindent {\bf  HIP 91751} (HD 172831, BD-07$^\circ$4670).
The 3 new HERMES measurements have been used to put  the  reference orbit  \citep{2013AJ....145...41K}, based on iodine-cell measurements, on the IAU velocity scale.  
The combined astrometric-spectroscopic orbit (Table~\ref{tab:astrometry}) confirms the original Hipparcos orbital model (already noted in Jancart et al. 2005). \nocite{2005A&A...442..365J}
\medskip\\ 
\noindent {\bf  HIP 92016} (HD 173282, BD-21$^\circ$5131).
There are a few new HERMES data points which do improve the reference orbit, which is important in prevision of the expected astrometric data that will give access to accurate masses. 
\begin{figure}[t]
\includegraphics[width=\linewidth]{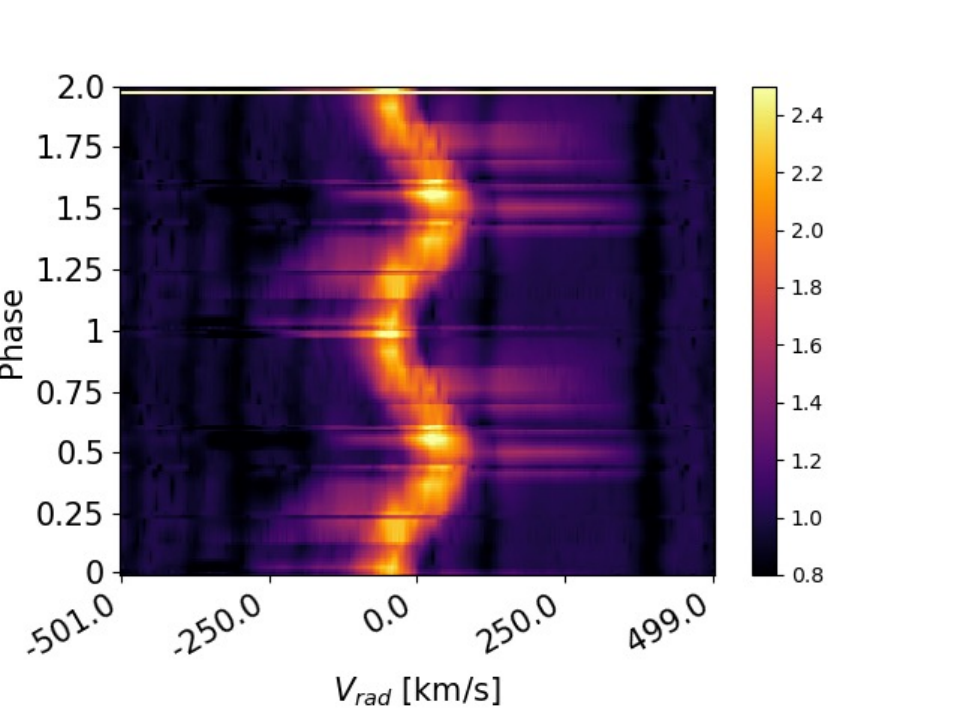}
\caption[]{\label{Fig:stream}
The spectral profile around the H$_\alpha$ line of  HIP~95176. The spectra have been corrected from the Doppler shift of the F component, which produces the lines at rest in the chosen frame (seen as dark vertical features). The H$\alpha$ line thus moves with the B companion. All spectra have been re-ordered according to their orbital phase, and interpolations between them have been performed to smooth the features. Phase 0 was set arbitrarily at JD~2455639.78, and roughly corresponds to quadrature, with the B star coming towards the observer. The color map has been normalized to unity at the continuum.
}
\end{figure}
\medskip\\
\noindent{\bf HIP 92726} (HD~175039, BD-05$^\circ$4798) is listed as a triple star in MSC \citep{2018ApJS..235....6T} with the inner pair Aa-Ab having a period of 50.51~d  
 \citep{2002AstL...28..393S} and the outer pair AB, which is the visual double star ADS~11791~AB (WDS~J18537-0533AB), has an orbital period of 258~yr and a separation on the order of 0.3 -- 1.1~arcsec \citep{1998ApJS..117..587H}. Therefore, both visual components fall in the HERMES input fiber. We collected 13 new HERMES spectra which reveal a double-peaked CCF. The main CCF component yields velocities in perfect agreement with the existing SB orbit   \citep{2002AstL...28..393S}. However, the second, fainter component shows a trend incompatible with the Ab motion since it does not vary in opposition with the Aa component. An orbit of period  13.3~d could be derived for that system (Fig.~\ref{fig:revorbits1}). The contrast of that second CCF component is about twice smaller than that of the main component, in agreement with the 1 mag difference between the A and B components of the visual pair, which has a semi-major axis of 0.63 arcsec, not resolved by the HERMES input fibre. Therefore, we must conclude that the second CCF component belongs to a Ba component rather than to the Ab component, so that the system must be considered quadruple rather than triple, with two short-period SB1 pairs orbiting each other in the 258~yr-long visual orbit.
\medskip\\
\noindent {\bf  HIP 95066} (26 Aql, HD~181391).
The new HERMES observations confirm the reference orbit \citep{1952ApJ...116..383F}. Since that orbit is old, the combined (HERMES + SB9) orbit is published to update the epoch of periastron passage. 
A combined astrometric/spectroscopic orbit for HIP~95066 was already obtained by \citet{2005A&A...442..365J}.  
\medskip\\
\begin{figure}
\includegraphics[width=\linewidth]{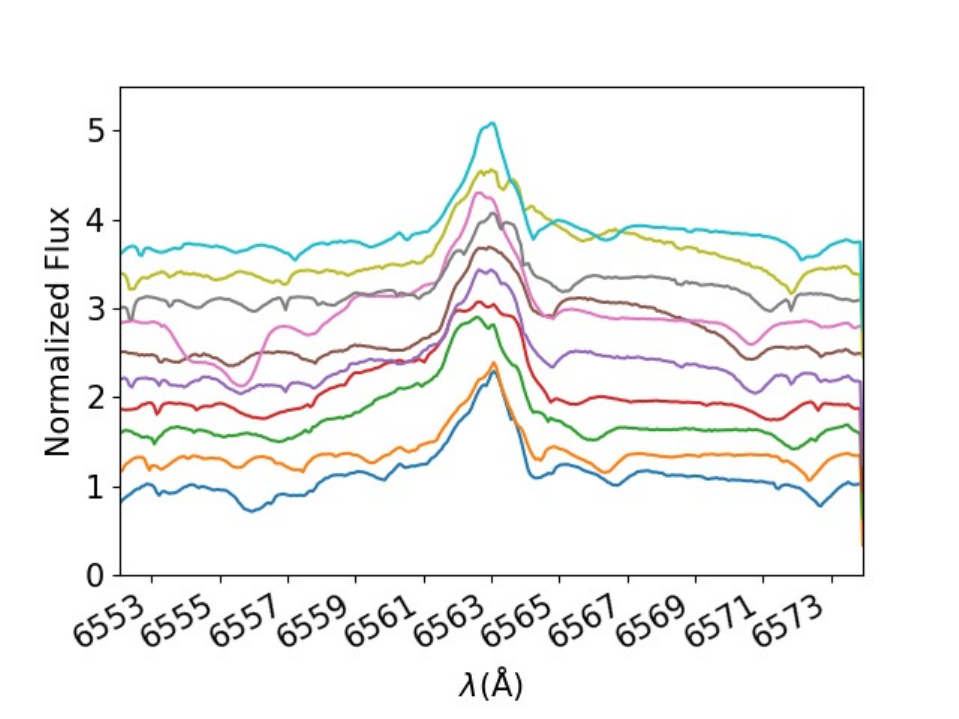}
\caption[]{\label{Fig:stream2}
Same as Fig.~\ref{Fig:stream} for the normalized spectral profiles, for 10 equidistant orbital phases (0., 0.1, 0.2,... 0.9, from top to bottom) extracted from Fig.~\ref{Fig:stream}. They have been shifted vertically by 0.33 units for the sake of clarity. The very extended red tail present at phase 0.5 is well visible, whereas the spectrum around phase 0.6 suddenly exhibits a broad absorption feature shortwards of H$\alpha$. }
\end{figure}
\noindent {\bf  HIP 95176} ($\upsilon$ Sgr, HD 181615,  HD 181616). This SB2 system consists in a F2 component dominating the visible spectrum, and a B star only visible in the far ultraviolet \citep{2006A&A...459..849K}. This semi-detached  system is a very rare hydrogen-deficient binary produced by a stable Roche-lobe overflow phase and possibly a progenitor of type-Ib and type-IIb supernovae \citep{1990MNRAS.247..400D, 2022ApJ...933...27H,2022Jeffery}.
The B star, despite not dominating the visible light, seems to be more massive than the F component. HIP~95176 seems therefore to bear some similarity to $\beta$~Lyr, because the component that would normally be called primary on the basis of its relative brightness could also be called secondary on the basis of its mass. There is a large offset (by 5.1~\kms) of the systemic velocities between the HERMES-only orbit and the 1929 orbit \citep{1931PAAS....6..278S}. Therefore, Table~\ref{tab:revorbits_old_new}  lists the HERMES-only orbit.
The HERMES data set  collects spectra with the highest resolution ever obtained for this system, and contains the largest number of spectra available so far, which make them very valuable \citep[compare with Table 1 of ][listing all data available so far for this system]{2006A&A...459..849K}. We have therefore performed an analysis going beyond the simple computation of the orbital elements of this system, which is complex. There are absorption features shifted by $\sim$+70 and $\sim$+130~\kms\ with respect to the stellar lines of the F component (Fig.~\ref{Fig:95176_CCF}), visible at all phases, a fact already noticed by \citet{1926PDAO....4....1P} and \citet{2006A&A...459..849K}. This must be due to re-absorption of light from the F component in material flowing towards that component at these velocities. Moreover, very wide emission tails are seen  alternating on each side  of the H$\alpha$ line (Figs.~\ref{Fig:stream} and \ref{Fig:stream2}), receding when the B star is moving in the hemisphere facing us, and approaching when the B star is moving in the hemisphere opposed to us. This behaviour may correspond to a stream of matter flowing from the B star towards the F star through the inner Lagrangian point, and emitting H$\alpha$ light. 
The H$\alpha$ stream moves along with the companion B star, with a relative velocity as large as 300~\kms. Spectral profiles are shown in a more conventional manner in Fig.~\ref{Fig:stream2}, to be compared with Figs.~3--8 of \citet{2006A&A...459..849K}.
\medskip\\ 
\noindent {\bf  HIP 104785} (HD 202020).  The new HERMES observations for this subgiant CH star confirm the reference orbit \citep{McClure-1997:a} and allow us to reduce the uncertainty on the orbital period.
\medskip\\
\noindent {\bf  HIP 110388} (KT Aqr, HD 212009).
The orbital solution exhibits an $O-C$ standard deviation of 0.35 km~s$^{-1}$, representative of the jitter existing on radial-velocities of giant stars \citep{2009A&A...498..627F}.
\medskip\\
\noindent {\bf  HIP 115126} (94 Aqr A, HD 219834A).
The new HERMES observations confirm the reference orbit \citep{2013AJ....145...41K} and do not improve its accuracy. It is nevertheless listed in Table~\ref{tab:revorbits_old_new} since the systemic velocity (not available from Kato's data obtained with an iodine cell) is now firmly established. This system was also recently analyzed by \cite{2022AJ....163..220V} using Bayesian inference. Improvement of orbits and determination of masses were also performed recently by \cite{2018AJ....156...85D}. An asteroseismic analysis using TESS photometric data shows that the rotation period of the primary is lower than expected from standard models of angular momentum evolution \citep{2020ApJ...900..154M}. This binary forms a triple with HIP~115125 (from MSC \citealt{2018ApJS..235....6T}) separated by 12.3 arcsec and on an orbit with a period estimated around 2700 y.
\medskip\\ 
\noindent {\bf  HIP 115142 B} (96 Aqr B, HD 219877B). Component B is a M3V star for which a preliminary orbit is provided by \citet{2007A&A...465..257T}.  The HERMES observations and the unique velocity from \citet{2015AJ....149....8T} confirm the existing SB1 orbit (Table~\ref{tab:revorbits_old_new}), without allowing us to substantially improve its quality. The eccentricity is still poorly constrained. 

\end{appendix}

\end{document}

%% file: tab/table1data.tex
183&2966&B4III&noRV&SB1&Bal&8&95&$-$1.30&1.10&1.71&(1)&R&CES& $-0.71\pm0.79$\\
443&1&K0IIIb&brightG&SB1&Arc&17&25&$-$1.24&4.71&1.25&(2)&R&Lick/Cape/DAO&$-0.21\pm0.25$\\
12390&135&F5V&--&SB2&F0&14,13&18,18&0.41,0.92&0.56,1.62&1.02,1.20&(3)&R&  & \phantom{}\tablefootmark{f}\\
14124&153&G0IV$-$V&noRV&SB2&G2&29,25&57,15&$-$2.06,$-$1.14&2.86,3.68&1.10,1.50&(4)&R & CamCOR & $-1.17\pm0.20$\\
14157&2373&K0V&noRV&SB2&Arc&16,14&31,31&0.36,$-$0.04&0.53,0.28&0.14,0.31&(5)&R&KPNO & $-2.00\pm0.06$\\
18216&2016&B5V&brightG&SB1&Bal&19&74&$-$2.30&16.71&1.30&(6)&R&CAT-CES & $-0.39\pm0.39$\\
22000&270&Am+sgG5&SB2&SB2&F0&16,15&27,10&$-$0.31,$-$1.60&3.36,5.12&5.19,2.83&(7)&R & Mount Wilson & $-1.15\pm0.20$\tablefootmark{g}\\
22701&285&F4III&&Single&Bal&6&14&7.28&9.80&9.58&(8)&\\
25912&2798&K1III&Acc7&SB1&Arc&6&12&0.36&0.06&0.37&(9)&R&CORAVEL & $+0.15\pm0.41$\\
26563&350&A4V&&Single&Bal&14&12&1.81&12.61&6.56&(8)&\\
28816&380&A1V+M6II&noRV&SB1&M4&46&55&$-$0.09&1.06&1.17&(10)&R& KPNO & $+0.05\pm0.16$\\
31205&407&G9.5IIIBa&astrom&SB1&Arc&8&68&0.51&0.42&0.40&(11)&R & CORAVEL & $+0.02\pm0.05$\\
32426&2884&F8IV-V+B9.5V\tablefootmark{a,e}&&SB2?\tablefootmark{a}&Arc&9&56&$-$0.16&0.19&0.86&(12)&R&CORAVEL&$+0.01\pm0.13$ \\
32467&2885&G9III&SB1/Acc9&SB1&Arc&3&37&$-$2.01&4.36&0.28&(12)&R&CORAVEL&$+0.07\pm0.06$\\
33168&2622&K1III+B\tablefootmark{b}&&SB1\tablefootmark{b}&Arc&2&71&$-$0.85&0.03&0.47&(13)&R&CORAVEL& $-0.85\pm0.10$\\
34316&2888&K0&nRV$<$10&SB1&Arc&14&22&$-$1.36&1.04&0.45&(12)&R&CORAVEL& $+0.07\pm0.11$\\
34935&2577&K3II+B7.5III&nRV$<$10&SBc\tablefootmark{e}&Arc&2&28&0.07&0.14&0.27&(14)&R&CORAVEL& $+0.25\pm0.17$\\
36419&3095&ApSrCrEu+F/G &&SB2&F0&2,2&78,78&$-$1.33,2.05&0.02,0.99&0.75,0.75&(15)&\\
36652&2765&Fm&&SB1&F0&6&29&0.02&0.19&0.34&(16)&\\
37041&2579&F5&SB1&SB1&F0&7&28&0.56&0.35&0.38&(14)&R & CORAVEL & $+0.20\pm0.08$\\
38217&1553&S3.5/3&Acc7&SB1&Arc&2&14&0.67&0.48&0.38&(17)&R&CORAVEL& $+0.14\pm0.19$\\
39198&2845&G5&astrom&SB1&Arc&22&17&0.42&0.72&0.43&(18)&R&CfA& $-0.01\pm0.09$\\
40944&2749&G8III+A3V&brightG&SB1&Arc&2&22&0.38&0.02&0.28&(19)&R&CORAVEL& $+0.36\pm0.12$\\
42368&2533&G+A1V&&SBc\tablefootmark{e}&Arc/Bal&2&100&$-$1.18&0.12&0.39&(20)&R& CamCOR &$-1.06\pm0.08$\\
43041&2750&K3III+B/A&brightG&SB1&Arc&2&15&0.11&0.51&0.32&(19)&R&CORAVEL& $-0.25\pm0.16$\\
43099&1708&A5&--&SB1&Bal&19&135&10.08&4.17&1.90&(21)&R&CfA& $-0.16\pm0.62$\\
45527&564&K2III&brightG&SB1&Arc&2&61&0.11&7.87&1.45&(22)&R&Cape& $+0.03\pm0.38$\\
50006&1588&K1IIIBa&&SB1&Arc&3&23&$-$0.20&0.49&0.44&(17)&R&CORAVEL& $+0.47\pm0.17$\\
50796&2610&K+MV&nRV$<$10&SB1&Arc&17&38&0.31&0.61&0.59&(23)&R&CfA& $+0.44\pm0.10$\\
51533&1564&KIIIBa&nRV$<$10&SB1&Arc&2&24&0.15&0.52&0.39&(17)&R&CORAVEL &  $+0.91\pm0.19$\\
53717&1565&G8Ib/IIBa&Acc7&SB1&Arc&2&18&0.20&0.39&0.26&(17)&R&CORAVEL & $+0.61\pm0.69$\\
55505A&2449&K4$-$5&noRV&SB1&Arc&23&152&0.18&7.16&1.14&(24)&R& CfA & $+0.11\pm0.19$\\
55505B&2450&K4$-$5&noRV&SB2+vis&Arc&8,16&152,152&$-$3.04,$-$0.29&10.33,9.54&1.91,3.90&(24)&R& CfA & $-0.38\pm0.01$\\
57341&2751&G8III+A&&SBc\tablefootmark{e}&Arc&2&13&0.13&0.10&0.34&(19)&\\
57791&693&K0III&astrom&SB1&Arc&6&9&0.81&1.19&0.17&(18)&R& CfA & $+0.01\pm0.07$\\
59750&1692&F9VFe$-$1.7CH$-$0.7&--&SB1&Arc&4&39&0.20&0.19&0.44&(21)&R& CamCOR & $+0.30\pm0.11$\\
63742&3198&G5V&--&SB1&Arc&3&42&$-$1.34&0.18&0.38&(25)&R&CamCOR & $+0.31\pm0.03$\\
69929&809&A0p&nRV$<$10&SB1&F0&2&39&$-$2.59&2.43&1.12&(26)&R&Stromlo & $-1.6\pm1.0$\\
69974&811&Am&noRV&SB2&F0&13,13&130,130&1.18,1.20&1.11,0.29&1.14,0.50&(27)&R& CfA & $+1.1\pm0.10$\\
73007&1566&K0IIIBa&Acc7&SB1&Arc&2&55&0.47&0.09&0.36&(17)&R&CORAVEL &  $+1.02\pm0.24$\\
73182&1475&M1.5V&--&SB2&Arc&8,7&75,10&0.11,0.07&0.13,0.29&0.81,0.92&(28)&R&ELODIE/Coralie & $+0.4\pm0.8$\\
75379&841&F4V&nRV$<$10&SB1&F0&3&42&0.64&0.05&0.08&(29)&R&DAO &  $+0.49\pm0.14$\\
75718&1638&G9V+M0V\tablefootmark{d}&--&SB2\tablefootmark{c}&Arc&3&97,10&0.57&0.07&0.30,2.02&(30)&\\
84402&2851&K1III&brightG&SB1&Arc&2&8&$-$1.51&0.70&0.03&(18)&R&CfA &  $-1.46\pm0.11$\\
90135&1046&K0III&nRV$<$10&SB1&Arc&12&7&0.34&0.30&0.003&(18)& R &CfA &  $-1.15\pm0.04$\\
90692&1063&G5&nRV$<$10&SB1&Arc&6&67&$-$0.21&3.23&1.00&(31)&\\
91751&1072&K0.5III&astrom&SB1&Arc&3&19&$-$1.76&0.07&0.01&(29)&R&Kato & $+31.83\pm0.02$\tablefootmark{i}\\
92016&1075&F5V&nRV$<$10&SB2&F0&5,2&65,14&$-$0.56,$-$0.25&0.48,0.57&1.80,0.66&(32)&R & KPNO & $-1.87\pm0.24$\\
92726&2059&G5&noRV&SB1&Arc&13&28&1.92&6.88&0.48&(33)&R&Tokovinin COR & $+0.86\pm0.21$\\
95066&1144&G8III$-$IV&brightG&SB1&Arc&6&51&$-$0.52&0.97&1.06&(34)&R& Lick & $-0.79\pm0.19$\\
95176&1147&B8p+F2pe&noRV&SB1&F0&30&&$-$4.84&8.97&$-$&(35)&R &  & \phantom{}\tablefootmark{f}\\
100738&1239&K3V&&SB1&Arc&3&93&0.13&0.05&0.15&(36)& \\ 
104785&2149&G0&Acc9&SB1&Arc&4&19&0.22&0.12&0.47&(37)&R& DAO & $+0.60\pm0.37$\\
110388&2957&M0III&&SB1&Arc&14&9&$-$1.41&1.49&0.17&(38)&R& CORAVEL & $-0.23\pm0.15$\\
111170&1479&F7V&nRV$<$10&SB2&G2&6,6&139,28&0.66,0.70&4.22,7.47&1.11,2.93&(3)&R& CORAVEL & $-0.32\pm0.10$\\
115126&1438&G5IV+K2V&--&SB1&Arc&2&27&0.56&0.07&0.02&(29)&R& HIDES & $+0.65\pm0.08$\\
115142A&2860&F3IV$-$V&nRV$<$10&SB2&F0ROT&4,4&64&$-$3.30,10.93&8.72,21.75&1.35&(39)&R& CamCOR & $-0.83\pm0.28$\\
115142B\tablefootmark{h}&2859&M3V&nRV$<$10&SB1&F0&6&25&$-$1.31&3.78&1.44&(40)&R& Simeis & $+0.29\pm0.54$\\
\hline

%% file: tab/combinedorbits.tex
183 & 40 & 6 & 0.29 & 0.03 & 1742 & 11 & 56842 & 23 & 4.3 & 0.8 & 11.9 & 0.4 &  &  & 1.7 & \\
 & 44 & 7 & 0.32 & 0.04 & 1740 & 22 & 53381 & 37 & 5.3 & 0.3 & 12.4 & 0.6 &  &  & 1.7 & \\
443 & 329.4 & 0.6 & 0.249 & 0.002 & 72.9407 & 0.0005 & 46089.0 & 0.1 & -6.42 & 0.02 & 16.52 & 0.03 &  &  & 1.1 & \\
 & 338 & 5 & 0.27 & 0.02 & 72.93 &  & 22530.3 & 0.8 & -6.6 & 0.3 & 16.4 & 0.3 &  &  & 1.3 & \\
12390\tablefootmark{c} & 215 & 1 & 0.249 & 0.008 & 970 & 3 & 57012 & 3 & 15.61 & 0.02 & 6.39 & 0.04 & 6.1 & 0.3 & 0.069 & 0.26\\
 & 221 & 3 & 0.230 & 0.006 & 968.3 & 0.7 & 45391 & 4 & 15.5 & 0.2 & 5.6 & 0.3 & 8.7 & 0.3 & 1.0 & 1.2\\
14124 & 301.1 & 0.4 & 0.689 & 0.003 & 362.955 & 0.004 & 57588.1 & 0.1 & 14.3 & 0.2 & 24.5 & 0.1 & 29.8 & 0.2 & 0.98 & 1.1\\
 & 300 & 1 & 0.688 & 0.005 & 363.1 & 0.1 & 43068.4 & 0.4 & 16.2 & 0.2 & 24.0 & .2 & 30.0 & .6 & 1.1 & 1.5\\
14157 & 174.63 & 0.06 & 0.760 & 0.001 & 43.32050 & 0.00004 & 49754.50& 0.004 & 29.33 & 0.06 & 54.24 & 0.06 & 60.7 & 0.1 & 0.44 & 0.51\\
 & 174.6 & 0.2 & 0.76 & .002 & 43.3206 & 0.0005 & 51487.5 & .01 & 30.75 & 0.09 & 54.3 & 0.3 & 60.4 & 0.3 & 0.14 & 0.31\\
18216 & 106 & 3 & 0.21 & 0.01 & 454.0 & 0.1 & 50497 & 3 & 20.1 & 0.3 & 22.3 & 0.3 &  &  & 1.9 & \\
 & 106 & 5 & 0.18 & 0.02 & 459 & 4 & 50954 & 7 & 20.5 & 0.2 & 21.7 & 0.4 &  &  & 1.3 & \\
22000 & 136.1 & 0.8 & 0.371 & 0.005 & 39.2826 & 0.0003 & 46274.1 & 0.1 & 41.7 & 0.2 & {52.6\tablefootmark{(a)}} & 0.5 & 48 & 1 & 3.8 & 3.7\\
 & 132 & 3 & 0.35 & 0.02 & 39.2824 & {Fixed} & 35470.6 & 0.3 & 43 & 1 & 51 & 1 & 49 & 1 & 5.2 & 2.8\\
22000.DR3 & 322.4&0.4&  0.349 &  0.002 &  39.284 &  0.002 &  58218.12 &  0.03 &  41.24 &  0.08 &  54.1 &  0.3 &   52.3 &  0.1 & \\
25912 & 107 & 17 & 0.17 & 0.04 & 2208 & 15 & 57765 & 180 & 1.6 & 0.5 & 5.2 & 0.2 &  &  & 0.31 & \\
 & 111 & 44 & 0.15 & 0.06 & 2221 & 44 & 48956 & 304& 1.5 & 0.4 & 5.2 & 0.3 &  &  & 0.37 & \\
28816 & 52 & 11 & 0.012 & 0.002 & 260.38 & 0.01 & 50762 & 8 & 18.69 & 0.16 & 22.15 & 0.06 &  &  & 1.0 & \\
 & 203.7 & 0.4 & 0.024 & 0.005 & 260 & 2 & 48528.8 & 0.4 & 18.7 & 0.1 & 21.3 & 0.2 &  &  & 1.2 & \\
31205 & 140 & 34 & 0.006 & 0.003 & 457.20 & 0.03 & 53707 & 40 & -2.93 & 0.02 & 8.93 & 0.04 &  &  & 0.4 & \\
 & 74 & 32 & 0.013 & 0.008 & 457.4 & 0.1 & 47677 & 41 & -3.45 & 0.05 & 9.03 & 0.07 &  &  & 0.4 & \\
32426 & 122.0 & 0.6& 0.352 & 0.004 & 1337.6 & 0.2 & 49541 & 2 & 23.23 & 0.13 & 18.3 & 0.1 &  &  & 0.95 & \\
 & 121 & 2 & 0.37 & 0.01 & 1338 & 2 & 44188 & 6 & 23.4 & 0.1 & 18.2 & 0.2 &  &  & 0.86 & \\
32467 & 4 & 1 & 0.585 & 0.005 & 1221.8 & 0.2 & 52836 & 3 & 23.44 & 0.04 & 8.23 & 0.08 &  &  & 0.32 & \\
 & 1.4 & 1.2 & 0.575 & 0.007 & 1212.6 & 2.6 & 44273 & 3 & 23.34 & 0.05 & 8.2 & 0.1 &  &  & 0.28 & \\
32467.DR3&5.8& 1.0 &0.56 & 0.01 & 1200 & 33 & 57727.0 & 1.4 & 23.56 & 0.07 & 8.04 & 0.08\\
33168 & 98.6 & 0.5 & 0.458 & 0.003 & 1540.9 & 0.2 & 46301.3 & 1.4 & 21.23 & 0.08 & 25.3 & 0.1 &  &  & 0.63 & \\
 & 98.8 & 0.6 & 0.459 & 0.004 & 1540.9 & 0.9 & 50925 & 2 & 22.1 & 0.07 & 25.3 & 0.1 &  &  & 0.47 & \\
34316 & 49 & 33 & 0.008 & 0.006 & 300.37 & 0.03 & 47380 & 28 & 21.69 & 0.05 & 7.57 & 0.06 &  &  & 0.43 & \\
 & {$-$} & {$-$} & 0 & {Fixed} & 300.8 & 0.2 & 44933 & 2 & 21.8 & 0.1 & 7.9 & 0.2 &  &  & 0.45 & \\
34935 & 212.2 & 1.2 & 0.216 & 0.004 & 1353.5 & 0.7 & 53940 & 3 & 16.6 & 0.2 & 18.10 & 0.08 &  &  & 0.27 & \\
 & 212.1 & 1.3 & 0.217 & 0.005 & 1353 & 1 & 47173 & 6 & 16.4 & 0.06 & 18.09 & 0.08 &  &  & 0.27 & \\
37041 & 230.8 & 0.5 & 0.154 & 0.002 & 601.38 & 0.04 & 47999 & 1 & 32.79 & 0.03 & 20.87 & 0.07 &  &  & 0.4 & \\
 & 234 & 3 & 0.152 & 0.005 & 601.5 & 0.3 & 48004 & 4 & 32.62 & 0.09 & 21.1 & 0.1 &  &  & 0.38 & \\
 37041.DR3 & 227.9 & 1.4 & 0.158 & 0.002 & 601.7 & 0.5 & 57615.6 & 2.1 & 33.21 & 0.05 & 21.00 & 0.06\\
38217 & 172 & 8 & 0.23 & 0.03 & 1155 & 1 & 56410 & 26 & 2.0 & 0.2 & 6.1 & 0.2 &  &  & 0.28 & \\
 & 168 & 9 & 0.23 & 0.03 & 1161 & 9 & 45991 & 43 & 1.9 & 0.1 & 6.1 & 0.2 &  &  & 0.38 & \\
39198 & 107.0 & 0.2 & 0.520 & 0.001 & 365.70 & 0.03 & 56752.65 & 0.07 & -4.18 & 0.01 & 9.57 & 0.02 &  &  & 0.25 & \\
 & 108 & 3 & 0.52 & 0.02 & 365.4 & 0.6 & 53827 & 1 & -4.2 & 0.2 & 9.6 & 0.2 &  &  & 0.43 & \\
40944 & 294.0 & 0.7 & 0.718 & 0.009 & 1212.1 & 0.3 & 53340.5 & 0.9 & -4.40 & 0.1 & 19.8 & 0.1 &  &  & 0.28 & \\
 & 294 & 1 & 0.72 & 0.01 & 1212 & 1 & 49704 & 3 & -4.8 & 0.1 & 19.7 & 0.6 &  &  & 0.28 & \\
42368 & 235.5 & 0.7 & 0.248 & 0.002 & 284.741 & 0.006 & 55299.3 & 0.5 & 33.83 & 0.07 & 20.24 & 0.06 &  &  & 0.92 & \\
 & 238 & 1 & 0.248 & 0.004 & 284.73 & 0.03 & 49890.7 & 0.8 & 34.87 & 0.07 & 20.2 & 0.1 &  &  & 0.39 & \\
43041 & 105 & 1 & 0.41 & 0.01 & 1496 & 2 & 55497 & 5 & 14.81 & 0.13 & 14.1 & 0.2 &  &  & 0.31 & \\
 & 105 & 2 & 0.41 & 0.02 & 1494 & 5 & 51006 & 9 & 15.0 & 0.2 & 14.1 & 0.3 &  &  & 0.32 & \\
43099 & 320 & 29 & 0.05 & 0.03 & 388.4 & 0.3 & 47664 & 30 & 42.7 & 0.6 & 9.6 & 0.3 &  &  & 1.9 & \\
 & 303 & 22 & 0.08 & 0.03 & 384 & 1 & 47255 & 22 & 41.9 & 0.3 & 9.5 & 0.3 &  &  & 1.9 & \\
45527 & 88 & 7 & 0.27 & 0.04 & 916.4 & 0.1 & 32313 & 14 & -7.7 & 0.3 & 9.9 & 0.3 &  &  & 1.4 & \\
 & 92.3 &  & 0.29 &  & 922 &  & 18549.2 &  & -7.7 &  & 1 &  &  &  & 1.45 & \\
50006 & 349 & 5 & 0.21 & 0.02 & 1450 & 1 & 47208 & 22 & 12.2 & 0.2 & 7.0 & 0.2 &  &  & 0.38 & \\
 & 353 & 8 & 0.20 & 0.02 & 1445 & 9 & 45782 & 36 & 11.8 & 0.1 & 7.0 & 0.1 &  &  & 0.44 & \\
50796 & 316.5 & 0.2 & 0.600 & 0.003 & 570.70 & 0.05 & 55781.3 & 0.2 & 25.09 & 0.03 & 20.42 & 0.08 &  &  & 0.57 & \\
 & 314.0 & 0.8 & 0.611 & 0.005 & 571.0 & 0.5 & 52355 & 1 & 24.9 & 0.1 & 20.8 & 0.2 &  &  & 0.59 & \\
51533 & 71 & 8 & 0.18 & 0.02 & 1773 & 5 & 54429 & 53& 1.1 & 0.2 & 5.0 & 0.1 &  &  & 0.35 & \\
 & 79 & 10 & 0.17 & 0.02 & 1754 & 13 & 45628 & 54 & 0.19 & 0.08 & 5.1 & 0.1 &  &  & 0.39 & \\
53717 & 283.8 & 7.5 & 0.14 & 0.02 & 1651 & 8 & 55995 & 48 & -6.9 & 0.7 & 5.36 & 0.12 &  &  & 0.25 & \\
 & 283 & 8 & 0.14 & 0.02 & 1654 & 9 & 46084 & 38 & -7.32 & 0.07 & 5.4 & 0.1 &  &  & 0.26 & \\
55505A & 65 & 3 & 0.47 & 0.02 & 264.53 & 0.06 & 55618 & 2 & 12.92 & 0.16 & 6.7 & 0.1 &  &  & 1.1 & \\
 & 64 & 2 & 0.48 & 0.02 & 262.2 & 0.5 & 48738 & 2 & 12.8 & 0.1 & 6.8 & 0.2 &  &  & 1.1 & \\
55505B & 108.9 & 0.9 & 0.783 & 0.005 & 314.77 & 0.08 & 52487 & 1 & 5.6 & 0.1 & 22.9 & 0.3 & 28.4 & 0.4 & 1.9 & 3.7\\
 & 110 & 1 & 0.78 & 0.06 & 315.2 & 0.4 & 48710 &  & 5.7 & 0.1 & 22.5 & 0.4 & 27.0 & 0.6 & 1.9 & 3.9\\
57791 & 113 & 2 & 0.234 & 0.003 & 490.36 & 0.06 & 53128 & 2 & 15.46 & 0.03 & 13.19 & 0.03 &  &  & 0.2 & \\
 & 102 & 3 & 0.220 & 0.007 & 490 & 1 & 53114 & 4 & 15.5 & 0.1 & 12.5 & 0.2 &  &  & 0.17 & \\
59750 & 196 & 15 & 0.05 & 0.01 & 844.3 & 0.2 & 49271 & 35 & 4.6 & 0.1 & 7.9 & 0.1 &  &  & 0.39 & \\
 & 199 & 19 & 0.05 & 0.02 & 844 & 1 & 47589 & 43 & 4.36 & 0.09 & 7.9 & 0.1 &  &  & 0.44 & \\
63742 & 294.1 & 1.6 & 0.254 & 0.005 & 216.48 & 0.03 & 57525.2 & 0.9 & -0.72 & 0.06 & 12.87 & 0.08 &  &  & 0.46 & \\
 & 295 & 2 & 0.261 & 0.007 & 216.48 & 0.06 & 53845 & 1 & -0.66 & 0.07 & 13.0 & 0.1 &  &  & 0.38 & \\
69929 & 86 & 9 & 0.23 & 0.04 & 1627 & 1& 37969 & 35 & -10.6 & 0.2 & 7.1 & 0.2 &  &  & 1.1 & \\
 & 82.4 &  & 0.21 &  & 1607 &  & 31475 &  & -10.9 &  & 7.5 &  &  &  & 1.1 & \\
69974 & 254 & 2 & 0.051 & 0.002 & 206.709 & 0.002 & 49959.1 & 0.8 & -7.1 & 0.1 & 25.59 & 0.05 & 27.23 & 0.06 & 1.6 & 0.6\\
 & 272.3 & 0.5 & 0.061 & 0.004 & 206.732 & 0.004 & 53070.3 & 0.3 & -8.05 & 0.05 & 24.8 & 0.2 & 27.31 & 0.07 & 1.1 & 0.5\\
73007 & 46 & 5 & 0.17 & 0.01 & 2941 & 8 & 8139 & 42 & -24.3 & 0.2 & 5.2 & 0.1 &  &  & 0.4 & \\
 & 50 & 5 & 0.16 & 0.01 & 2930 & 12 & 46405 & 45 & -25.13 & 0.05 & 5.16 & 0.08 &  &  & 0.36 & \\
73182 & 128.3 & 0.2 & 0.758 &  0.001 & 308.86 & 0.01 & 50270.86 &0.04 & 28.8 & 0.8 & 18.23 & 0.06 & 27.2 & 0.1 & 0.63 & 0.4\\
 & 127.6 & 0.4 & 0.764 & 0.002 & 308.81 & 0.01 & 51197.3 & 0.1 & 28.74 & 0.02 & 18.47 & 0.07 & 27.4 & 0.1 & 0.61 & 0.6\\
75379 & 337.7 & 0.8 & 0.672 & 0.008 & 226.944 & 0.005 & 42244.8 & 0.3 & -8.85 & 0.06 & 14.0 & 0.2 &  &  & 1.7 & \\
 & 339.8 & 0.2 & 0.665 & 0.001 & 226.944 & 0.002 & 53593.00 & 0.04 & -9.4 &  & 14.18 & 0.04 &  &  & 0.078 & \\
84402 & 122.1 & 0.9 & 0.430 & 0.003 & 2392 & 5 & 53201 & 3 & -8.81 & 0.08 & 8.95 & 0.03 &  &  & 0.07 & \\
 & 119 & 1 & 0.413 & 0.004 & 2349 & 8 & 53191 & 4 & -7.41 & 0.04 & 9.11 & 0.07 &  &  & 0.03 & \\
90135 & 269.2 & 1.3 & 0.0636 & 0.0006 & 2325.3 & 0.8 & 54157.5 & 8.3 & -6.53 & 0.04 & 7.00 & 0.02 &  &  & 0.40& \\
 & 314 & 2 & 0.068 & 0.001 & 2303.3 & 0.9 & 54441 & 11 & -5.017 & 0.019 & 7.636 & 0.033 &  &  & 0.003 & \\
91751 & 75.5 & 0.1 & 0.2017 & 0.0005 & 485.20 & 0.02 & 57480.6 & 0.3 & 25.75 & 0.03 & 9.646 & 0.006 &  &  & 0.03 & \\
 & 75.5 & 0.2 & 0.2019 & 0.0006 & 485.21 & 0.02 & 54084.3 & 0.2 & 27.505 & 0 & 9.648 & 0.006 &  &  & 0.011 & \\
92016 & 291.5 & 0.3 & 0.731 & 0.001 & 33.1610 & 0.0001 & 46671.58 & 0.01 & -3.05 & 0.24 & 51.0 & 0.1 & 69.2 & 0.2 & 1.9 & 0.7\\
 & 290.8 & 0.2 & 0.726 & 0.001 & 33.1607 & 0.0003 & 45079.90 & 0.03 & -2.7 & 0.1 & 49.8 & 0.2 & 69.5 & 0.3 & 1.8 & 0.66\\
92726A & 134.9 & 0.6 & 0.585 & 0.004 & 50.5155 & 0.0009 & 52688.73 & 0.04 & -16.6 & 0.2 & 22.0 & 0.2 &  &  & 0.46 & \\
 & 135.6 & 0.8 & 0.592 & 0.005 & 50.514 & 0.004 & 50011.5 & 0.1 & -17.5 & 0.1 & 22. & .2 &  &  & 0.48 & \\
92726B & 356.7 & 0.8 & 0.350 & 0.004 & 13.29 & 0.01 & 57626.80 & 0.04 & -20.53 & 0.05 & 11.28 & 0.06 &  &  & 2.1 & \\
95066 & 153.2 & 0.3& 0.840 & 0.003 & 266.528 & 0.003 & 30221.9 & 0.1 & -18.78 & 0.06 & 30.1 & 0.3 &  &  & 1.1 & \\
 & 152.7 &  & 0.8 &  & 266.544 &  & 33420.2 &  & -18 &  & 29.9 &  &  &  & 1.1 & \\
{95176\tablefootmark{(b)}} & 18 & 20 & 0.07 & 0.029 & 138.2 & 0.2 & 55644 & 8 & 7.5 & 0.98 & 47 & 1.1 &  &  & 4 & \\
 & 16.8 &  & 0.06 &  & 138.0 &  & 19644 &  & 13.3 & 0 & 49.1 &  &  &  &  & \\
104785 & 279 & 18 & 0.06 & 0.01 & 2068 & 4 & 47120 & 100 & -23.9& 0.3 & 6.4 & 0.1 &  &  & 0.45 & \\
 & 279 & 19 & 0.08 & 0.04 & 2064 & 10 & 47122 & 105 & -24.5& 0.1 & 6.5 & 0.2 &  &  & 0.47 & \\
110388 & 92 & 22 & 0.025 & 0.015 & 795.5 & 0.4 & 48025 & 49 & -18.6 & 0.1 & 6.90 & 0.07 &  &  & 0.35 & \\
 & 70 & 6 & 0.17 & 0.04 & 792 & 1 & 53529 & 15 & -17.7 & 0.2 & 6.6 & 0.2 &  &  & 0.17 & \\
111170 & 172 & 1 & 0.373 & 0.007 & 632.47 & 0.06 & 56641 & 1 & -9.5 & 0.1 & 11.34 & 0.08 & 16.8 & 0.1 & 1.2 & 3.6\\
 & 172 & 2 & 0.38 & 0.01 & 630.1 & 0.3 & 43995 & 2 & -9.7 & 0.1 & 11.4 & 0.2 & 21.0 & 0.6 & 1.1 & 2.9\\
115126 & 215 & 2 & 0.174 & 0.007 & 2302 & 2 & 58402 & 14 & 10.58 & 0.06 & 5.89 & 0.08 &  &  & 0.17 & \\
 & 212.1 & 0.4 & 0.162 & 0.002 & 2298 & 3 & 53748 & 3 & 9.9 & 0 & 6.025 & 0.008 &  &  & 0.021 & \\
115142A & 292.1 & 0.8 & 0.500 & 0.005 & 21.23870 & 0.00002 & 43886.40 & 0.03 & -4.9 & 0.3 & 36.7 & 0.2 & 73.5 & 0.64 & 3.3 & 0.85\\
 & 293.4 & 1.2 & 0.515 & 0.008 & 21.2387 & 0.0001 & 53825.6 & 0.1 & -5.1 & 0.2 & 36.1 & 0.4 &  &  & 1.4 & \\
115142B & 51 & 33 & 0.80 & 0.18 & 657.6 & 0.2 & 57352 & 4 & -7 & 1 & 12 & 5 &  &  & 1.3 & \\
& 59 & 11 & 0.7 &  & 660 & 4 & 51429 & 7 & -7.2 & 0.4 & 8 & 1 &  &  & 1.4 & \\
\hline

%% file: tab/astromOrbits.tex
31205 & 8.2 & 0.9 & -15.8 & 0.9 & 6.0 & 0.8 & 3.3 & 0.5 & 73 & 12 & 307 & 14\\
31205.HIP& 8.3 & 0.9 & -16.0 & 0.9 & 6.2 & 0.8 & -3.9 & 0.9 & 74 & 12 & 134 & 13\\
31205.DR3& 7.22	& 0.07 & -15.72 & 0.08 & 4.96 & 0.09 &  2.87 &	0.06 & 75 & 2 &  135	& 2  \\
\\
45527 & 11.4 & 0.9 & 24 & 1 & 4.3 & 0.8 & 9.2 & 0.7 & 94 & 5.6 & 284 & 6\\
45527.HIP& 11.7 & 1.0 & 25 & 3 & 4.3 & 0.8 & 10.6 & 2.8 & 92.6 & 4.7 & 283 & 5  \\
\\
50796 & 21 & 3 & -75 & 2 & -26 & 2 & 18 & 3 & 81 & 12 & 169 & 8\\
50796.HIP& 29 & 3 & -80 & 2 & -22 & 2 &  \\
\\
57791 & 11 & 1 & 0.1 & 0.8 & 0.6 & 0.6 & 6.8 & 0.7 & 76 & 7 & 112 & 5\\
57791.HIP& 13 & 1 & 0.9 & 0.8 & 0.9 & 0.6 & 7.7 & 1.2 & 87 & 6 & 108 & 5\\
57791.DR3& 10.6 & 0.3 & 0.0 & 0.2 &	-0.5 &	0.1 & 6.6 & 0.2 &   73.7 &   0.6 & -71.8 &   0.5 \\
\\
59750 & 44.7 & 0.8 & 32.5 & 0.7 & -1013 & 0.5 & 27.6 & 0.6 & 75 & 2 & 192 & 2\\
59750.HIP& 44 & 1 & 32.0 & 0.7 & -1012.4 & 0.8 & 29.0 & 0.9 & 74 & 2 & 13 & 3 \\
\\
63742 & 44 & 1 & -188 & 1 & -224.7 & 0.9 & 12.1 & 0.7 & 115 & 6 & 198 & 7\\
63742.HIP& 45 & 1 & -190 & 1 & -220 & 1 & 11 & 2 & 126 & 11 & 33 & 14 \\
\\
69929 & 11.3 & 0.8 & -58 & 1 & -33 & 2 & 11.9 & 0.9 & 82 & 7 & 262 & 15\\
69929.HIP& 11.1 & 0.9 & -65.6 & 0.9 & -35.4 & 0.6\\
69929.DR3&11.5 & 0.5 &-52.2 & 0.5 &-46.0&0.5 &  \\
\\
75379 & 30 & 1 & -72 & 1 & -153.5 & 0.7 & 9 & 1 & 51 & 9 & 214 & 7\\
75379.HIP& 31 & 1 & -69 & 2 & -152.7 & 0.8 & 8 & 2 & 49 & 19 & 145 & 16 \\
\\
75718 & 49 & 1 & 75 & 1 & -357.8 & 0.9 & 38 & 1 & 59 & 2 & 93 & 4\\
75718.HIP& 50 & 1 & 73 & 1 & -363.4 & 0.8 & 38.6 &1.0 &60.3 & 1.8& 95.8& 3.5\\
\\
84402 & 11 & 1 & -10 & 3 & -19 & 1 & 23 & 2 & 122 & 5 & 183 & 8\\
84402.HIP& 11 & 1 & -19 & 1 & -5.9 & 0.9 & \\
84402.DR3& 10.2 & 0.3 & -8.1 & 0.3 & -12.6 & 0.2 &\\
\\
90135 & 16.3 & 0.8 & 41 & 4 & 41 & 4 & 23 & 2 & 90 & 13 & 224 & 6\\
90135.HIP& 17.1 & 0.9 & 51.9 & 0.8 & 52.3 & 0.6&\\
\\
91751 & 13.2 & 0.8 & -6.8 & 0.6 & -8.9 & 0.5 & 6.5 & 0.5 & 59 & 5 & 293 & 7\\
91751.HIP& 13.4 & 0.9 & -6.8 & 0.7 & -9.0 & 0.5 & 6 & 1 & 57 & 8 & 297 & 8\\
91751.DR3& 12.7& 0.1& -5.71 &	0.09 & -8.95 &	0.05 & 6.33 & 0.18 &  61.9 &  0.9 & -81 &   5 \\
\\
95066 & 21.2 & 0.8 & 113.5 & 0.8 & 44.9 & 0.5 & 8.8 & 0.5 & 76 & 9 & 130 & 7\\
95066.HIP& 21.2 & 0.8 & 113.4 & 0.8 & 44.8 & 0.6 & 7.7 & 1.6 & 71 & 13 & 133 & 9\\
\hline